\documentclass{article}

\usepackage{amsmath,amssymb,amsfonts,amsthm}
\usepackage[all]{xy}
\usepackage{color}

\usepackage{rotating}
\usepackage{fullpage}

\renewcommand{\(}{\begin{equation}}
\renewcommand{\)}{\end{equation}}
\newcommand{\bea}{\begin{eqnarray}}
\newcommand{\eea}{\end{eqnarray}}

\newcommand{\Z}{{\mathbb Z}}

\theoremstyle{plain}

\newtheorem{proposition}{Proposition}[subsection]

\newtheorem{theoremapp}{Theorem A.\!\!}
\newtheorem{propositionapp}{Proposition A.\!\!}

\theoremstyle{definition}

\newtheorem{remark}{Remark}[subsection]
\newtheorem{observation}{Observation}[subsection]

\newtheorem{definitionapp}{Definition A.\!\!}

\newtheorem{remarkapp}{Remark A.\!\!}

\numberwithin{equation}{subsection}

\begin{document}

\title{
Multiple M5-branes, String 2-connections,\\ 
and 
7d 
nonabelian 
 Chern-Simons theory
}

\author{Domenico Fiorenza, Hisham Sati, Urs Schreiber}
\maketitle

\begin{abstract}
  The worldvolume theory
  of coincident M5-branes is expected to contain a nonabelian 2-form/nonabelian gerbe
  gauge theory that is a higher analog of self-dual Yang-Mills theory.
  But the precise details  -- in particular the global moduli / 
  instanton / magnetic charge structure -- have remained elusive.
  Here we deduce from anomaly cancellation a natural candidate for the 
  holographic dual of this nonabelian 2-form field, under $\mathrm{AdS}_7/\mathrm{CFT}_6$ duality. 
  We find this way a 7-dimensional nonabelian Chern-Simons theory of 
  String 2-connection
  fields, which, in a certain higher gauge, are given locally by non-abelian 2-forms with values
  in an affine Kac-Moody Lie algebra.
  We construct the corresponding action functional on the entire smooth moduli
  2-stack of field configurations, thereby defining the theory globally,
  at all levels and with the full instanton structure, which is nontrivial
  due to the twists imposed by the quantum corrections. 
  Along the way we explain
  some general phenomena of higher nonabelian gauge theory that we need.

\end{abstract}

 \newpage
 
\tableofcontents

\newpage

\section{Introduction}

  The quantum field theory (QFT) on the worldvolume of M5-branes
  is known \cite{Witten04, HNS, He} to be a 6-dimensional $(0, 2)$-superconformal theory
  that contains a 2-form potential field $B_2$, 
  whose 3-form field strength $H_3$ is self-dual
   (see \cite{Moore} for a recent survey). 
  Whatever it is precisely and in generality, 
  this QFT has been argued to be the source of deep 
  physical and mathematical phenomena, such as
  Montonen-Olive S-duality \cite{Witten04}, 
  geometric Langlands duality \cite{Witten09}, and 
  Khovanov homology \cite{Witten11}. Yet, and despite this interest, 
  a complete description of the precise details of this QFT is still lacking. 
  In particular, as soon as one considers the worldvolume theory of 
  \emph{several coincident} M5-branes, the 
  2-form appearing locally in this 6d QFT is expected to be
  \emph{nonabelian} (to take values in a nonabelian Lie algebra).
  But a description of this \emph{nonabelian gerbe theory}
  has been elusive (a \emph{gerbe} is a ``higher analog'' of a gauge bundle,
  discussed in detail below in section \ref{Nonabelian}). 
  See \cite{Ha, B08, Sati10}
  for surveys of the problem and recent developments.
  Here we add another piece to the scenario, by proposing a 7d Chern-Simons theory which appears to be a natural candidate for the holographic dual of the multiple M5-branes 6d QFT via  $\mathrm{AdS}_7/\mathrm{CFT}_6$-duality, and by identifying the nonabelian 2-form fields appearing in the theory as local data of (twisted) String 2-connections.
%  Here we present a consistent formulation
%  of nonabelian 2-form fields and propose dynamics for them under holography.

\vspace{3mm}
Namely, for a single M5-brane, the Lagrangian of the theory has been formulated 
in \cite{HSe, PeS, Sch, PST, APPS} and
%Furthermore,
 in  
  this 
  %abelian 
  case
  there is, due to \cite{Witten96}, a \emph{holographic dual} description of the 6d theory
  by 7-dimensional abelian Chern-Simons
  theory, as part of $\mathrm{AdS}_7/\mathrm{CFT}_6$-duality
  (reviewed for instance in \cite{AGMOO}).
  We give here an argument, following \cite{Witten96, Witten98} but taking  
  the quantum anomaly cancellation of the M5-brane in 
  11-dimensional supergravity into account, that 
  in the general case: the $\mathrm{AdS}_7$/$\mathrm{CFT}_6$-duality
  involves a 
  7-dimensional \emph{nonabelian} Chern-Simons
  action that is evaluated on higher nonabelian gauge fields which 
  we identify as 
  \emph{twisted 2-connections over the $\mathrm{String}$-2-group}, 
  as considered in \cite{SSSI, FSS}. 
  Then we give a precise description of a certain canonically existing 7-dimensional 
  nonabelian gerbe-theory on boundary values of 
  quantum-corrected supergravity field configurations in terms of nonabelian 
  differential cohomology. We show that this has the properties expected 
  from the quantum anomaly structure of 11-dimensional supergravity.
  In particular, 
  we discuss that there is a higher gauge in which these field configurations
  locally involve non-abelian 2-forms with values in the 
  Kac-Moody central extension of the loop Lie algebra of the special 
  orthogonal Lie algebra $\mathfrak{so}$
  and of the exceptional Lie algebra $\mathfrak{e}_8$.
  We also describe 
  %and explicitly construct 
  the global structure of the moduli 2-stack of 
  field configurations, which is more subtle.

  \vspace{3mm}
Most of the ingredients of the 7d theory that we present are implicit in
  earlier publications of the authors, 
  notably \cite{SSSI}, \cite{SSSIII} and \cite{FSS}. There, however, we focused
  on viewing the ``indecomposable'' 7d \emph{Lagrangian} 
  (section \ref{InfinCS7CS}) 
  as a \emph{differential twist} that controls the
  magnetic dual heterotic Green-Schwarz anomaly cancellation, in direct analogy of
  how the ordinary 3-dimensional Chern-Simons Lagrangian serves, 
  as discussed in these references,
  as a differential twist that controls the direct heterotic Green-Schwarz mechanism. The present
  article serves to make the corresponding Chern-Simons theory and its role in
  11-dimensional supergravity explicit.  
  Its relation to M-branes is also discussed in section 3.5 of \cite{Sati10}.
  Further connections between String structures (and their variants) and M-branes 
  are given in \cite{Sati10,tw,II}.

\vspace{3mm}

The description of higher degree form fields in string theory via {\it abelian} 
differential cohomology has been deeply influenced by the works of 
Freed \cite{Freed} and Hopkins-Singer \cite{HopkinsSinger}.
Such a description is very convenient when the gauge fields involved
are, indeed, abelian; this includes Maxwell fields, Kalb-Ramond $B$-fields, 
the self-dual field on a single M5-brane (see \cite{BM}), 
and Ramond-Ramond fields. 
However, when considering systems such as given by multiple 
M5-branes, the theory becomes nonabelian -- in some appropriate sense --
and hence one needs to describe in a mathematically precise 
way the corresponding nonabelian higher degree gauge fields. 
This requires using 
{\it nonabelian} differential cohomology, a generalization not just of the 
theory of line bundles/circle bundles with connection, but a generalization 
of the theory of general gauge bundles with connection, 
hence of general Yang-Mills fields. 
A theory that accomplishes this has been laid out in \cite{survey},
based on earlier work that includes \cite{SchreiberWaldorfIII}, \cite{SSSI} and
\cite{FSS}.
The use of such a formalism is 
not only to set up the correct language  -- which of course is desirable --
but also to obtain a machinery that produces the dynamics of the fields,
as well as their relation and their consistent coupling to other fields,
in a systematic way, constrained by suitable general principles of higher
gauge theory. 
As a result, couplings and higher order gauge transformations 
which otherwise have to be guessed may now be derived systematically. 
Furthermore,  the result is typically more subtle than what could have been
-- and in some cases has been --  guessed.
The twisted String 2-connections 
and their canonical 7-dimensional action functional are an example of
this, which we will explain in detail. 

\vspace{3mm}
The systems discussed here are just special cases of an 
infinite hierarchy of higher nonabelian gauge fields 
with higher Chern-Simons type action functionals that canonically arise
in higher nonabelian differential cohomology in a canonical way that we briefly indicate
in section \ref{NaturalCSFunctional} below. 
The full 11-dimensional Chern-Simons term of 11d-supergravity is another
example.
In \cite{FRSI} we show that another class of examples is 
given by globalizations of AKSZ $\sigma$-models, such as the
Courant $\sigma$-model that is induced from generalized Calabi-Yau spaces.
There is a whole zoo of further examples; see section 4.6 of \cite{survey}.
Moreover,
to each such theory in dimension $(n+1)$ is associated a corresponding
generalized higher WZW model in dimension $n$.  In particular there is a
6-dimensional WZW-type theory associated with the boundary of the 7-dimensional
String-connection theory discussed here. But details of this are beyond the 
scope of the present article.

\vspace{3mm}
What we do in this article can be summarized in the following main points:
\begin{enumerate}

\item As a warm-up, we provide in sections \ref{DeterminantLineBundles}
and \ref{Section D-branes}
a description of 
the familiar case of gauge fields on multiple D-branes, but formulated in 
terms of the nonabelian differential cohomology of stacks 
of $\mathrm{U}(n)$-bundles in a way that has by direct analogy
a generalization to multiple M5-branes considered afterwards.

\item We discuss, in section \ref{7dTheory}, a refinement
 $\hat {\mathbf{I}}_8$ of the anomaly 
 8-class or one-loop polynomial 
 $I_8$ of 11-dimensional supergravity to nonabelian differential cohomology
 by a higher stacky Chern-Weil construction. This refinement 
 is a universal differential characteristic map that is naturally 
 defined on the moduli 2-stack
 of boundary supergravity $C$-field configurations 
 %from above.
constructed in \cite{FiSaSc}. Note that an elliptic refinement of the one-loop
term is given in \cite{S-String}. 

\item We consider the 7-dimensional nonabelian Chern-Simons action functional
 canonically induced by $\hat {\mathbf{I}}_8$ on boundary $C$-field configurations
 in section \ref{7dCSInSugraOnAdS7}.
% {7dChernSimons}.
 We demonstrate that locally
 -- or globally in the trivial instanton sector -- this reduces to 
 the functional that is implied by the one-loop correction in 11-dimensional
 supergravity, as discussed in section \ref{DualityAnd7dCS}.
 
\item 
  Throughout the article we discuss various aspects of this 7d theory. 
  %to take note of.  
  We comment on the  relation to loop groups 
  and the reason for passing to 2-groups in 
  \ref{EvidenceFromTheWorlvolumeTheory},
 % \ref{String2GroupOn5Brane},
  point out the role of Lie $n$-algebras in 
  \ref{LInfinityAlgebras},
  explain in what sense the 2-form on the M5-brane worldvolume is nonabelian
  in sections \ref{Section stacks} and \ref{TheFields}.
\end{enumerate}

\vspace{3mm}
%This paper is organized as follows.
The discussion is separated into three parts. 
First, in section 
\ref{Evidences} we provide 
heuristic physical arguments aimed at characterizing 
the properties that the sought-after mathematical objects should satisfy. 
This involves looking at the problem form various angles, 
which we outline below. 
We present an argument for why 
the spaces of states of the 7-dimensional nonabelian Chern-Simons 
theory are a plausible candidate for the
conformal blocks of the 6d theory on worldvolume of coincident fivebranes. 
This argument is necessarily non-rigorous, but it seems to be 
as trustworthy as the argument in \cite{Witten98}, of which it is 
a direct extension.
Then in section \ref{Nonabelian} we review aspects of higher nonabelian
gauge theory in a way that prepares the ground for our main construction.
Finally in Section \ref{7dTheory} we give a precise definition and discussion
of action functionals of 7-dimensional Chern-Simons theories whose fields are 
twisted 2-connections with values in the $\mathrm{String}$ 2-group.
In conjunction with 
the physical arguments of section \ref{Evidences}, this can be regarded
as a proposal for how to make aspects of the physical heuristics involved 
there precise.

\vspace{3mm}
We hope to study the supersymmetric extension of the current constructions
 in a separate article. 

%\newpage

\vspace{3mm}
This article makes use of mathematical concepts in the theory of 
\emph{higher stacks} and \emph{nonabelian (differential) cohomology}.
In section \ref{Nonabelian} we offer some introduction and explanation
that should be sufficient for an appreciation of section \ref{7dTheory}.
However, the reader who wishes to dig deeper into the mathematics to which we appeal
should look at \cite{SSSIII}, \cite{FSS} and \cite{survey} (perhaps in that order).
For ease of reference, in the tables below we list mathematical objects
that we will mention frequently, together with their physical meaning.
These tables are also useful in the description of the C-field and its
dual in \cite{FiSaSc}.

\medskip
We will use various  notions of cohomology, starting with differential forms 
and working our way up through refinements. These are summarized in the table 
\vspace{3mm}
\begin{center}
\begin{tabular}{|l||l|l|}
\hline
\phantom{$\bigl(^{a}_{b}$}{\bf cohomological notion} & {\bf gauge theoretic notion}
\\
\hline
\hline
\phantom{$\bigl(^{a}_{b}$}differential forms $\Omega_{\mathrm{cl}}$ 
& field strengths / classical description
\\
  \hline
\phantom{$\bigl(^{a}_{b}$}cohomology $H(-)$ & instanton configurations / magnetic charges 
\\ 
  \hline
\phantom{$\bigl(^{\int}_{b}$}differential cohomology $\hat{H}(-)$ & equivalence classes of gauge fields
 \\
   \hline
\phantom{$\bigl(^{a}_{b}$}cocycle $\infty$-groupoid ${\bf H}(-)$ & actual gauge fields with (higher) gauge transformations
\\
\hline
\end{tabular}
\end{center}

\vspace{3mm}
We will also use various (higher) \emph{moduli stacks} in order to
precisely capture the global nature of (higher) gauge fields and their
(higher) gauge transformations. One may think of these as 
\emph{integrated BRST complexes} or \emph{integrated Lie $n$-algebroids}, 
see section \ref{HigherSmoothStacks}.
To guide the reader through
the various stacks, here is a table that should serve to set some
notation and also as a dictionary between stacky notions and
the corresponding bundle structures appearing in relation to the physics of 
M5-branes and M-theory.  We have
\vspace{3mm}
\begin{center}
\begin{tabular}{|l||l|l|}
\hline
 \phantom{$\bigl(^{a}_{b}$} {\bf symbol} & {\bf (higher) moduli stack of...}
  \\
  \hline
  \hline
\phantom{$\bigl(^{a}_{b}$}  $\mathbf{B}{\mathrm{U}(1)}$ & circle bundles / Dirac magnetic charges
  \\
  \hline
\phantom{$\bigl(^{a}_{b}$}  $\mathbf{B}{\mathrm{U}(1)}_{\mathrm{conn}}$ & ${\mathrm{U}(1)}$-connections / abelian Yang-Mills fields
  \\
    \hline
\phantom{$\bigl(^{a}_{b}$}  $\mathbf{B}\mathrm{Spin}_{\mathrm{conn}}$ & Spin connections / field of gravity
  \\
    \hline
\phantom{$\bigl(^{a}_{b}$}  $\mathbf{B}E_8 $ & $E_8$-instanton configurations
    \\
      \hline
 \phantom{$\bigl(^{a}_{b}$} $(\mathbf{B}E_8)_{\mathrm{conn}} $ & $E_8$-Yang-Mills fields
  \\
   \hline 
 \phantom{$\bigl(^{a}_{b}$}  $\mathbf{B}^2 {\mathrm{U}(1)}_{\mathrm{conn}}$ & $B$-field configurations (without twists)
  \\
   \hline 
\phantom{$\bigl(^{a}_{b}$}   $\mathbf{B}^3 {\mathrm{U}(1)}_{\mathrm{conn}}$ & $C$-field configurations (without twists)
  \\
    \hline
\phantom{$\bigl(^{a}_{b}$}  $\mathbf{B}\mathrm{String}_{\mathrm{conn}}$ & String 2-connections / nonabelian 2-form connections
  \\
    \hline
\phantom{$\bigl(^{\int}_{b}$}  $\mathbf{B}\mathrm{String}^{2\mathbf{DD}_2}$ & first Spin characteristic class 
$\lambda=\tfrac{1}{2}p_1$ divisible by 2
  \\
    \hline
 \phantom{$\bigl(^{\int}_{b}$} $\mathbf{B}\mathrm{String}^{2\mathbf{a}}$ & $E_8$-twisted $\mathrm{String}$-2-connections
  \\
    \hline
\phantom{$\bigl(^{a}_{b}$}  $\mathbf{CField}$ & bulk configurations of supergravity $C$-fields (and gravity)
  \\
    \hline
 \phantom{$\bigl(^{\int}_{b}$} $\mathbf{CField}^{\mathrm{bdr}}$ & $C$-field configurations on (5-brane) boundaries
  (and $E_8$-gauge fields)
  \\
  \hline
\end{tabular}
\end{center}

\medskip 

These concepts combine to give actual configuration spaces of (higher) gauge fields
by evaluating cohomology on spacetime \emph{with coefficients in} a (higher) moduli stack.
Let $G = {\mathrm{U}(1)}, \mathrm{Spin}, \mathrm{String}, \mathbf{B}{\mathrm{U}(1)}, \ldots$ be a (higher)
gauge group with Lie $n$-algebra $\mathfrak{g}$ (see \cite{SSSI}), 
and let $X$ be a (spacetime) manifold. 
Then we have
%\vspace{3mm}
\begin{center}
\begin{tabular}{|l||l|l|}
  \hline
\phantom{$\bigl(^{a}_{b}$}  {\bf symbol} & {\bf gauge theoretic meaning}
  \\
  \hline
  \hline
  \phantom{$\bigl(^{\int}_{b}$} $[X, \mathbf{B}G_{\mathrm{conn}}]$ & moduli stack of $G$-gauge fields on $X$
  \\
  \hline
\phantom{$\bigl(^{\int}_{b}$}  $\mathbf{H}(X, \mathbf{B}G_{\mathrm{conn}})$ & collection of gauge fields with $G$-gauge transformation on $X$
  \\
  \hline
\phantom{$\bigl(^{\int}_{b}$}  $H(X, \mathbf{B}G_{\mathrm{conn}})$ &equivalence classes of $G$-gauge fields on $X$
 \\
  \hline
\phantom{$\bigl(^{\int}_{b}$}  $\hat H^n(X) \simeq H(X, \mathbf{B}^n {\mathrm{U}(1)}_{\mathrm{conn}})$ & 
    equivalence classes of abelian $n$-form gauge fields on $X$
  \\
  \hline
\phantom{$\bigl(^{\int}_{b}$}  $H(X, \mathbf{B}G)$ & set of underlying instanton sectors
  \\
  \hline
\phantom{$\bigl(^{\int}_{b}$}  $\Omega(X,\mathbf{B}G)$ & $\mathfrak{g}$-valued (higher) field strengths
  \\
  \hline
\end{tabular}
\end{center}

%\vspace{3mm}

%%%%%%%%%%%%%%%%%%%%%%%%%%%%%%%%%%%%%%%%%%%%%%%%%%%%%%%%%
\section{Evidence for and ingredients of the 7d nonabelian gerbe theory}
% from
%$\mathrm{AdS}_7$/$\mathrm{CFT}_6$}
%holography}
\label{Evidences}
%%%%%%%%%%%%%%%%%%%%%%%%%%%%%%%%%%%%%%%%%%%%%%%%%%%%%%%%

In this section we present physical arguments for why one should expect
the nonabelian gerbe theory to be the right description of the system of 
multiple fivebranes. Along the way, we provide our own interpretations which 
help us identify the relevant ingredients in that theory. 
There are (at least) two aspects to this
\begin{enumerate}
  \item In one regime, the M5-brane worldvolume is to be thought of 
  as  embedded into an ambient 11-dimensional spacetime that carries a supergravity
  $C$-field which has a direct restriction to the brane. 
  The restriction at the level of de Rham cohomology and differential forms 
is discussed in \cite{Hodge} from the point of view of boundary conditions.
  We provide another description in section \ref{EvidenceFromTheWorlvolumeTheory}.

  \item
     In another regime, we have ``black'' 5-branes identified as the 
	 asymptotic boundary 
	 \cite{Ma}
	 of a compactification on $S^4$ to an 
	 $\mathrm{AdS}_7$-solution \cite{PvNT}
	 of 11d supergravity. Here the boundary space of states of the
	 $C$-field holographically induces the conformal blocks
	 of the 5-brane superconformal theory. 
	 	 This is discussed in section \ref{DualityAnd7dCS}.
\end{enumerate}

We should stress that our use of holography, and the particular 
configuration related to AdS/CFT, serves as a motivation 
and indeed our constructions will work in full generality.

%%%%%%%%%%%%%%%%%%%%%%%%%%%%%%%%%%%%%%%%%%%%%%%%%%%%%%%%%
\subsection{The M5-brane worldvolume theory: loop groups and the String group}
\label{EvidenceFromTheWorlvolumeTheory}
%%%%%%%%%%%%%%%%%%%%%%%%%%%%%%%%%%%%%%%%%%%%%%%%%%%%%%%%%

%\vspace{3mm}
Various ingredients and aspects of 5-brane physics have been 
conjectured or argued for before in the literature,
see \cite{Sati10} for an outline. Among them are the ones we discuss below,
all of which are subsumed by the proposal we make.
We will also provide heuristic interpretations and connections to 
the String 2-group and loop groups,
appropriate for the description of  M5-branes.

\paragraph{Nonabelian gerbes and 2-gerbes with connections.}
%\noindent{\bf Nonabelian gerbes and 2-gerbes with connections.}
%\attn{\color{blue} "instanton YM does not seem standard term?}
The notion of a \emph{gerbe} or \emph{2-bundle} with connection is a higher analog
of the notion of a \emph{principal bundle} with connection, hence of 
(instanton) Yang-Mills field configurations, where  the term ``higher'' is as in 
``higher degree differential forms'': just as a connection on a gauge bundle 
is locally given by a 1-form (the gauge potential),  a connection on a 
2-bundle/1-gerbe is locally given by a 2-form.
Several known arguments imply that the worldvolume theory
of multiple coincident 5-branes contains a field that is a 2-form connection 
on a nonabelian gerbe in analogy to how Yang-Mills theory is the theory
of a field that is a 1-form connection on a principal bundle.

\vspace{3mm}
\noindent {\bf (i)} The first argument invokes the lift of string-D4-brane systems
from string theory to M-theory. It is well known that the
open string ending on $N$ coincident D-branes couples to
a $\mathrm{SU}(N)$-valued 1-form connection field on 
the Chan-Paton bundle on the D-brane, 
whose line holonomy
along the boundary of the string worldsheet provides the boundary
term in the action. To be precise, 
there is in addition
the $B$-field in the ambient spacetime whose restriction to the 
D-brane \emph{twists} this 1-form connection field.
The lift of this configuration to M-theory through the inverse of the 
double dimensional 
reduction is an M2-M5 brane
configuration with open membranes 
\cite{St,To} 
ending on $N$ coincident 
M5-branes.  
Now, the membrane boundary $\partial {\rm M}2 \subset {\rm M}5$ 
couples to a 2-form connection on the 5-brane
and the restriction of the ambient supergravity $C$-field
induces a twist. For this situation to be compatible with its
reduction to string theory, the 2-form must in some way
take values in a nonabelian Lie algebra. Hence it should be
a 2-connection on a nonabelian gerbe on the 5-brane, which 
is twisted by the 2-gerbe on which the $C$-field is a 3-connection.

\medskip
We can deduce precisely this phenomenon also from the 
nature of the gauge-invariant field strengths
on branes that are familiar from the literature:
The gauge-invariant 2-form field strength on a D-brane in the 
background of a $B$-field is, locally, the combination
\(
  \mathcal{F}= B +F
  \,,
  \label{Combined2FormOnBrane}
\)
where $F$ is the trace of the curvature of the 
connection on the worldvolume of the D-brane.
This phenomenon (and its global generalization)
is explained (as we discuss in section \ref{Section D-branes}) 
by the fact that the gauge field on the D-brane is a kind of
trivialization of the 2-bundle/1-gerbe underlying the restriction of the 
$B$-field to the brane: a twisted Chan-Paton bundle is a kind of 
trivialization of a circle 2-bundle/1-gerbe.
This phenomenon has a higher generalization. In general 
\emph{twisted} $n$-bundles / $(n-1)$-gerbes may serve as a kind of 
trivialization of an $n+1$-bundle. (A detailed mathematical 
discussion of this is given in section 1.3.1 and 4.4 of \cite{survey}.)

\vspace{3mm}
Now, it is known that the invariant self dual 3-form field strength on the M5-brane 
is accordingly locally of the form 
\(
  \mathcal{H}= H+ C
  \label{Combined3FormOnBrane}
  \,,
\)  
where $C$ is the restriction of the ambient $C$-field to the brane, and where
$H=dB_2$ is the curvature of the 2-form potential on the 
M5-brane. 
%{\color{green}
This can be seen from the M2-brane as follows. 
Consider the part of the action given by 
$S_C= \int_{\Sigma_3} C_3$, where $\Sigma_3$ is the 
worldvolume of the M2-brane. 
This action is not invariant for an open M2-brane
unless we introduce a two-form gauge field $B$ 
coupled to the boundaries of the M2-brane, 
with $S_B= \int_{\partial \Sigma_3} B_2$
and require that it transforms as $B_2 \to B_2 - \Lambda_2$
under the $C$-field gauge transformation
$C\to C + d\Lambda_2$. Here $\Lambda_2$
is a two-form field. 
Then gauge invariance requires considering the 
combination $\mathcal{H}=C + H$. 
%}
Therefore this is another reason to expect that the 2-form field on the M5-brane
is the local connection on a \emph{twisted 2-bundle} whose twist is
given by the $C$-field. 

\vspace{3mm}
\noindent {\bf (ii)} 
The second argument (see \cite{Witten04}) proceeds by 
a similar dimensional reduction, but now from 6 to 4 dimensions.
One finds that compactifiying the conformally invariant 
worldvolume theory of a single fivebrane with its 
abelian 2-form on a torus yields abelian Yang-Mills theory (electromagnetism)
in 4-dimensions, such that the residual conformal transformations
on the compactified space becomes the Montonon-Olive electric-magnetic
duality of 4-dimensional Yang-Mills theory. Since this gives,
in the abelian case,
a natural  geometric explanation for the otherwise more 
mysterious S-duality of (super) Yang-Mills theory, it is natural 
to expect that the same mechanism is the source of 
electric-magnetic duality also generally in nonabelian 
(super) Yang-Mills theory.
Motivated by these arguments a definition of 
twisted nonabelian 2-gerbe with connection has been 
proposed in \cite{AJ} and argued to be 
relevant for the description of 5-brane physics.
The notion of higher twisted
gerbes with higher connections 
has been fully formalized
in \cite{survey}.
By appealing to theorems about these structures that we 
provided in \cite{FSS, SSSIII},
as well as to a study of topological effects within
$\mathrm{AdS}/\mathrm{CFT}$-duality, we argue in 
section \ref{DualityAnd7dCS} below
that indeed these structure on 5-branes are \emph{implied} 
by quantum anomaly cancellation of M5-branes in M-theory.

\vspace{3mm}
There are further connections to other branes which 
highlight some of the topological and geometric considerations 
that we consider. We mention that 
the connection of Fivebrane structures in relation to the 
NS5-branes in type IIA is given in \cite{SSSIII, NS5}.

%%%%%%%%%%%%%%%%%%%%%%%%%%%%%
\paragraph{Loop group and String 2-group degrees of freedom on the M5-brane.}
%\label{String2GroupOn5Brane}
%%%%%%%%%%%%%%%%%%%%%%%%%%%%
There have been put forward various arguments 
that mean to identify gauge \emph{loop groups} 
(see \cite{PresslySegal} for mathematical background)
controlling the gauge theory on M5-branes. We recall these arguments, 
recasting them on firm mathematical ground within our perspective, 
and indicating how they will be refined in sections  \ref{Nonabelian} and \ref{7dTheory}.

\vspace{3mm}
First, consider again the twisted Chan-Paton bundles that appear on the D4-brane
in 10-dimensional type IIA string theory.
These are controlled topologically by the obstruction theory of lifts through the
universal circle extension
\(
  {\mathrm{U}(1)} \to U(\mathbb{H}) \to \mathrm{P U} ( \mathbb{H})
\label{U U}
\)
of the group of projective unitary operators $\mathrm{P U} ( \mathbb{H})$ on any
separable Hilbert space $\mathbb{H}$
(see section \ref{Section D-branes} for details): they are 
projective unitary bundles whose obstruction to lift to genuine
unitary bundles is the class of the ambient B-field gerbe 
(related to the third integral Stiefel-Whitney class).
Now, under the double dimensional reduction from M-theory to 
type IIA string theory, the D4-brane
in ten dimensions comes from an M5-brane in eleven dimensions.
In the spirit of  \cite{MaSa} it has, essentially, been argued in \cite{AJ} 
that reversing double dimensional reduction goes along with 
a \emph{delooping} of the above sequence, and that this should involve
the M-theory $E_8$-degrees of freedom. 
Notice that the homotopy type of the 
topological space $\mathrm{PU}(\mathcal{H})$ is that of an 
Eilenberg-MacLane space $K(\mathbb{Z},2)$, characterized by the fact that 
its only nontrivial homotopy group is $\pi_2(\mathrm{PU}(\mathbb{H}))  \simeq \mathbb{Z}$,
and that the only nonvanishing homotopy group of the topological space underlying
the Lie group $E_8$ in degree $< 15$ is $\pi_3(E_8) \simeq \mathbb{Z}$, 
so that the only nonvanishing homotopy group of the \emph{loop group} $\Omega E_8$
in degree $< 14$ is $\pi_2(\Omega E_8) \simeq \mathbb{Z}$. Therefore, 
over the manifolds of dimension $\leq 11$ that appear in the string theoretical
context,
projective unitary bundles have the same classification as $\Omega E_8$-bundles
\cite{MaSa},
hence in these dimensions the group $E_8$ is a model for the delooping of 
$\mathrm{PU}(\mathbb{H})$
\(
  \mathrm{PU}(\mathbb{H}) \simeq_{14} \Omega E_8
  \,.
\)
Therefore, still in these dimensions,
the sequence \eqref{U U} is homotopy equivalent to a sequence for a central 
extension $\hat \Omega E_8$ of the loop group $\Omega E_8$
\(
  {\mathrm{U}(1)} \to \hat \Omega E_8 \to \Omega E_8
  \,.
\)
So far this is the argument from \cite{MaSa, AJ}. 
Note that one has to be a bit careful with this, because it concentrates on homotopy types 
and ignores the
\emph{geometric} (gauge) structure, which is different for projective unitary bundles
and for $\Omega E_8$-principal bundles. 

\vspace{3mm}
In order to fully deloop the whole sequence, we may observe next that 
a space of the homotopy type of a $K(\mathbb{Z},2)$ is a delooping for
${\mathrm{U}(1)}$. We will find it useful to make this explicit by writing $B {\mathrm{U}(1)}$ for $K(\Z, 2)$. 
Then a delooping of the sequence (\ref{U U}) over spaces of dimension
$< 14$ can be written in the form
\(
   B {\mathrm{U}(1)} \to~ ?~ \to E_8
   \,.
\)
This means that the middle term here is
analogous to an ordinary group extension of $E_8$ by a circle group, only that the 
circle group ${\mathrm{U}(1)}$ is replaced by a \emph{higher} or \emph{shifted} circle group
$B {\mathrm{U}(1)}$. 
If we keep arguing from the point of view of sufficiently low-dimensional spaces, then 
we notice the truncated homotopy equivalence $\Omega E_8 \sim_{14} B {\mathrm{U}(1)}$
and observe that for any group $G$ the group $L G$ of \emph{free} loops 
(without fixed basepoint) forms a split extension 
\(
  \Omega G \to L G \to G
  \,.
\) 
Hence in sufficiently low dimension and ignoring geometry, 
the above question mark ``?" could be filled
by the \emph{trivial} extension
\(
  \Omega E_8 \to L E_8 \to E_8
\)
of $E_8$ by its free loop group. 
While this may serve as a guide, it is too simplistic, because there is
no reason to expect a trivial extension here. At the opposite extreme is the
\emph{universal} non-trivial such extension, which is such that every other one is a multiple of
it. This universal shifted central extension of $E_8$ is known as the
\emph{String group} of $E_8$, denoted
\(
   B {\mathrm{U}(1)} \to \mathrm{String}(E_8) \to E_8
   \,.
\)
Its homotopy type is that of a certain topological group, but as a \emph{geometric}
(smooth) object it is not a Lie group. Instead it is a \emph{higher} analog
of a Lie group called a \emph{smooth 2-group} 
(technical details of which we review in section \ref{Section stacks}
below). 

\vspace{3mm}
In \cite{BCSS} it is shown that $\mathrm{String}(G)$
for any simply connected compact simple Lie group $G$ has a presentation by what is
called a \emph{crossed module} of two ordinary Lie groups, namely by the 
Kac-Moody central extension $\hat \Omega G$ of the loop group of $G$
(but now regarded as a genuine Lie group) and the based path group of $G$.
In this sense, the $\mathrm{String}(E_8)$ 2-group remembers the loop group degrees of freedom 
even after delooping to the 5-brane. 
Notice, however, that the notion of (gauge) equivalence
for higher groups is considerably richer than for ordinary groups, so that 
one and the same 2-group may be presented by rather different looking constructions,
some of which do not manifestly involve loop groups. A review of this
phenomenon  we append in the Appendix. 
%section \ref{TheFields}.

\medskip

There is also a different way to arrive at the
conclusion that (twisted) $\mathrm{String}$-2-connections
are the right fields in one-loop-corrected supergravity and on 5-branes.
This is the perspective of \cite{SSSIII, FSS}, which is at the heart of
our development in sections \ref{Nonabelian} and \ref{7dTheory}.
The argument requires some background ideas in gauge theory and higher
stacks that we survey in \ref{HigherSmoothStacks}. The reader
not familiar with this relation
may want to come back to the following argument after having looked
at those sections. Here is the argument, central to our main point.
The fields of supergravity over an M5-brane boundary $\Sigma$ locally look like
a $\mathrm{Spin}$-connection (gravity) and an $E_8$-gauge-field.
Crucially, these are subject to a \emph{constraint} \cite{Witten96}
which demands that the Pontrjagin class $\tfrac{1}{2}p_1$
of the former equals twice the canonical 4-class $a$ of the latter.
We recall this in more detail below in section \ref{TheSugraChernSimonsTerm}.

\medskip
Consider, for the moment,
the set of \emph{gauge equivalence classes} of $\mathrm{Spin}$-structures on $X$,
which we  write $H(\Sigma, \mathbf{B}\mathrm{Spin})$,
and the set of gauge equivalence classes of $E_8$-instantons, which we
write $H(\Sigma, \mathbf{B}E_8)$, and finally the set of degree-4 integral
cohomology classes, which we write $H(X, \mathbf{B}^3 U(1))$.
Then the first Spin class $\tfrac{1}{2}p_1$ is a map
\(
  \tfrac{1}{2}p_1 : H(\Sigma, \mathbf{B}\mathrm{Spin}) \to H(\Sigma, \mathbf{B}^3 U(1))
\)
and the $E_8$ class $a$ is a map
\(
  a : H(\Sigma, \mathbf{B}E_8) \to H(\Sigma, \mathbf{B}^3 U(1))
\)
and the \emph{set of pairs} $P(\Sigma)$ of gauge equivalence classes of 
field configurations that satisfy the quantization
condition constraint is the \emph{fiber product} 
(or ``pullback'') of these two maps,
the set that universally completes a diagram of maps like this:
\(
  \raisebox{20pt}{
  \xymatrix{
    P(\Sigma) \ar[r] \ar[d] & H(\Sigma, \mathbf{B}E_8) \ar[d]^{2a}
	\\
	H(\Sigma, \mathbf{B}\mathrm{Spin})
	\ar[r]^{\tfrac{1}{2}p_1}
	&
	H(\Sigma, \mathbf{B}^3 U(1))~~\;.
  }
  }
\)
From this point of view it seems as if the supergravity quantization condition
simply restricts the configuration space of fields.
However,  there is a problem with this argument. For \emph{quantization} of a gauge theory,
the set of gauge equivalence classes of field configurations is an invalid 
starting point. What one instead needs to consider is the \emph{BRST complex}
of field configurations, or rather, its integrated version, the \emph{moduli stack}
of field configurations. 
We write $\mathbf{B}\mathrm{Spin}_{\mathrm{conn}}$ for the universal moduli
stack of $\mathrm{Spin}$-connections. Then 
$\mathbf{H}(\Sigma, \mathbf{B}\mathrm{Spin}_{\mathrm{conn}})$ denotes the
integrated BRST complex, containing the genuine $\mathrm{Spin}$-connection fields on 
$\Sigma$, \emph{and} the gauge transformations between them. This is no longer
a set, but is now a \emph{groupoid}. Its set of \emph{connected components}
recovers the set $H(X, \mathbf{B}\mathrm{Spin}_{\mathrm{conn}})$ of gauge 
equivalence classes of fields. Similar comments and notation apply to 
$E_8$ and $\mathbf{B}^2 U(1)$. 

\medskip
These structures are now a valid starting point for quantization. Therefore, 
the above constraint should be imposed on these structures. The crucial difference
now is that when we ask for the structure
that universally completes this fiber product diagram
\(
  \raisebox{20pt}{
  \xymatrix{
    ? 
	\ar[r] \ar[d]^>{\ }="t" 
	& 
	\mathbf{H}(\Sigma, \mathbf{B}E_8) 
	\ar[d]^{2\hat {\mathbf{a}}}_<{\ }="s"
	\\
	\mathbf{H}(\Sigma, \mathbf{B}\mathrm{Spin}_{\mathrm{conn}})
	\ar[r]_{\tfrac{1}{2}\hat{\mathbf{p}}_1}
	&
	\mathbf{H}(\Sigma, \mathbf{B}^3 U(1))
	\ar@{=>}_{\hat B} "s"; "t"
  }
  }
  \label{HomotopyPullbackFirstAppearance}
\)
then going around the square in the two possible ways no longer
needs to yield genuinely equal field configurations. It suffices
that the two field configurations obtained are connected by a gauge transformation
$B$, as indicated. This is indeed the only way to make gauge-invariant sense of this diagram.
In the following, \emph{all} square diagrams (always of higher smooth stacks) that we display 
are implicitly filled by a gauge transformation this way, but only sometimes
do we display it explicitly.

Such fiber products ``up to gauge transformation'' are well known in 
homotopy theory. They are called \emph{homotopy fiber products}
or \emph{homotopy pullbacks}. If we compute them in the full context of higher
gauge theory, we find two crucial differences to the above naive idea
of imposing the quantization constraint.

\medskip
First, the object in the top left is no longer the simple restriction
of the direct product of $\mathrm{Spin}$- and $E_8$-connections that satisfy
the quantization constraint. The reason is that the \emph{choice} of
gauge transformation $B$ on each pair is now part of the field content data.
The new field that appears this way is well known in string theory, at least for 
Ho{\v r}ava-Witten boundaries of 11-d supergravity \cite{HoravaWitten}: it is the field strength
of a twisted \emph{2-form} field (``$B$-field'') with twisted Bianchi identity
\(
  d \mathcal{H}_3 = \langle F_\omega \wedge F_\omega \rangle - 2 \langle F_A \wedge F_A\rangle
  \,,
  \label{eq Bia}
\)
where $F_\omega$ and $F_A$ are the curvatures of the Spin connection $\omega$ and the gauge 
connection $A$, respectively, and
\(
\mathcal{H}_3=dB+\mathrm{CS}_3(\omega)-2\mathrm{CS}_3(A).
\) 
Equation \eqref{eq Bia} manifestly exhibits the structure of the square diagram
\eqref{HomotopyPullbackFirstAppearance}
presented by de Rham cocycles.

\medskip
The second effect is that the object denoted ``?'' above is not itself
a groupoid anymore. It turns out to be a higher groupoid, here a \emph{2-groupoid}
that contains not just gauge transformations, but
\emph{gauge-of-gauge transformations} (coming from 
\emph{ghosts-of-ghosts} in the corresponding BRST complex). In
section \ref{SmoothStringC2} below we identify the question mark here
with the 2-groupoid $\mathbf{H}(\Sigma, \mathbf{B}\mathrm{String}^{2\mathbf{a}})$
of \emph{$E_8$-twisted String-2-connections}. If the $E_8$-twist here
vanishes, then this involves the genuine $\mathrm{String}$-2-group
which we had motivated already via loop groups 
in section \ref{EvidenceFromTheWorlvolumeTheory}
above.
This effect of imposing the relation $\tfrac{1}{2}p_1 = 2 a$ not
on gauge equivalence classes but on moduli stacks / integrated BRST complexes of fields
is the key step that leads us to nonabelian higher form fields
in the following discussion. In the closely related context of 
anomaly cancellation in heterotic string theory, this very
phenomenon has been discussed in some detail in \cite{SSSIII}.

\vspace{3mm}
The theory of such nonabelian 2-form connections has been developed in 
\cite{SchreiberWaldorfIII, FSS, survey}. We review String 2-connections
in section \ref{Section stacks} below and discuss
twisted $\mathrm{String}$-2-connections in section \ref{SmoothStringC2}. 
In section \ref{7dTheory} we 
systematically derive the
\emph{twisted} String 2-connections on M-branes 
 in 11-dimensional supergravity (or M-theory) 
that were anticipated in \cite{AJ}
in higher analogy with the twisted unitary bundles on boundaries/D-branes in 10-dimensional
string theory.

%%%%%%%%%%%%%%%%%%%%%
\subsection{M-branes and $L_\infty$-algebras, Lie $n$-algebras and ``3-algebras''}
\label{LInfinityAlgebras}
%%%%%%%%%%%%%%%%%%%%%
The \emph{higher Lie groups} that appear in higher gauge theory -- 
such as the $\mathrm{String}$-2-group already mentioned - have an 
infinitesimal approximation by a higher analog of Lie algebras.
These higher Lie algebras are known as \emph{Lie-infinity algebras}
or \emph{$L_\infty$-algebras}, for short. 
(A description in the context that we need here
is in \cite{SSSI}). While an ordinary Lie algebra is a vector space 
equipped with a binary skew bracket that satisfies the Jacobi identity,
an $L_\infty$-algebra is a chain complex of vector spaces, which is 
equipped with $k$-ary skew graded brackets for all $k \in \mathbb{N}$, such that 
these
satisfy a certain joint higher analog of the Jacobi identity. 

\vspace{3mm}
If the underlying chain complex of an $L_\infty$-algebra 
is concentrated in the lowest $n$ degrees,
then we also speak of an \emph{$n$-term $L_\infty$-algebra} or \emph{Lie $n$-algebra}. 
These are the infinitesimal approximations to Lie $n$-groups. For instance, the
smooth 2-group $\mathrm{String}$ has a Lie 2-algebra: $\mathfrak{string}$.
This degree ``$n$" of Lie $n$-groups and Lie $n$-algebras is directly related to the 
dimension of the branes that can be charged under them. A brane with $n$-dimensional
worldvolume can be charged under a Lie $n$-group / Lie $n$-algebra. For instance,
there is an abelian Lie 2-algebra $b\mathbb{R}$, 
given by the chain complex concentrated on $\mathbb{R}$
in degree 1 and having trivial bracket. It is the Lie 2-algebra of the
Lie 2-group $\mathbf{B}{\mathrm{U}(1)}$, which is the smooth incarnation of the topological group
$B {\mathrm{U}(1)}$ that we encountered before. A 2-connection on a 2-bundle whose gauge 2-group is
$\mathbf{B}{\mathrm{U}(1)}$ and whose Lie 2-algebra is $b \mathbb{R}$ is precisely a 
Kalb-Ramond $B$-field. Indeed, this has a holonomy over 2-dimensional worldsheets
and the string with its 2-dimensional worldvolume is charged under it.

\vspace{3mm}
In direct analogy of this situation, there is a Lie 3-algebra $b^2 \mathbb{R}$ with Lie 3-group
$\mathbf{B}^2 U (1)$. A 3-connection with values in this is essentially what the
supergravity $C$-field is (we give the details in section \ref{CFieldConfigurations}).
A detailed discussion of Lie 3-algebras related to $C$-fields and Chern-Simons
couplings of 2-branes is in \cite{SSSI}, \cite{SSSIII} and \cite{FSS}, as are discussions of 
further higher analogs, such as the Lie 6-algebras related to the magnetic dual 
$C$-field and Chern-Simons couplings of the 5-brane, which we will consider
in section \ref{SecondPontrjaginOnStacks}. 
The description of supergravity theories by D'Auria and Fr{\'e} \cite{CDF}
can also be formulated in terms of higher gauge field with values in super $L_\infty$-algebras.
Notably there is a super Lie 3-algebra  
and a super Lie 6-algebra extension of the
super Poincar{\'e}-Lie algebra such that the action functional of 11-dimensional 
supergravity is a variant of a higher Chern-Simons action for these. Moreover,
the infinitesimal automorphism $L_\infty$-algebra of these contains in degree 0 the M-theory super
Lie algebra (\cite{SSSI} and section 4.3.2.2 of \cite{survey}).

\vspace{3mm}
There is therefore ample  theory and examples for the role of 
Lie 2-algebra in string theory, the role of Lie 3-algebras
in membrane theory, the role of Lie 6-algebras in 5-brane theory and generally
of Lie $(n+1)$-algebras in $n$-brane theory, 
a large part of which we discussed before (starting in \cite{SSSI})  and some part of which will concern us here.  

\medskip

A different and conjectural proposal for a role of 
higher Lie algebraic structures in membrane theory has been proposed in \cite{BL},
motivated from a supersymmetric extension of the M2-brane action. 
There a certain trilinear term appears, satisfying an
invariant condition which the authors called a ``3-algebra'' structure,
a terminology subsequently picked up by many publications.
In the process, the term transmuted sometimes into ``3-Lie
algebra"  and sometimes even into ``Lie 3-algebra".
Unfortunately, the Bagger-Lambert ``3-algebra" is not a Lie 3-algebra
in the established sense of an $L_\infty$-algebra structure on a graded
vector space $V$. 
The reason is that for the notion of an $L_\infty$-algebra it is crucial
that $V$ is an $\mathbb{N}$-graded (or $\Z$-graded) vector space and that the $n$-ary
brackets respect the degree in a certain way. But in the
Bagger-Lambert proposal, $V$ is all concentrated in a single degree (is
regarded as ungraded). One immediately finds that in this case the
$L_\infty$-respect of the trinary bracket for the grading would implies that $V$ is taken
to be in degree $\tfrac{1}{2}$. Since this is not in $\mathbb{N}$, it does not yield an 
$L_\infty$-algebra.
But the $\mathbb{N}$-grading (or $\Z$-grading) of $L_\infty$-algebras is crucial
for the homotopy theoretic interpretation of $L_\infty$-algebras as higher
Lie algebras. None of the good theory of $L_\infty$-algebras survives when
this grading is dropped. This grading has its origin in the Dold-Kan
correspondence, which establishes integral graded homological
structures as models for structures in homotopy theory
(see section 2.1.7 in \cite{survey} for a discussion of this in the context
of higher gauge theory).
Notably, a higher Lie algebra is supposed to have a Lie integration to
a smooth $n$-groupoid. Under this process, the elements in degree $k$ of
the higher Lie algebra become tangents to the space of $k$-morphisms of
this smooth $n$-groupoid. Clearly, here only integer $k$ make any sense.

\vspace{3mm}
On the other hand, it is of course possible to consider the structure
of ``$L_\infty$-algebras without grading", even if these will not have a good
theory. This notion has once been introduced by Filippov 
\cite{Fi}
 under the name ``$n$-Lie algebra''.
The innocent-looking difference between the terms
``Lie $n$-algebra" and  ``$n$-Lie algebra"
corresponds, unfortunately, to a major difference in the behavior of
the concepts behind these terms.
It was argued in \cite{LP,LR} that these ``3-algebras'' might also 
play a role in the description of multiple M5-branes. 
For that, a nonabelian generalization of the 
field content given by the $(0,2)$ tensor multiplet is proposed; 
this involves nonabelian versions of the fields in that supermultiplet, 
namely the scalars, the fermions and the antisymmetric 3-form. In addition, 
a nonabelian gauge field and a non-propagating vector are 
introduced. In the construction, however, the nonabelian two-form 
$B_{\mu \nu}^a$ never appears, which seems to be a problem for the 
quantum theory. 

\vspace{3mm}
On general grounds, it is clear from our point of view  that 2-brane physics is governed by Lie
3-algebraic structures, but it is not yet clear how the trinary
operation highlighted in \cite{BL} would be an example.
In view of this, it might be noteworthy that the equivalent
reformulation and generalization of the BLG model by the ABJM model \cite{ABJM}
does not involve any ``3-algebras" at all.
On the other hand, comparison with other structures suggests that
possibly the trinary operation is indeed a structure in higher Lie
theory, but not the trinary bracket on an $L_\infty$-algebra. Instead, it
can be seen to be in analogy with a higher \emph{symplectic structure}, a
``2-plectic structure''. This is argued in \cite{SS}, and this would make sense also in
homotopy theory (see the section 4.5 on higher symplectic geometry in \cite{survey}).

\vspace{3mm}

In conclusion, the reader expecting to see higher Lie algebraic structures
on the M5-brane will find them play a pivotal role in our discussion in 
sections \ref{Nonabelian} and \ref{7dTheory}. 
It is not, however, quite the kind of algebraic structure that 
 \cite{BL, LP,LR} propose, but one that has a good homotopy-theoretic
 interpretation.

%{\color{green}
%The supersymmetric equations of motion for the 
%3-Lie algebra $(0, 2)$ tensor multiplet derived in \cite{LP}
% can be interpreted as gauge field equations on loop space
% \cite{PaS}.
%}

%%%%%%%%%%%%%%%%%%
\subsection{Holography and Chern-Simons theory}
%Evidence from holography}
\label{DualityAnd7dCS}
%%%%%%%%%%%%%%%

In this section we  give   a physical argument that 
the 7-dimensional nonabelian gauge theory, to be discussed more fully 
below in section \ref{7dTheory}, is the Chern-Simons part of 
11-dimensional supergravity on $\mathrm{AdS}_7 \times S^4$
with 4-form flux on the $S^4$-factor and 
with quantum anomaly cancellation conditions taken into account.
We, moreover, argue that this implies that the 
states of this 7-dimensional CS theory over a 7-dimensional
manifold encode the conformal blocks 
of the 6-dimensional worldvolume theory of coincident M5-branes. 
The argument is based on the available but incomplete
knowledge about $\mathrm{AdS}/\mathrm{CFT}$-duality,
as reviewed in \cite{AGMOO}, and cohomological
effects in M-theory as discussed in \cite{Sati10}.

\medskip

We start in section \ref{AdSCompactifications} with some remarks
about the relevant compactifications of 11d sugra to $\mathrm{AdS}_7$. Then in 
\ref{TheSugraChernSimonsTerm} we discuss the subtleties
of quantum anomaly corrections to the 7-dimensional Chern-Simons theory
inside 11-dimensional supergravity.

%
%\paragraph{Boundary of AdS space.}
%The boundary of AdS space is the conformal boundary. 
%The space AdS${}_n$ is conformal to a compact solid cylinder with 
%conformal boundary having the topology $\R \times S^{n-2}$. 
%The conformal factor, which vanishes at infinity has 
%a nonzero normal derivative there. 
%AdS space itself has the general topology $S^1 \times \R^{n-1}$. 
%Let $M^{n+1}$ be an $(n+1)$-dimensional space-time, i.e. a connected
%time-oriented Lorentzian manifold, with metric $g$. $M$ is an 
%asymptotically locally anti-de Sitter space-time provided that there exists a spacetime-with-boundary
%$M'$, with 
%Lorentzian metric $g'$, such that (see \cite{HE} \cite{AM})
%\begin{enumerate}
%\item the boundary at infinity $\mathcal{I}=\partial M'$ is timelike, i.e. is a Lorentzian manifold in the 
%metric induced from $g'$; 
%\item $M$ is the interior of $M'$;
%\item The physical metric falls at a reasonable rate: 
%$g'=f^2 g$, where $f$ is a smooth function on $M'$ such that $\Omega>0$ 
%on $M$ and $\Omega=0$ along $\mathcal{I}$ 
%\end{enumerate}
%The canonical example is universal anti-de Sitter space-time which conformally 
%embeds into the static Einstein universe of topology $\R \times S^n$, so that the 
%closure $M'$ is $\R \times \mathbb{B}^n$, where $\mathbb{B}^n$ is a closed hemisphere of 
%$S^n$ and  $\mathcal{I}=\R \times S^{n-1}$. 
%

%%%%%%%%%%%%%%%%%%%%%%%%%%%%%%%%%%%%%%%%%%%%%%%%%
\paragraph{$\mathrm{AdS}$-compactifications.}
\label{AdSCompactifications}
%%%%%%%%%%%%%%%%%%%%%%%%%%%%%%%%%%%%%%%%%%%%%%%%%%

There are 6d theories with different amount of
supersymmetry, whose duals under $\mathrm{AdS}_7/\mathrm{CFT}_6$
have, in particular, different boundary behaviors of the $C$-field. 
(We analyze the moduli for different boundary conditions 
in detail in \cite{FiSaSc}).
While the  maximally supersymmetric $(0,2)$-theory is dual to supergravity on 
$\mathrm{AdS}_7 \times S^4$, there is also a $(0,1)$-superconformal 
theory in 6d, and it is dual to a compactification on a 
$\mathrm{AdS}_7 \times \mathbb{C}^2 /\!/\mathbb{Z}_k$-orbifold, 
with the 5-branes sitting at an orbifold fixed point. 
Whereas in the first case the supergravity fields are otherwise
unconstrained, in the second case they
satisfy boundary conditions as in Ho{\v r}ava-Witten theory
\cite{HoravaWitten}.

\paragraph{\it The minimal supergravity in seven dimensions.}

The field content of the massless representations of the minimal 
$D=7$, $N=2$ supergravity coupled to vector multiplets, written 
in terms of 2-form potential is \cite{BKS}
$
(g, B_2, A_1^I, \phi^\alpha, \phi, \psi_1^i, \chi^i, \theta^{\alpha i})$,
 where the first five fields are bosonic and the last three are fermionic. 
The reduction to six dimensions is as follows:
\begin{enumerate}
\item This minimal gauged supergravity 
compactified on $S^1$ leads to non-chiral $N=(1,1)$ 6d supergravity
\cite{GPv}.
\item The theory reduces on the orbifold $S^1/\Z_2$ to 6d, $N=(0,1)$, 
chiral theory  \cite{AK}. Only fields of even $\Z_2$-parity survive on the 
two orbifold planes: $(g, B_2, A^I, \phi^\alpha, \phi, \xi)$, which 
includes the chiral multiplet 
$(g, B_2^+, \psi_1^{i-})$ with $B_2=B_2^+ + B_2^-$, a sum of self-dual and 
anti-self-dual parts.
This is a Ho{\v r}ava-Witten-like construction in seven dimensions
implementing a form of Green-Schwarz anomaly cancellation.
\end{enumerate}

\paragraph{\it The (0, 1) theory as dual to AdS${}_7$.}
The AdS${}_7$ vacuum with $N=2$ supersymmetry 
is the supergravity dual, in the context of the AdS/CFT correspondence, 
of the 6d, $N=(0, 1)$, SCFT \cite{FKPZ}.
%\paragraph{Remark. The appearance of gauge fields and gauging 
%of the 2-form potential.} 
The nonabelian gauging of the self-dual tensor fields 
can be performed by introducing tensor gauge degrees of freedom
with $p$-form gauge parameters, $p=\{0, 1,2\}$ . 
This is used in \cite{SSW} to build a 2-form potential which carries 
a representation of the structure group. This is possible due to the 
3-form potential which mediates couplings between the tensor and vector multiplets.  
 The full nonabelian field strengths of the gauge field $A$ and the 2-form 
 gauge potential $B$ are proposed there to be 
\(
 \mathcal{F}^r= F^r + h_I^r B^I\;, \qquad
 \mathcal{H}^I= d_A B^I + \mathrm{CS}^I + g^{Ir}C_r\;,
   \label{SSWProposalForCurvatures}
 \)
 where $F$ is the ordinary curvature 2-form of the gauge field, 
 $d_A$ is the $A$-covariant exterior derivative, $\mathrm{CS}^I$ is a
Chern-Simons 3-form of $A$ for some bilinear form, and 
 $g^{Ir}$ and $h_I^r$ are couplings of St\"uckelberg type.
 
Our approach provides a systematic way of obtaining consistent  
couplings that include terms of this kind of form.\footnote{
See the discussion around equation (\ref{2FormCurvature})
for the case of untwisted String-2-connections, 
and then section \ref{7dCSInSugraOnAdS7}
for the general case of twisted String-2-connections.}
%{\color{green}
In particular, within the heuristic model presented in 
section \ref{EvidenceFromTheWorlvolumeTheory},
we interpret the appearance of a 3-form potential $C$ as
an indication of the presence of a Lie group  $G$, in addition to the 
appearance of a 2-form potential $B_2$ which indicates the presence of a 
based loop group (of that Lie group).

%%%%%%%%%%%%%%%%%%%%%%%%%%%%%%%%%%%%%%%%%%%%%%%%%%%%%%%%%%%%%%%%%%%%%%
\paragraph{\it The anomaly-corrected nonabelian 7d Chern-Simons term.}
\label{TheSugraChernSimonsTerm}
%%%%%%%%%%%%%%%%%%%%%%%%%%%%%%%%%%%%%%%%%%%%%%%%%%%%%%%%%%%%%%%%%%%%%%

Generally, there are two, seemingly different, realizations of the
\emph{holographic principle} in quantum field theory.  
On the one hand, Chern-Simons theories in dimension $4k+3$
have spaces of states that can be identified with spaces
of correlators of $(4k+2)$-dimensional conformal field theories
(spaces of ``conformal blocks'') on their boundary. 
For the case $k = 0$ this was discussed in \cite{WittenCS}, 
for the case $k = 1$ in \cite{Witten96, HNS, He}, and the case $k=2$ 
in \cite{BM}. On the other hand, 
$\mathrm{AdS}/\mathrm{CFT}$ duality (see \cite{AGMOO} for a review)  
identifies
correlators of $d$-dimensional CFTs with states of compatifications
of string theory, or M-theory, on asymptotically anti-de Sitter
spacetimes of dimension $d+1$ (see \cite{Witten98a}).

\medskip
However, in \cite{Witten98} it was pointed out that these two mechanisms are 
in fact closely related. A detailed analysis of the
$\mathrm{AdS}_5/\mathrm{SYM}_4$-duality shows that the
spaces of correlators of the 4-dimensional theory can be identified
with the spaces of states obtained by geometric quantization
just of the Chern-Simons term 
in the effective action of 
type IIB string theory on $\mathrm{AdS}_5$. The relevant part of this action 
locally
reads
\(
  (B_{\mathrm{NS}}, B_{\mathrm{RR}})
  \mapsto 
  N \int_{\mathrm{AdS}_5} B_{\mathrm{NS}} \wedge d B_{\mathrm{RR}}
  \,,
\)
where $B_{\mathrm{NS}}$ is the local Neveu-Schwarz 2-form field, 
$B_{\mathrm{RR}}$ is the local RR 2-form field, and where  $N$ is the RR 
5-form flux picked up from integration over the (internal) $S^5$ factor.

\paragraph{\it The abelian theory.}
As briefly indicated in \cite{Witten98}, 
the similar form of the Chern-Simons term of
11-dimensional supergravity (M-theory) on $\mathrm{AdS}_7$
suggests that an analogous argument shows that, under 
$\mathrm{AdS}_7$/$\mathrm{CFT}_6$-duality, the conformal blocks
of the $(0, 2)$-superconformal theory are identified with the 
geometric quantization of a 7-dimensional Chern-Simons theory.
%In \cite{Witten98} t
That Chern-Simons action is taken, 
locally on $\mathrm{AdS}_7$, to be (up to an overall numerical factor)
\(
\begin{aligned}
  C_3 &\mapsto \int_{\mathrm{AdS}_7 \times S^4}
    C_3 \wedge G_4 \wedge G_4
	& =
	N \int_{\mathrm{AdS}_7} C_3 \wedge d C_3
\end{aligned}
\,,
\)
where now $C_3$ is the local incarnation of the supergravity $C$-field, 
and where $G_4$ is its curvature 4-form locally equal to $d C_3$.
This is the $(4 \cdot 1 + 3 = 7) $-dimensional abelian Chern-Simons theory
shown in \cite{Witten96} to induce on its 6-dimensional boundary the self-dual
2-form, in the \emph{abelian} case.

\paragraph{\it The nonabelian theory.}
We may notice, however, that there is a term that is missing from (or that can be added to) 
the above
Lagrangian. The quantum anomaly cancellation via M5-branes in 11-dimensional supergravity
is known to require instead a Lagrangian whose Chern-Simons
term locally reads
\(
\begin{aligned}
  (\omega, C_3) &\mapsto \int_{\mathrm{AdS}_7 \times S^4}
    C_3 \wedge \left(\tfrac{1}{6} G_4 \wedge G_4 - I_8^{\mathrm{dR}}(\omega)\right)
	\label{SugraChernSimonsTerm}
	\,,
\end{aligned}
\)
where $\omega$ is the Spin connection form, locally, and where
$I_8^{\mathrm{dR}}(\omega)$ is a de Rham representative of the integral 
cohomology class \cite{DLM,VW} 
\(
  I_8 = \tfrac{1}{48}\left(p_2 - \lambda^2\right)\;,
  %(\frac{1}{2}p_1) \cup (\frac{1}{2}p_1)\right)
  \label{I8}
\)
where 
\(
  \lambda := \frac{1}{2}p_1
  \label{Lambda}
\)  
with $p_1$ and $p_2$  the first and second Pontrjagin classes,
respectively, of the given $\mathrm{Spin}$ bundle over 11-dimensional
spacetime $X$.
This means that after passing to the effective theory on 
$\mathrm{AdS}_7$, this corrected Lagrangian picks up another
7-dimensional Chern-Simons term, now one which depends on 
\emph{nonablian} fields. Locally, this reads 
\(
  \begin{aligned}
     S_{7d\mathrm{CS}}
	 :
     (\omega, C_3) & \mapsto 
	 \frac{N}{6}\int_{\mathrm{AdS}_7} C_3 \wedge d C_3
	 -
	 N\int_{\mathrm{AdS}_7} \mathrm{CS}_{I_8}(\omega)\;,
  \end{aligned}
  \label{The7dLagrangianInMotivation}
\)
where 
$
  N := \int_{S^4} G_4
$
is the $C$-field flux on the 4-sphere factor and
$\mathrm{CS}_{I_8}(\omega)$ is some Chern-Simons form for 
$I_8^{\mathrm{dR}}(\omega)$, defined locally by (see also \cite{SSSII,Sati10})
\(
  d \mathrm{CS}_{I_8}(\omega) = I_8^{\mathrm{dR}}(\omega)
  \,.
\)
However, the above action functional, which is locally a functional of a
3-form and a $\mathrm{Spin}$ connection, cannot globally be of this form, as
even the field that looks locally like a $\mathrm{Spin}$ connection 
cannot globally be a $\mathrm{Spin}$ connection.
To see this, we first notice that there
is a quantization condition on the supergravity fields on the 
11-dimensional $X$
\cite{WittenFluxQuantization}, which in cohomology requires the
identity
\(
  2[G_4] = \tfrac{1}{2}p_1 +  2 a \quad
  \text{in }\quad H^4(X,\mathbb{Z})
  \,,
  \label{QuantizationCondition}
\)
where on the left we have the integral 
class underlying the $C$-field, and
on the right we have the sum of the first fractional Pontrjagin class
of the $\mathrm{Spin}$-connection and 
the canonical class $a$ of an `auxiliary' or `topological'
$E_8$ bundle 
on the 11-dimensional spacetime $X$. 
%This is 
%essentially the datum of the second Chern 
%class of a $\mathrm{U}(n)$-bundle on $X$.

\vspace{3mm}
Moreover, by the arguments in \cite{Sati10Twist} we expect that the 
integral class of the $C$-field vanishes on (a vicinity of) the 
5-brane. This means that on an asymptotic neighbourhood 
%$\sim (\partial X) \times i$
of the asymptotic boundary $\partial X$, the above quantization condition becomes
\(
 \tfrac{1}{2}p_1 + 2 a=0 \quad
  \text{in }\quad H^4(\partial X, \mathbb{Z})
\label{cond}
 \,.
\)
Notice that requiring $[G_4] = 0$ at the boundary means that the $C$ field
is still there, but given by a globally defined differential 3-form $C_3$.

\vspace{3mm}
\noindent 
As we have indicated around (\ref{HomotopyPullbackFirstAppearance}), 
imposing condition \eqref{cond} in a gauge equivariant
way involves refining it from an equation between cohomology classes
(hence gauge equivalence classes) to a \emph{choice of coboundary} between cocycles
for $\tfrac{1}{2}p_1$ and $2a$.
Doing so has two effects. 
\begin{enumerate}
\item 
The first is that, according to 
\cite{SSSIII,Sati10Twist,FSS}, what locally looks like
a Spin connection is globally instead a 
\emph{2-connection on a twisted 
$\mathrm{String}$-principal 2-bundle}, or equivalently 
a \emph{twisted differential String structure}, where the twist is given by
the class $2 a$.  
The total space of such a principal 2-bundle may be
identified \cite{survey} with a (twisted) \emph{nonabelian bundle gerbe}.
Therefore, the configuration space of fields of the effective 7-dimensional
nonabelian Chern-Simons action above should involve not just Spin connection forms,
but  also \emph{$\mathrm{String}$-2-connection} form data. By \cite{SSSI} this
is locally given by nonabelian 2-form field data. 

\item 
The second effect is that on the space of twisted String-2-connections,
the differential 4-form $\mathrm{tr}(F_\omega \wedge F_\omega)$,
which under the Chern-Weil homomorphism represents the image of 
$\frac{1}{2}p_1$, locally satisfies \cite{SSSIII,FSS}
\(
  d \mathcal{H}_3 = \mathrm{tr}(F_\omega \wedge F_\omega) + 2 \mathrm{tr}(F_A \wedge F_A) - 2 d C_3
  \,,
\)
where $\mathcal{H}_3$ is the 3-form curvature component of the twisted $\mathrm{String}$-2-connection,
and where $F_\omega$ and $F_A$ are the curvatures of 
the connection $\omega$ on the Spin bundle and 
of a connection $A$ on the auxiliary $E_8$ bundle, respectively.
This is the twisted Bianchi identity of the curvature 3-form,
or equivalently 
the de Rham refinement of equation
(\ref{QuantizationCondition}), whose form is unaffected by the 
integral constraint (\ref{cond}).

\end{enumerate}

\noindent Therefore the quantum correction term in the supergravity Lagrangian
\eqref{The7dLagrangianInMotivation} now becomes (still for local data)
\(
   - N \int
   \mathrm{CS}_{I_8}(\omega)
   = 
   \frac{N}{48}
   \int 
   \left(
   \left(
     \mathcal{H}_3 + 2 C_3 - 2 \mathrm{CS}_3(A)
   \right)
   \wedge
   \left(
     d \mathcal{H}_3 + 2 d C_3 - 2\langle F_A \wedge F_A\rangle
   \right)
   -
   \mathrm{CS}_7(\omega)
   \right)
  \label{main action}
  \,,
\)
%Note that, despite the $C$-field has cohomologically vanished, we have a reminiscence of the 7-dimensional abelian Chern-Simons action term  $\frac{1}{6}\int_{\mathrm{AdS}_7} C_3\wedge dC_3$ from the original action in the term $\frac{1}{48}\int_{\mathrm{AdS}_7} H_3\wedge dH_3$ in equation \eqref{main action}. Yet, it should be remarked that despite both the fields $C_3$ and $H_3$ are locally given by abelian three forms, they have a very different global nature: $C_3$ is the local datum of a connection on a 2-gerbe, while $H_3$ is the local datum for the curvature of a (twisted) String-connection. 
where $\mathrm{CS}_3(A)$ is the ordinary Chern-Simons term for the $E_8$-connection 
\(
  \mathrm{CS}_3(A) = 
  \mathrm{tr}(A \wedge d A) 
  +
  \tfrac{2}{3}\mathrm{tr}(A \wedge A \wedge A)
  \,,
\)
and where $\mathrm{CS}_7(\omega)$ is the degree 7 Chern-Simons term
for the Spin-connection, given by 
\(
\begin{aligned}
  \mathrm{CS}_7(\omega)
  = & 
  \langle \omega \wedge d \omega \wedge d\omega \wedge d\omega \rangle
  +
  k_1\langle \omega \wedge [\omega \wedge \omega] \wedge d\omega \wedge d\omega \rangle
  \\
  & + 
  k_2\langle \omega \wedge [\omega \wedge \omega] \wedge [\omega \wedge \omega] \wedge d\omega \rangle
  +
  k_3\langle \omega \wedge [\omega \wedge \omega] \wedge [\omega \wedge \omega] 
  \wedge [\omega \wedge \omega] \rangle \;,
 \end{aligned}
 \label{Degree7ChernSimonsForm}
\)
for suitable scalar constants $k_i$  (see \cite{SSSII}).
Notice that when multiplying out the brackets, a term proportional to $\int C_3 \wedge d C_3$
appears. Since, by the boundary condition, $C_3$ is here a globally defined form,
this term may be rescaled by rescaling the 3-form. Therefore, we can absorb the
first summand of (\ref{The7dLagrangianInMotivation}) into the quantum correction
Lagrangian and take the 7d Chern-Simons action to be a multiple of
$\int \mathrm{CS}_{I_8}$.

\vspace{3mm}
In \cite{SSSIII} we have discussed the local data of 
such 7-dimensional nonabelian Chern-Simons Lagrangian 
on String-2-connections. In \cite{FSS} we have
provided the global description of the action functional
on the full moduli 2-stack of $\mathrm{String}$-2-connections.
There we had concentrated on the role of this 7-dimensional action 
in the definition of \emph{twisted differential Fivebrane structures}.
In section \ref{7dTheory} below we discuss its role as a 
fully-fledged 7-dimensional Chern-Simons theory on 
nonabelian 2-form fields.

%\end{enumerate}

\medskip
\noindent We therefore see that 
including the $I_8$-correction term and the refined quantization condition in 
$\mathrm{AdS}_7/\mathrm{CFT}_6$ leads to a 7-dimensional 
theory that contains orthogonal and unitary nonabelian 
gauge field degrees of freedom and which is globally controled by
a twisted nonabelian gerbe structure. As we discuss in detail below
in section \ref{Sec CS elements},
%2Connections},
this kind of data has several rather different looking
equivalent incarnations, due to the fact that these higher connections
have a much richer gauge structure that ordinary connections.
In particular, one can pass from descriptions that locally look
only slightly nonabelian but have complicated global transformation laws,
to equivalent descriptions that have global transformation laws
more like ordinary connections but which, in compensation, exhibit
a richer nonabelian structure locally:
it is locally given \cite{SchreiberWaldorfII}\cite{SchreiberWaldorfIII} 
by a
1-form $A \in \Omega^1(-,P_* \mathfrak{so})$ with values in the Lie algebra
of paths in the Lie algebra $\mathfrak{so}$ of the orthogonal group 
 and a 2-form 
$B \in \Omega^2(-,\hat \Omega \mathfrak{so})$ with values in 
the Kac-Moody central extension of the loop Lie algebra of $\mathfrak{so}$.

\medskip
%In summary, these considerations suggest the following 
%statement.
%\medskip
%\emph{
%The spaces of conformal blocks of the 6-dimensional $(0, 2)$-theory
%should be the spaces of states of a 7-dimensional Chern-Simons theory
%that locally involves two abelian Chern-Simons terms, as well
%as two 7-dimensional nonabelian Chern-Simons terms for 
%nonabelian 2-form gauge fields with values in Kac-Moody algebras.}

\medskip

The above line of arguments suggests that the Chern-Simons term
that governs 11-dimensional supergravity on $\mathrm{AdS}_7 $
is an action functional on 
fields that are twisted $\mathrm{String}$-2-connections
such that the action functional is locally given by expression
(\ref{The7dLagrangianInMotivation}). 
In Sections \ref{CupProductTheoryOfTwo3DCSTheories} 
and \ref{InfinCS7CS} we discuss a precise formulation of 
higher Chern-Simons theories satisfying these properties
and which extend beyond Anti-de Sitter spaces.

%{\color{red}

%%%%%%%%%%%
\subsection{Which gauge group(s)?}
%%%%%%%%%%%%

We ask the natural question: Is/are there (a) particular 
Lie group(s) that is/are associated with the fivebrane theory?
We provide several further arguments which favor the exceptional 
groups, including $E_8$. 
\footnote{This section expands in part on the discussion in section \ref{EvidenceFromTheWorlvolumeTheory}.}

\medskip

First, note that the anomaly cancellation in the 
ambient seven-dimensional theory with boundary,
leading to chiral $N=(0, 1)$ 6d theory, is 
worked out in \cite{GK}. 
The resulting admissible groups from the anomaly argument 
 are indeed the exceptional groups. 

\medskip
 The system of multiple M5-branes, generalizing the
system of $n$ D-branes leading to U($n$) nonabelian 
gauge symmetry, can be described by twisted $\mathrm{String}(G)$-
gerbes, whose cocycle data involves the universal central extension 
$\hat \Omega G$ of the
based loop group $\Omega G$, 
where $G$ is any of the Lie groups 
%${\rm Spin}(n)$, $n \geq 7$, 
$E_6$, $E_7$, $E_8$, $F_4$, and $G_2$, as was also 
 proposed in \cite{AJ}.
Indeed, it has been argued in \cite{Urs-thesis} 
that classical membrane fields are loops.

\medskip
The form of the action of the fivebrane suggests working in eight
dimensions, where the interpretation of the terms becomes transparent.
This also suggests the existence of an $E_8$ gauge theory on this
eight-dimensional extension $Z^8$ of the worldvolume $M^6$. 
%This will be our starting point in this section.
The topological part of the action 
extended to eight dimensions looks like \cite{Witten96} 
\(
\langle ~[G_4] \cup [G_4] - \lambda \cup [G_4]~,~[Z^8]~ \rangle\;, 
\)
with $[G_4]$ a degree four characteristic class
and $[Z^8]$ the fundamental class on which the composite degree 
eight cohomology class is evaluated. 
Now we observe that, up to dimension 8, all exceptional Lie
groups have the same homotopy type as the Eilenberg-MacLane space
$K(\Z, 3)$, so that $G_4$ can be viewed essentially as the
characteristic class of any bundle with structure group an 
exceptional Lie group $G_2$, $F_4$, $E_i$, $i=6,7,8$. 
Therefore, while $E_8$ is not singled out by this argument, it is
certainly the case that it could be assumed to describe the 
extension of the fivebrane worldvolume to eight dimensions.

\vspace{3mm}
The space of conformal blocks, in the sense 
of Witten \cite{Witten09}, has dimension bigger than one for 
groups other than $E_8$. This means that the theory does not have  a distinguished
partition function. Therefore, this quantum field theory argument favors $E_8$.

\vspace{3mm}
We hence start with an $E_8$ bundle on the 8-dimensional 
extension $Z^8$. This extension is the 
total space of the 2-disk bundle over $M^6$ (in the sense of \cite{MaSa,S-gerbe}), 
and the process of extension 
can be done in two different ways. The first way is to take $M^6$ 
to be the boundary of $W^7$ and then take this 7-dimensional space
to be the base space of a circle bundle with total space $Z^8$. 
The second way is to take $M^6$ itself as the base space of a circle
bundle with total space a 7-dimensional manifold $Y^7$, which 
we take to be the boundary of $Z^8$. So the two-disk bundle 
$
\mathbb{D}^2 
\to
Z^8
\to
M^6
$ 
can be viewed as 
\begin{center}
\begin{minipage}{2in}
\begin{equation}
\begin{split}
\xymatrix{
S^1 
\ar[r]
&
Z^8
\ar[d]
\\
& 
W^7
&
~M^6=\partial W^7~
~\ar@{_{(}->}[l]_{\hspace{-5mm} i}
%\ar[u]^i
}
\end{split}
\nonumber
\end{equation}
\end{minipage}
\hspace{1.5cm}
or
\hspace{1.5cm}
\begin{minipage}{3in}
\xymatrix{
Z^8
&
~~Y^7=\partial Z^8
\ar@{_{(}->}[l]_{\hspace{-5mm}j}
\ar[d]
&
S^1
\ar[l]
\;.
\\
&
M^6
&
}
\end{minipage}
\end{center}

\paragraph{Why loop bundles?}
Proceeding with the second case, we have 
to some extent a seven-dimenional analog of the 
eleven-dimensional Ho{\v r}ava-Witten set-up. 
Here will will consider the M5-brane worldvolume 
$M^6$ to be disconnected. That is we will take $M^6$ to be composed
of the two boundary components for  the manifold with boundary $W^7$.
Taking an $E_8$ bundle on $Z^8$ gives us an $LE_8$ bundle on 
$W^7$, in a process which is similar to the one relating 
M-theory to type IIA string theory (see \cite{Ev,MaSa}). 
Then we consider the reduction of this $LE_8$ bundle to 
$M^6$, in a process analogous to Horava-Witten, except that it is 
for loop bundles instead of finite-dimensional bundles. 
We get this way an $LE_8$ bundle on 
each boundary component $M_i^6$, $i=1, 2$. 

\vspace{3mm}
Let us now go with the restriction of the $E_8$ bundle on 
$Z^8$ to its boundary $Y^7$. This will give an $E_8$ bundle,
whose reduction over the circle fiber down to $M^6$ leads again 
to an $LE_8$ bundle, in this case in exact analogy to the case
of going from $Z^{12}$ to the M-theory boundary and then 
reducing on the M-theory circle to get type IIA string theory (see 
\cite{MaSa,S-loop,S-gerbe}). 
Therefore, we have the following statement
%\begin{proposition}

\medskip
{\it For any exceptional Lie group $G$,
 there is a loop $G$ bundle on the fivebrane worldvolume. 
 This arises from the dimensional reduction of a $G$ bundle on the 
extended worldvolume. }
%\end{proposition}

\paragraph{Reduction to the D4-brane.}
 Assuming we have a bundle with structure group an exceptional 
 Lie group  $G$ or its loop group on the M5-brane worldvolume 
 $M^6$, it is natural to ask to what structure this reduces 
 on the D4-brane worldvolume. Since such a reduction would be 
 over a circle, one possibility is to perform a reduction similar to that 
 of going from M-theory to type IIA string theory, as in \cite{DMW}.
 For example, the inclusion $E_8 \supset \left(\mathrm{SU}(5) \times \mathrm{SU}(5)\right)/\Z_5$
 provides a way of getting unitary groups from exceptional groups. 
 Such unitary groups will be in the stable range due to the relatively
 low dimension of our spaces, so that we would effectively get unitary groups 
 of all ranks. 
 Breaking  loop bundles to
finite-dimensional (unitary) bundles is considered in 
a similar context in \cite{S-gerbe}.

\paragraph{Center-of-mass and noncommutativity.}
In the absence of a $B$-field, the field theory of $N$ coincident D-branes is a 
supersymmetric U($N$) gauge theory. The ${\mathrm{U}(1)}$ part represents the 
interactions of D-brane open string with the bulk closed strings, representing the 
supergravity fields. This ${\mathrm{U}(1)}$ center-of-mass part decouples from the 
open string dynamics leading to an effective $\mathrm{SU}(N)$ gauge theory. 
However, we will explain, in Section \ref{DeterminantLineBundles} below,
that this ${\mathrm{U}(1)}$ part does have a role to play
which is interesting in its own right. We will then generalize this discussion 
to the case of multiple M5-branes. 
%\paragraph{With $B$-field.}
%\vspace{3mm}
However, in the presence of a constant $B$-field background,
non-commutativity of the resulting gauge theory \cite{SW} makes it impossible to 
separate the center-of-mass part. The reason is that the $B$-field makes the 
left and right mover of the open string sector not to be treated on equal footing.  
Indeed, gauge transformations on SU($N$) do not close \cite{Ar}.
Noncommutative SO($N$) and Sp($N$) theories can also be constructed
with a similar phenomenon occurring \cite{BSST}.
We anticipate that multiple M5-branes in the presence of a $C$-field will 
lead to a similar inability to separate this analog the center-of-mass part. 
What this means, from the point of view of the description of the String group
in section \ref{EvidenceFromTheWorlvolumeTheory},
 is that 
we cannot separate the based loop group part $\Omega G$ 
from the finite-dimensional
part that corresponds to the underlying Lie group $G$, as their twisted
product is what forms the String group. Recall that in the above-mentioned model: 
%that we discussed in section \ref{String2GroupOn5Brane}: 
$S^1$ is replaced by $K(\Z,2)$, which is 
homotopy equivalent to a loop group of an exceptional Lie group in 
our range of dimensions. The $S^1$ is the center-of-mass in 
the case of D-branes; analogously, one may interpret the 
factor $K(\Z,  2)\sim_8 \Omega G$, for $G$ exceptional,
 as the `center-of-mass' in the 
case of M5-branes.

%
%\attn{We might need a stacky description of PU(H) bundles in the 
%twisted case}
%
%
%%
%%\vspace{3mm}
%%We consider differential refinement of twisted Spin${}^c$ structures. 
%
%
%
%
%\begin{theorem}
%... Multiple D-branes are described by the above system...
%\end{theorem}
%

%%%%%%%%%%%%%%%%%%%%%%%%%%%%%%%%%%%%%%%%
%\subsection{The twisted String structure on the M5-brane (extended) worldvolume}
%\label{ArgumentForTheTwistedStringStructure}
%%%%%%%%%%%%%%%%%%%%%%%%%%%%%%%%%%%%%%%%
%We now outline the argument, using \cite{Sati10Twist}, on the role of 
%twisted String structure on the 6-dimensional fivebrane worldvolume, as well as
%on its extension to seven and eight dimensions via the Chern-Simons construction.
%There it was shown how this arises both from the extrinsic as well as the intrinsic 
%description of the fivebrane, that is either from looking at the worldvolume theory
%with the fields in it, or from studying the mapping of the fivebrane to the target 
%11-dimensional manifold. The first argument stems from the form of the action, which 
%in eight dimensions looks like \cite{Witten96} \cite{HS}
%$\int_{Z^8} G_4 \cup G_4 - \lambda G_4$. The second argument relies on 
%roughly a two-out-three principle, that is if we have three vector bundles 
%$E_i$, $i=1,2,3$ sitting in a a sequence $E_1 \to E_2 \to E_3$, then if two of the
%bundles admit a structure then the third does.
%\footnote{Of course such a principle does not hold for just any structure, and 
%should satisfy a nice behavior under Whitney sum.}

%%%%%%%%%%%%%%%%%%%%%%%%%%%%%%%%%%%%%%%%%%%%%%%%%%%%%%%%
\subsection{Generalizations: Nontrivial normal bundle and the role of the ADE groups}
\label{Generalizations}
%%%%%%%%%%%%%%%%%%%%%%%%%%%%%%%%%%%%%%%%%%%%%%%%%%%%%%%%

We used in section \ref{DualityAnd7dCS} 
the comparatively good available understanding of 
holography over asymptotically $\mathrm{AdS}$ spaces
in order to give a plausibility argument that suggests that the
7-dimensional quantum field theory 
%which exists by theorem \ref{TheTheorem} 
encodes holographically the
nonabelian 6-dimensional $(0 , 2)$-theory. 
Once one accepts this, there are evident generalizations of
this 7-dimensional theory to more general setups than
covered by $\mathrm{AdS}/\mathrm{CFT}$. Here we indicate 
some of these generalizations and make connections to 
higher structures.

\vspace{3mm}
The M5-brane background in the eleven-manifold $X^{11}$ breaks the structure group 
Spin$(10, 1)$ 
of the Spin bundle to the (twisted) product bundle with structure group 
Spin$(5,1)\times$Spin$(5)$ corresponding to the breaking 
of the tangent bundle $TX^{11}|_{M^6}= TM^6 \oplus \cal{N}$, where
$\cal{N}$  is the normal bundle. 

\vspace{3mm}
The ${{N}}=(0, 2)$ theory associated with arbitrary group $G$ is suggested
to have an anomaly of the following general form when coupled to the 
${\rm Spin}(5)_R$ normal bundle $\cal{N}$ \cite{In}
\(
I_8(G)=r(G) I_8(1) + \frac{c(G)}{24}p_2(\cal{N})\;,
\label{I8G}
\)
where 
\(
I_8(1)=\tfrac{1}{48}\left(p_2({\cal{N}}) - p_2(M) + (\lambda({\cal{N}}) - \lambda(M))^2 \right)
\label{I81}
\) 
is the anomaly polynomial for a single free $(0, 2)$ 
supermultiplet 
%\cite{AgW} 
\cite{Witten96}.
Here $r(G)={\rm rank} (G)$ is the rank of the group and $c_G=\dim G\cdot h_G$,
where $h_G$ is the dual Coxeter number of $G$.

\vspace{3mm}
Now, motivated by our discussion on String structure associated to the 
M5-brane (see the discussion around expression \eqref{cond}),
 we impose our condition leading to a twisted String structure.
Inspecting the formulae \eqref{I8G} and \eqref{I81} we see that 
imposing first the condition $\frac{1}{2}p_1(M)- \frac{1}{2}p_1({\cal{N}})=0$ leads to 
a simplification. Indeed, with this twisted String condition
(with the twist being $\frac{1}{2}p_1(\cal{N})$), the general
anomaly polynomial \eqref{I8G} becomes
\(
I_8(G)_{\rm ts}=-\tfrac{1}{48}\left[p_2(M)-
\left( r(G) +2c(G)\right) p_2(\cal{N})
\right]\;.
\label{I8ts}
\)
Assuming the conjectural formula \eqref{I8G} holds, a couple of
 remarks are in order:

\begin{enumerate}

\item In the case of a twisted String structure, with the twist given by the
first Spin characteristic class of the normal bundle,
 we interpret the vanishing of the anomaly, given by expression 
 \eqref{I8ts}, as saying that we have a twisted Fivebrane 
 structure, with the twist again being due to the normal bundle,
that is the twist is given by a fractional second Pontrjagin class of the 
normal bundle. In order for this to serve as a twist
in the sense of \cite{SSSIII},
 it has to be an integral class. This imposes the condition
 \(
 \tfrac{1}{6}(r(G) + 2c(G)) \in \Z\;.
 \)
This might be considered as a condition analogous to 
%(or a generalization of) 
the condition $\frac{1}{6}c(G) \in \Z$, derived in \cite{In}.

\item We now would like to take expression 
\eqref{I8ts} as a basis for the nonabelian Chern-Simons 
term as in Section 
\ref{DualityAnd7dCS}. 
We can indeed apply the general formula as above (see
expression \eqref{main action}), 
but now we have both a Chern-Simons 7-form for the tangent 
bundle as well as one for the normal bundle. This then resembles the 
discussion in \cite{SSSIII} where the gauge bundle corresponding to 
the heterotic string plays an analogous topological role -- namely a twist -- 
that the normal bundle plays here. 
Our more general Chern-Simons theory will be given by expression
\eqref{main action} but with the last term there replaced by the Chern-Simons 
form corresponding to $I_8(G)_{\rm ts}$, given in 
expression \eqref{I8ts}.
\end{enumerate}
\noindent We summarize what we have as follows:

%\paragraph{Proposal}
\vspace{3mm}
{\it The Chern-Simons theory can be defined for a general ADE group. 
An ADE group $G$ induces the corresponding $\mathrm{String}(G)$-2-group
involving the centrally extended loop group $\hat \Omega G$, 
which serves as the structure 2-group  for 2-bundles that
underly the worldvolume two-form field. 
For $G$ nonabelian,
we have a nonabelian gerbe and hence a theory of multiple M5-branes.
A necessary condition for the existence of such theory is the presence of 
a twisted String structure, with the twist given by the normal bundle. This in 
turn leads to a twisted Fivebrane structure, with a similar type of twist 
arising from the normal bundle.}

%%%%%%%%%%%%%%%%%%%%%%%%%%%%%%%%%%%%%%%%%%%%%%%%%%%%%%%%
\section{Nonabelian higher gauge theory}
\label{Nonabelian}
%%%%%%%%%%%%%%%%%%%%%%%%%%%%%%%%%%%%%%%%%%%%%%%

We review here aspects of higher nonabelian connections and their higher gauge theory,
formulated on smooth higher stacks. We do so in a way that is 
tailored towards our application in 
section \ref{7dTheory} and should serve as a warmup for that application, but
the discussion is of independent relevance for higher nonabelian gauge theory
and for other of its applications in string theory. 
These structures and their applications have
been discussed in earlier work such as \cite{SSSI, SSSIII, FSS, survey}.
\footnote{The reader who feels reasonably  comfortable with the general ideas used there
might wish to skip ahead to section \ref{7dTheory}. }

%%%%%%%%%%%%%%%%%%%%%%%%%%%%%%%%%%%%%%%%%%%%
\subsection{Higher smooth moduli stacks -- integrated BRST Lie $n$-algebroids}
\label{HigherSmoothStacks}
%%%%%%%%%%%%%%%%%%%%%%%%%%%%%%%%%%%%%%%%%%%%

Physics has for a long time been concerned with and formulated in terms of \emph{geometry}.
But ever since the inception of gauge theory, it is secretly also concerned with and
formulated in terms of \emph{homotopy theory}. A \emph{gauge transformation}
is essentially what mathematically is called an \emph{isomorphism in a groupoid}
or \emph{homotopy in a space}. The central idea in physics that ``it is bad to quotient out
gauge transformations and good to remember them" is equivalently the idea in homotopy theory
that ``it is bad to force everything to be a discrete set (of equivalence classes) and
good to instead retain groupoids with their isomorphisms and spaces with their
homotopies".
Therefore, what \emph{really} matters in physics is the combination of both geometry
and homotopy theory. This is, then, a theory where we have \emph{geometric families of homotopies},
such as \emph{smooth families of homotopies}. For instance the configuration space of 
Yang-Mills theory is not just a smooth collection of field configurations, 
and is not just a groupoid of gauge transformations, but is a combination
of both, namely a \emph{smooth groupoid}. In this formulation, both the field 
configurations may vary smoothly,
as do their gauge transformations.

\vspace{3mm}
For a \emph{higher gauge theory} such as the 2-form theory of the $B$-field, 
we have the same situation, but with even richer structure. The $B$-field has a \emph{smooth 2-groupoid}
of field configurations, where in addition to the gauge transformations there are now
smooth families of gauge-of-gauge transformations. Next, for the supergravity $C$-field
the configuration space is correspondingly a \emph{smooth 3-groupoid}, and so on.

\paragraph{Why stacks?}
For historical reasons, a smooth groupoid is also called a 
\emph{stack on smooth manifolds}. This terminology is often used in the context 
of refined \emph{moduli spaces} which are then called \emph{moduli stacks}.
For instance, for $G$ any Lie group, there is a topological space $B G$
which is the ``moduli space of $G$-instantons'' in that for $X$ any manifold, 
the homotopy classes of maps $X \to B G$ correspond to equivalence classes of 
$G$-principal bundles ($G$-instantons) on $X$. The trouble with this concept
is that $B G$ does not know about the smooth gauge transformations 
given by $G$-valued functions, nor does it know about actual gauge fields,
namely about connections on $G$-principal bundles. 
This is where the moduli \emph{stacks} come in. There is a smooth groupoid / smooth
stack which we will write as $\mathbf{B}G$ and which is such that 
maps of smooth stacks $X \to \mathbf{B}G$ correspond to $G$-bundles on $X$,
and smooth homotopies of such maps correspond to smooth gauge transformations
of $G$-bundles. Furthermore, there is a \emph{differential refinement} to
a richer smooth stack which we denote $\mathbf{B}G_{\mathrm{conn}}$, and which
is such that maps $X \to \mathbf{B}G_{\mathrm{conn}}$ correspond to $G$-Yang-Mills 
gauge fields on $X$, and homotopies of such maps 
correspond 
to smooth gauge transformations.
Accordingly then, there is the smooth \emph{mapping stack} 
$[X, \mathbf{B}G_{\mathrm{conn}}]$
whose elements are gauge fields on $X$, and whose morphisms are gauge transformations.
This is the true ``configuration space" of Yang-Mills theory on $X$. 
If we forget the smooth structure on this, we write $\mathbf{H}(X, \mathbf{B}G_{\mathrm{conn}})$,
the \emph{cocycle groupoid} of nonabelian differential $G$-cohomology. Its connected components
$H(X, \mathbf{B}G_{\mathrm{conn}})$ is the set of gauge equivalence classes of field
configurations: the \emph{cohomology set} of nonabelian differential $G$-cohomology on $X$.

\paragraph{Importance in (higher) gauge theory.}
The above stack is in fact a global refinement of 
an object long familiar in gauge thery, namely the 
\emph{BRST-complex} for Yang-Mills fields on $X$. A BRST complex is,
in a precise sense, the infinitesimal approximation -- the \emph{Lie algebroid} -- 
of a smooth moduli stack of field configurations. 
The \emph{ghosts} of the BRST complex are the
cotangents to the spaces of morphisms / gauge transformations in the stack. 
For higher gauge theory, the order-$n$ ghosts-of-ghosts 
in the BRST complex are the cotangents
to the space of $n$-morphisms in the higher moduli stack
and exhibit a \emph{Lie $n$-algebroid} structure. 
This is one way to understand
the use of (higher) moduli stacks in physics, as the natural way to
incorporate the BRST quantization of (higher) gauge theories into a powerful
ambient mathematical context, and to refine it from infinitesimal
(higher) gauge transformations to finite ones.

\medskip
Therefore, similarly, there is for each natural number $n$ 
a \emph{higher} moduli stack $\mathbf{B}^n {\mathrm{U}(1)}_{\mathrm{conn}}$ of $n$-form gauge fields.
For instance $[X, \mathbf{B}^2 {\mathrm{U}(1)}_{\mathrm{conn}}]$ is the stacky configuration space
of the $B$-field on $X$, with its gauge transformations and gauge-of-gauge
transformations, whose infinitesimal approximation is the BRST complex
for a 2-form field with its ghosts and ghosts-of-ghosts. As opposed to the BRST
complex, the full stack of field configurations knows not just about the infinitesimal
gauge transformations,
but also of the finite gauge transformations. It therefore contains genuinely the full
information about the gauge field configurations.

\medskip

For all $n \in \mathbb{N}$, there is the smooth moduli $n$-stack $\mathbf{B}^n U(1)_{\mathrm{conn}}$
of $n$-form fields, discussed in more detail in section \ref{HigherU1Connections}.
Cohomology $H(X, \mathbf{B}^n U(1)_{\mathrm{conn}})$ with coefficients in this 
is \emph{ordinary differential cohomology}. More generally, for $G$ any higher smooth
group, one can consider higher moduli stacks of nonabelian $G$-connections. The corresponding
cohomology $H(X, \mathbf{B}G_{\mathrm{conn}})$ is \emph{nonabelian differential cohomology}.

\paragraph{The Dold-Kan correspondence.} 
Handling higher stacks is a bit more subtle than handling just manifolds or just
topological spaces, but there are a handful of simple but powerful tools 
that allow one to efficiently work with them in a way that is very close to 
common operations in physics. 
One such tool is, for instance, the \emph{Dold-Kan correspondence}. In
simplified terms this establishes that the homological algebra of 
\emph{chain complexes of sheaves}, 
which is familiar in string theory mostly from the study of the 
topological string, presents a sub-class of higher stacks, namely the
``strictly abelian'' higher stacks. For instance, for every sheaf $A$
of abelian groups on all smooth manifolds -- e.g. the
sheaf $A = {\mathrm{U}(1)}$ of smooth circle-valued functions -- there is
a chain complex of sheaves
\(
  A[n] = (A \to 0 \to 0 \to \cdots \to 0)
\)
concentrated on $A$ in degree $n$, and the Dold-Kan correspondence identifies
this (up to a suitable notion of equivalence) with a moduli $n$-stack
$\mathbf{B}^n A$ that classifies instanton configurations for 
$A$-valued gauge fields of higher order $n$. For instance, $\mathbf{B}^2 {\mathrm{U}(1)}$
classifies instanton configurations of $B$-fields and $\mathbf{B}^3 {\mathrm{U}(1)}$
classifies instanton configurations of $C$-fields. This we come to in
section \ref{HigherU1Connections}. For technical details on the 
Dold-Kan correspondence and higher stacks see section 2.1.7 in \cite{survey}.

\paragraph{Geometric realizations and smooth refinement.}
While smooth higher stacks have richer structure than topological spaces, 
there is a map called \emph{geometric realization} that sends
any smooth higher stack to the topological spaces which is the ``best approximation''
to it, in a precise sense.
This is an ``$\infty$-functor'' 
\footnote{See sections 3.2.2 and 3.3.3 of \cite{survey}. }
\(
  \vert-\vert : \mathrm{Smooth}\infty\mathrm{Grpd} \to \mathrm{Top}
  \,,
  \label{GeometricRealization}
\)
For instance the geometric realization of the moduli stack 
$\mathbf{B}\mathrm{Spin}$ of $\mathrm{Spin}$-principal bundles 
is the ordinary classifying space $B \mathrm{Spin}$
\(
  \vert \mathbf{B} \mathrm{Spin}\vert \simeq B \mathrm{Spin}
\)
(all up to weak homotopy equivalence).
And the geometric realization of the $n$-stack $\mathbf{B}^n {\mathrm{U}(1)}$ 
is the 
Eilenberg-MacLane space $K(\mathbb{Z}, n+1)$ (notice the degree shift)
which classifies integral cohomology
\(
  \vert \mathbf{B}^n {\mathrm{U}(1)}\vert \simeq K(\mathbb{Z}, n+1)
  \,.
\)
%(These statements are shown in section 3.2.2 of \cite{survey}). 
Geometric realization necessarily forgets crucial geometric information
and information about the nature of gauge transformations.
But for a large class of higher moduli stacks (not for all, but for all
non-differentially refined stacks that are of interest to us here), it 
remembers the information about gauge equivalence classes. 
For instance equivalence classes of morphisms of smooth stacks
$X \to \mathbf{B}E_8$ from a smooth manifold $X$ are in bijection
with homotopy classes of continuous maps $X \to B E_8$.
This is important for the present discussion, as very different
looking smooth higher moduli stacks may become equivalent 
after geometric realization. For instance the equivalence
\(
  \vert \mathbf{B} \mathrm{PU}(\mathcal{H})\vert 
 \simeq \vert \mathbf{B}^2 U(1)\vert
\) 
controls the 
 nonabelian cohomology of the restriction of the $B$-field to
 D-branes (section \ref{Section D-branes}) and the equivalence
\(
  \vert \mathbf{B}E_8\vert \simeq_{15}
  \vert \mathbf{B}^3 U(1)\vert
\)
controls the higher nonabelian cohomology of the restriction of the
$C$-field to M5-branes (section \ref{CFieldConfigurations}).

Using the notion of geometric realization, we may say that an ordinary
universal characteristic class $c \in H^{n+1}(B G, \mathbb{Z})$
has a \emph{smooth refinement} to a morphism of $n$-stacks
\(
  \mathbf{c} : \mathbf{B}G \to \mathbf{B}^n U(1)
\)
if the geometric realization $B G \to K(\mathbb{Z}, n+1)$ of 
this morphism represents $c$. For $G$ a compact Lie group, such
smooth lifts exist uniquely, up to equivalence, by theorem 3.3.29 in
\cite{survey}. (This is how we obtain the String 2-group in section
\ref{Section stacks}.) Since $\mathbf{B}^n U(1)$ is the moduli stack
for circle $n$-bundles / $(n-1)$-gerbes, $\mathbf{c}$ also constructs
an $(n-1)$-gerbe on the moduli stack $\mathbf{B}G$. The looping
$\Omega \mathbf{c}$ is there fore an $(n-2)$-bundle gerbe
over $G$ itself. For $n = 3$ this bundle-gerbe perspective 
on smooth refinements is spelled out in \cite{WaldorfString}.

\medskip

There are more such tools for handling higher stacks, 
but here we will not further dwell on recalling these. 
The interested reader can find explanations in  \cite{FSS} and in more
details in \cite{survey}. 
\footnote{The reader who does not know
yet and who cannot be bothered to go through 
these details should nevertheless be able to follow the discussion below,
if only he or she keeps the intuitive idea of a higher stack as a
collection of higher smooth families of higher gauge transformations in mind.
}

%%%%%%%%%%%%%%%%%%%%%%%%%%%%%%%%%%%%%%%%%%%%%%%%%%%%%%%%%%%%%%%%%
\subsection{Determinant line bundles and $\mathbf{c}_1$-twists}
\label{DeterminantLineBundles}
%%%%%%%%%%%%%%%%%%%%%%%%%%%%%%%%%%%%%%%%%%%%%%%%%%%%%%%%%%%%%%%%%

A central role in our discussion to follow is played by universal characteristic
maps refined to smooth stacks, and of their homotopy fibers. As a warmup and general
motivation, we expose here the simplest non-trivial example of this general concept, 
whose component ingredients are all still familiar from traditional theory. 
This is the example induced by the smooth refinement $\mathbf{c}_1$ of the 
first Chern class on unitary bundles.
In particular we describe how to 
 regard \emph{principal $\mathrm{SU}(N)$-bundles} (with $\mathfrak{su}_N$-connections) 
 equivalently as \emph{trivially $c_1$-twisted principal ${\mathrm{U}(N)}$-bundles
  (with $\mathfrak{u}_N$-connections)}.
\footnote{Our discussion of $\mathrm{String}$-2-connections in section \ref{Section stacks}
will proceed by close analogy with the constructions here.
  }
  
\vspace{3mm}
  Recall from the theory of characteristic classes that the obstruction to reducing the structure group of a principal $\mathrm{U}(N)$-bundle to $\mathrm{SU}(N)$ is the first Chern class of the bundle. The stacky perspective on this says: the smooth universal first Chern class is the homotopy class of the morphism of stacks
\(
  \mathbf{c}_1=\mathbf{B}\!\det:\mathbf{B}{\mathrm{U}(N)}\to \mathbf{B}{\mathrm{U}(1)}
  \label{DeterminantAsSmoothC1}
\) 
from the moduli stack of principal ${\mathrm{U}(N)}$-bundles to the moduli stack of principal ${\mathrm{U}(1)}$-bundles induced by the determinant $\det:{\mathrm{U}(N)}\to {\mathrm{U}(1)}$. Furthermore, 
 the fact that $c_1$ represents the obstruction to the ${\mathrm{U}(N)}$-to-$\mathrm{SU}(N)$ reduction becomes the following statement: the stack $\mathbf{B}\mathrm{SU}(N)$ of principal $\mathrm{SU}(N)$-bundles is the homotopy pullback
\(
  \raisebox{20pt}{
  \xymatrix{
    \mathbf{B}\mathrm{SU}(N)\ar[r]\ar[d]&{*}\ar[d]
	\\
    \mathbf{B}{\mathrm{U}(N)}\ar[r]^{\mathbf{c}_1}&\mathbf{B}{\mathrm{U}(1)}~
	\;.
  }
  }
  \label{SUNHomotopyPullback}
\)
By the definition of homotopy pullbacks (see also the discussion in the 
introduction around (\ref{HomotopyPullbackFirstAppearance})), this says that a morphism of 
stacks $X \to \mathbf{B} \mathrm{SU}(N)$, hence an $\mathrm{SU}(N)$-principal
bundle over $X$, is equivalently a morphism $X \to \mathbf{B}\mathrm{U}(N)$,
hence a ${\mathrm{U}(N)}$-principal bundle $P$, together with a choice of trivialization of 
the composite morphism $X \to \mathbf{B}{\mathrm{U}(N)} \stackrel{\mathbf{c}_1}{\to} \mathbf{B}{\mathrm{U}(1)}$,
hence of the determinant bundle $\mathbf{c}_1(P)$.
Moreover, the whole groupoid of $\mathrm{SU}(N)$-principal bundles 
on a manifold $X$ is \emph{equivalent} to the groupoid of $\mathrm{U}(N)$-principal bundles 
on $X$ that are equipped with a trivialization of their associated 
determinant ${\mathrm{U}(1)}$-principal bundle. 

\vspace{3mm}

Let us describe a morphism of stacks $X \to \mathbf{B} \mathrm{SU}(N)$, with
$\mathbf{B} \mathrm{SU}(N)$ identified with a homotopy pullback as above, in explicit
detail, following \cite{FSS}. 
For this we first need to choose an open cover $\mathcal{U}=\bigcup_i U_i$ of $X$,
which is ``good'', meaning that all non-empty finite intersections of the $U_i$ are
contractible.
In terms of this choice, a map of stacks from $X$ into the above homotopy pullback
is given by the following data.
\begin{itemize}
\item ${\mathrm{U}(1)}$-valued functions $\rho_i$ on the patch $U_i$~;
\item ${\mathrm{U}(N)}$-valued functions $g_{ij}$ on the double intersection 
$U_{ij}:=U_i \cap U_j$, with $g_{ii}=1$,
\end{itemize}
subject to the constraints
\begin{itemize}
\item $\det(g_{ij})\rho_j=\rho_i$ on $U_{ij}$;
\item $g_{ij} g_{jk} g_{ki}=1$ on the triple intersection $U_{ijk}:=U_i\cap U_j\cap U_k$.
\end{itemize}
Morphisms between $(\rho_i,g_{ij})$ and $(\rho_i',g_{ij}')$ are the gauge transformations locally given by ${\mathrm{U}(N)}$-valued functions $\gamma_{i}$ on $U_{i}$ such that $\gamma_i g_{ij}=g_{ij}' \gamma_j$ and $\rho_i\det(\gamma_i)=\rho_i'$.

\vspace{3mm}
Note that the classical description of objects in $\mathbf{B}\mathrm{SU}(N)$ corresponds to the \emph{gauge fixing} $\rho_i\equiv 1$; at the level of morphisms, imposing this gauge fixing constrains the gauge transformation $\gamma_i$ to satisfy $\det(\gamma_i)=1$, i.e. to take values in $\mathrm{SU}(N)$. From a categorical point of view, this amounts to saying that the embedding of the groupoid of $\mathrm{SU}(N)$-principal bundles into the homotopy fiber of $\mathbf{c}_1$ given by $(g_{ij})\mapsto (1,g_{ij})$ is fully faithful. It is also essentially surjective: use the embedding ${\mathrm{U}(1)}\to {\mathrm{U}(N)}$ given by $e^{it}\mapsto (e^{it},1,1,\dots, 1)$ to lift $\rho_i^{-1}$ to a ${\mathrm{U}(N)}$-valued function $\gamma_i$ with $\det(\gamma_i)={\rho_i}^{-1}$; then $(\gamma_i)$ is an isomorphism between $(\rho_i,g_{ij})$ and $(1,\gamma_ig_{ij}{\gamma_j}^{-1})$.

\vspace{3mm}
Next we turn to connections. It is a well known fact from Chern-Weil theory that the 
de Rham image of the first Chern class of a ${\mathrm{U}(N)}$-principal bundle can be realized 
as the de Rham cohomology class $[\mathrm{tr}(F_\nabla)]$, where $F_\nabla$ is the \emph{curvature} 2-form  of a $\mathfrak{u}_n$-connection $\nabla$. The cohomology equation 
\(
[\mathrm{tr}(F_\nabla)]=0
\)
is equivalent to $\mathrm{tr}(F_\nabla)=d\alpha$ for some 1-form $\alpha$, and we can therefore think of the choice of such an $\alpha$ as the choice of a trivialization of the characteristic form 
$\mathrm{tr}(F_\nabla)$. Since the 1-form $\alpha$ is naturally interpreted as a $\mathfrak{u}_1$-connection on a \emph{trivial} principal ${\mathrm{U}(1)}$-bundle, the trivialization of   $\mathrm{tr}(F_\nabla)$ becomes the equation
\(
\mathrm{tr}(F_\nabla)=F_\alpha\;,
\)
i.e., we are identifying the characteristic 2-form $\mathrm{tr}(F_\nabla)$ with the curvature 2-form of a connection on a trivial principal ${\mathrm{U}(1)}$-bundle. All this has 
%
%an extremely neat and 
% Urs: agreed!, but maybe we shouldn't say this ourselves?
%
a simple interpretation in terms of stacks: the 
smooth first Chern class $\mathbf{c}_1\colon \mathbf{B}{\mathrm{U}(N)}\to \mathbf{B}{\mathrm{U}(1)}$ has a \emph{differential refinement} to a morphism
\(
   \hat{\mathbf{c}}_1\colon \mathbf{B}{\mathrm{U}(N)}_{\mathrm{conn}}
     \to 
   \mathbf{B}{\mathrm{U}(1)}_{\mathrm{conn}}
   \label{DifferentialC1}
\)
from the moduli stack of ${\mathrm{U}(N)}$-principal bundles with 
$\mathfrak{u}_N$-connections to the moduli stack of 
${\mathrm{U}(1)}$-principal bundles with $\mathfrak{u}_1$-connections, induced by the Lie algebra morphism 
\(
\mathrm{tr}: \mathfrak{u}_N\to \mathfrak{u}_1\;.
\)
In terms of local data, the morphism $\hat{\mathbf{c}}_1$ maps the $\mathfrak{u}_N$-connection 1-form $A_i$ to the Chern-Simons 1-form 
$\mathrm{CS}_1(A_i)=\mathrm{tr}(A_i)$, and the identity
\(
\mathrm{tr}(F_{A_i})=d \mathrm{CS}_1(A_i)
\)
shows that the curvature characteristic 2-form of the $\mathfrak{u}_N$-connection $(A_i)$ can be identified with the curvature of the 2-form of the $\mathfrak{u}_1$-connection $(CS_1(A_i))$. This means that, as a morphism of stacks, the traced curvature
\(
\mathbf{B}{\mathrm{U}(N)}_{\mathrm{conn}}\xrightarrow{\mathrm{tr}\,\mathrm{curv}} \Omega^2_{\mathrm{closed}}
\)
actually factors as
\(
\mathbf{B}{\mathrm{U}(N)}_{\mathrm{conn}}\xrightarrow{\hat{\mathbf{c}}_1}\mathbf{B}{\mathrm{U}(1)}_{\mathrm{conn}}\xrightarrow{\mathrm{curv}} \Omega^2_{\mathrm{closed}}\;,
\)
where $\Omega^2_{\mathrm{closed}}$ is the stack whose smooth $U$-paramaterized 
families of objects 
are closed 2-forms on $U$ (with trivial morphisms).
Moreover, the stack of $\mathrm{SU}(N)$-principal bundles 
with $\mathfrak{su}_N$-connections is the (homotopy) pullback
\(
  \raisebox{20pt}{
  \xymatrix{
    \mathbf{B}\mathrm{SU}(N)_{\mathrm{conn}}
	\ar[r]
	\ar[d]
	&
	{*}
	\ar[d]
	\\
    \mathbf{B}{\mathrm{U}(N)}_{\mathrm{conn}}\ar[r]^{\hat{\mathbf{c}}_1}
	&
	\mathbf{B}{\mathrm{U}(1)}_{\mathrm{conn}}~\;.
 }
 }
 \label{SUConnHomotopyPullback}
\)
Again, we write out the data for objects and morphisms in the groupoid of  $\mathrm{SU}(N)$-bundles with $\mathfrak{su}_N$-connections over a fixed smooth manifold $X$, presented as a homotopy pullback
this way. For a fixed good open cover $\mathcal{U}$ of $X$, the objects of this groupoid are
 \begin{itemize}
%\item a ${\mathrm{U}(1)}$-valued function $\beta$ on $X$;
\item ${\mathrm{U}(1)}$-valued functions $\rho_i$ on $U_i$;
\item $\mathfrak{u}_N$-valued 1-forms $A_i$ on $U_i$; 
\item ${\mathrm{U}(N)}$-valued functions $g_{ij}$ on $U_{ij}$, with $g_{ii}=1$.
\end{itemize}
subject to the constraints
\begin{itemize}
%\item $\mathrm{tr}A_i+d\mathrm{log}\rho_i=d\mathrm{log}\beta$ on $U_i$;
\item $\mathrm{tr}A_i+d\mathrm{log}\rho_i=0$ on $U_i$~;
\item $\det(g_{ij})\rho_j=\rho_i$ on $U_{ij}$~;
\item $A_j=g_{ij}^{-1}A_i g_{ij} +g_{ij}^{-1} dg_{ij}$ on $U_{ij}$~;
\item $g_{ij}g_{jk}g_{ki}=1$ on $U_{ijk}$~,
\end{itemize}
and the classical description of $\mathfrak{su}_N$-connections on a principal $\mathrm{SU}(N)$-bundle corresponds to the gauge fixing $\rho_i\equiv 1$.
% and $\beta\equiv1$.
This should not be surprising: the data $(\rho_i)$
%$(\beta,\rho_i)$ 
are the data of the trivialization of the principal ${\mathrm{U}(1)}$-bundle with $\mathfrak{u}_1$-connection induced by $\hat{\mathbf{c}}_1$; fixing these data to $1$ is equivalent to requiring that this bundle with connection is trivially trivialized.

\vspace{3mm}
As far as concerns the morphisms, in the homotopy pullback description, a morphism between $
(\rho_i,A_i,g_{ij})$ and $(\rho_i',A_i',g_{ij}')$ is the datum of 
%(\beta,\rho_i,A_i,g_{ij})$ and $(\beta',\rho_i',A_i',g_{ij}')$ is the datum of 
\begin{itemize}
%\item a ${\mathrm{U}(N)}$-valued function $\phi$ on $X$;
 \item ${\mathrm{U}(N)}$-valued functions $\gamma_{i}$ on $U_{i}$
\end{itemize}
such that
\begin{itemize}
%\item $\beta'=\beta \det(\phi)$ on $X$;
%\item $A_i'=(\phi \gamma_{i})^{-1}A_i\phi \gamma_{i}+(\phi \gamma_{i})^{-1}d(\phi \gamma_{i})$;
\item $A_i'=\gamma_{i}^{-1}A_i\gamma_{i}+\gamma_{i}^{-1}d\gamma_{i}$~;
\item $\rho_i\det(\gamma_i)=\rho_i'$ on $U_i$~;
\item $\gamma_i g_{ij}=g_{ij}' \gamma_j$ on $U_{ij}$~.
\end{itemize}
Note that the gauge fixing 
%$\beta=1$, 
$\rho_i=1$ imposes 
%$\det(\phi)=1$ and 
$\det(\gamma_i)=1$, and  one recovers the classical description of isomorphisms between principal $\mathrm{SU}(N)$-bundles with $\mathfrak{su}_N$-connections. 

\vspace{3mm}
Notice that the curvature characteristic 2-form of a  $\mathfrak{su}_N$-connection (either in the classical or in the homotopy pullback description) is identically zero. 
This means that we actually went far beyond our original aim that was to kill only the cohomology class of the curvature characteristic form, and not the curvature characteristic 2-form itself. This is not unexpected: 
morphisms of principal bundles with connections are too narrow to capture the flexible nature of requiring something to be zero only in cohomology. A natural way to remedy this is to consider instead the moduli stack $\mathbf{B}\mathrm{U}(N)_{\mathrm{conn}, c_1=0}$ defined as the homotopy fiber of
the composite map
\(
   \mathbf{c}_1
     :
	 \mathbf{B}{\mathrm{U}(N)}_{\mathrm{conn}}
       \xrightarrow{\hat{\mathbf{c}_1}}
     \mathbf{B}{\mathrm{U}(1)}_{\mathrm{conn}}
       \rightarrow	 
	 \mathbf{B}{\mathrm{U}(1)}
	 \,,
	\label{CompositeMapc1}
\)
where the second morphism forgets the connection.
Since we have a homotopy pullback diagram
\(
  \raisebox{20pt}{
  \xymatrix{
     \Omega^1 \ar[r] \ar[d] & {*} \ar[d]
	 \\
	 \mathbf{B}{\mathrm{U}(1)}_{\mathrm{conn}}
	 \ar[r]
	 &
	 \mathbf{B}{\mathrm{U}(1)}
  }
  }
  \label{FiberOfForget}
\)
it follows by the pasting law 
for homotopy pullbacks that 
the homotopy fiber of (\ref{CompositeMapc1}) 
is equivalently described as the homotopy pullback
\(
  \raisebox{20pt}{
  \xymatrix{ 
    \mathbf{B}\mathrm{U}(N)_{\mathrm{conn},c_1=0}\ar[r]\ar[d]&{\Omega^1}
	  \ar[d]
	  \\
     \mathbf{B}{\mathrm{U}(N)}_{\mathrm{conn}}
	 \ar[r]^{\hat{\mathbf{c}}_1}
	 &
	 \mathbf{B}{\mathrm{U}(1)}_{\mathrm{conn}}~~\;.
  }
  }
\label{BSUconnVariant}
\)
In other words,  $\mathbf{B}\mathrm{U}(N)_{\mathrm{conn},c_1=0}$ is the collection of all the homotopy fibers of $\hat{\mathbf{c}}_1:\mathbf{B}{\mathrm{U}(N)}_{\mathrm{conn}}\to \mathbf{B}{\mathrm{U}(1)}_{\mathrm{conn}}$, with varying ``background field'' in $\Omega^1$, and so local data for a map from a manifold $X$
into this stack are
 \begin{itemize}
\item ${\mathrm{U}(1)}$-valued functions $\rho_i$ on $U_i$;
\item $\mathfrak{u}_1$-valued 1-forms $\mathcal{H}_{i}$ on $U_{i}$;
\item $\mathfrak{u}_N$-valued 1-forms $A_i$ on $U_i$; 
\item ${\mathrm{U}(N)}$-valued functions $g_{ij}$ on $U_{ij}$, with $g_{ii}=1$.
\end{itemize}
subject to the constraints
\begin{itemize}
\item $\mathcal{H}_i=d\mathrm{log}\rho_i+\mathrm{tr}A_i$ on $U_i$;
\item $\det(g_{ij})\rho_j=\rho_i$ on $U_{ij}$;
\item $A_j=g_{ij}^{-1}A_i g_{ij} +g_{ij}^{-1} dg_{ij}$ on $U_{ij}$;
\item $\mathcal{H}_i=\mathcal{H}_j$ on $U_{ij}$;
\item $g_{ij}g_{jk}g_{ki}=1$ on $U_{ijk}$.
\end{itemize}
In particular, the local curvature 2-form of a 
connection classified by $\mathbf{B}{\mathrm{U}(N)}_{\mathrm{conn},c_1=0}$ is 
\(
  \mathrm{tr}(F_{A_{i}}) = d H_i
  \,.
\)
Notice that since $\mathcal{H}_i=\mathcal{H}_j$ on $U_{ij}$, the $(\mathcal{H}_i)$ define a global 1-form on $X$ and so $\mathrm{tr}(F_{A_{i}})=d\mathcal{H}_i$ precisely says that the cohomology class of the curvature characteristic 2-form vanishes.
 The stringy analog of this we discuss below, around equation (\ref{StringConnVar}).

%%%%%%%%%%%%%%%%%%%%%%%%%%%%%%%%%%
\subsection{Higher ${\mathrm{U}(1)}$-bundles with connections}
\label{HigherU1Connections}
%%%%%%%%%%%%%%%%%%%%%%%%%%%%%%%%%%%%%%%%%%%%%%%%%%%%%%%%%

We discuss the moduli stack of ordinary circle bundles with connection
(abelian Yang-Mill fields) and then the higher analogs, the higher stacks of
circle $n$-bundles with connection ($B$-fields, $C$-fields, etc.).

\paragraph{The stack circle-bundles with connections.}
We start with the moduli stack $\mathbf{B}{\mathrm{U}(1)}_{\mathrm{conn}}$
of ordinary circle bundles with connection in a way that prepares for the
generalizations to follow. 

\medskip

The local data for a $\mathfrak{u}_1$-connection on a ${\mathrm{U}(1)}$-principal bundle over a 
smooth manifold $X$, locally trivialized over an open cover
$\{U_i \hookrightarrow X\}$, are given by ``vertices'' (objects) 
$A_i$, which are $\mathfrak{u}_1$-valued 
1-forms on $U_i$ and ``edges'' (morphisms/gauge transformations) 
$g_{ij}$, which are ${\mathrm{U}(1)}$-valued functions 
on the double intersections $U_{ij}$. 
This data is subject to two constraints: ``an edge 
$g_{ij}$ has to go from $A_i$ to $A_j$'':
\(
d\log g_{ij}=A_j-A_i\;;
\)
and ``going around the boundary of a 2-simplex is a trivial path'':
\(
g_{ij} g_{jk} g_{ik}^{-1} =1\;.
\)
\(
  \xymatrix{
    & 
    A_j 
	\ar[dr]^{g_{j k}}
    &
    \\
    A_i 
	\ar[ur]^{g_{i j}}
    \ar[rr]_{g_{ik}} 
    &&
    A_k~~\;.
  }
\)
This can be elegantly stated as follows: the stack $\mathbf{B}{\mathrm{U}(1)}_{\mathrm{conn}}$ is 
the image under the Dold-Kan correspondence 
(briefly discussed in section \ref{HigherSmoothStacks})
of the 2-term chain complex of sheaves
\(
  C^\infty(-;{\mathrm{U}(1)})\xrightarrow{d\log}\Omega^1(-)
  \,.
\)
Here in degree 0 we have the sheaf $\Omega^1(-)$ 
of 1-forms (which assigns to any smooth manifold $U$ the additively abelian group
of 1-forms on U), and in degree 1 similarly the sheaf of smooth ${\mathrm{U}(1)}$-valued functions.
The differential is the operation that takes a ${\mathrm{U}(1)}$-valued function, forms (locally)
any $\mathbb{R}$-valued lift and then produces the differential of that. 
This is just a rephrasing of the above explicit description of the local data for $\mathbf{B}{\mathrm{U}(1)}_{\mathrm{conn}}$, but it highlights an important aspect. Namely, there is no need to speak Stackish fluently to see the central point of the above sentence: ``$\mathbf{B}{\mathrm{U}(1)}_{\mathrm{conn}}$ is something built from the 2-term complex $C^\infty(-;{\mathrm{U}(1)})\xrightarrow{d\log}\Omega^1$.''. 

\paragraph{The stack of ${\mathrm{U}(1)}$-principal $n$-bundles with connection.}
The above immediately 
suggests the following generalization: 
for every $n \in \mathbb{N}$, the moduli $n$-stack of 
\emph{circle $n$-bundles} or equivalently \emph{${\mathrm{U}(1)}$-bundle $(n-1)$-gerbes}
is the image $\mathbf{B}^n {\mathrm{U}(1)}$ under the Dold-Kan map (see section \ref{HigherSmoothStacks})
of the chain complex of sheaves (over smooth manifolds)
\(
  C^\infty(-, {\mathrm{U}(1)}) \to 0 \to 0 \to \cdots \to 0
  \,,
\)
which is concentrated in degree $n$ on the sheaf of smooth ${\mathrm{U}(1)}$-valued functions.
Similarly, 
the moduli stack $\mathbf{B}^n {\mathrm{U}(1)}_{\mathrm{conn}}$ of principal ${\mathrm{U}(1)}$-$n$-bundles 
\emph{with connections} is the $n$-stack obtained 
via Dold-Kan from the $(n+1)$-term complex of sheaves ${\mathrm{U}(1)}[n]^\infty_D$ given by
\(
  C^\infty(-;{\mathrm{U}(1)})
   \xrightarrow{d\log}
   \Omega^1(-)
    \xrightarrow{d}
	\Omega^2(-)
	\xrightarrow{d}
	\cdots
	\xrightarrow{d}\Omega^n(-)
	\;,
\)
with the sheaf of $n$-forms, $\Omega^n(-)$, in degree 0, and 
the sheaf of ${\mathrm{U}(1)}$-valued smooth functions in degree $n$. This is
known as the \emph{Beilinson-Deligne complex}.
The morphism of chain complexes 
\(
\raisebox{20pt}{
\xymatrix{
  C^\infty(-;{\mathrm{U}(1)})
  \ar[r]^{\phantom{mm}d\log}
   \ar[d]
   &
   \Omega^1(-)
   \ar[r]^{d}\ar[d]
   &
   \Omega^2(-)
   \ar[r]^{d}
   \ar[d]
   &\cdots
   \ar[r]^{d}
   \ar[d]
   &
   \Omega^n(-)
   \ar[d]^{d}
   \\
   0
   \ar[r]
   &
   0\ar[r]
   &
   0
   \ar[r]
   &
   \cdots
   \ar[r]
   &
   \Omega^{n+1}_{\mathrm{closed}}(-)
}
}
\,,
\)
where in the top row we have the Deligne complex as before and in the bottom
row we have simply the sheaf on $n$-forms, extended by 0 to a chain complex,
induces the \emph{curvature} morphism
\(
  \mathbf{B}^n{\mathrm{U}(1)}_{\mathrm{conn}}\to \Omega^{n+1}_{\mathrm{closed}}\;,
\)
mapping a connection on a ${\mathrm{U}(1)}$-principal $n$-bundle to its curvature $(n+1)$-form. 
Also, the evident morphism of chain complexes 
\(
  \xymatrix{
    C^\infty(-;{\mathrm{U}(1)})\ar[r]^{\phantom{mm}d\log}
	\ar[d]
	&
	\Omega^1(-)\ar[r]^{d}\ar[d]
	&
    \Omega^2(-)
	\ar[r]^{d}\ar[d]
	&
	\cdots
	\ar[r]^{d}\ar[d]
	&
	\Omega^n(-)
	\ar[d]
	\\
    C^\infty(-;{\mathrm{U}(1)})
	\ar[r]
	&
	0
	\ar[r]
	&
    0
	\ar[r]
	&
	\cdots\ar[r]&0
}
\)
induces the map on moduli that forgets the connection 
on a circle $n$-bundle and induces the natural forgetful morphism
\(
\mathbf{B}^n{\mathrm{U}(1)}_{\mathrm{conn}}\to \mathbf{B}^n{\mathrm{U}(1)}
\)
to the moduli $n$-stack of principal ${\mathrm{U}(1)}$-$n$-bundles. 

\vspace{3mm}
For $X$ a smooth manifold, homotopy classes of maps of stacks 
$X \to \mathbf{B}^n {\mathrm{U}(1)}$ are in bijection with the integral cohomology
classes of $X$ in degree $n+1$. We write
\(
  \pi_0 \mathbf{H}(X, \mathbf{B}^n {\mathrm{U}(1)})
  \simeq
  H^{n+1}(X, \mathbb{Z})
  \,.
\)
Similarly, the homotopy classes of morphisms of smooth higher stacks
$X \to \mathbf{B}^n {\mathrm{U}(1)}_{\mathrm{conn}}$ are in bijection with the
\emph{differential cohomology} of $X$. We write
\(
  \pi_0 \mathbf{H}(X, \mathbf{B}^n {\mathrm{U}(1)}_{\mathrm{conn}})
  \simeq
  \hat H^{n+1}(X)
  \,.
\)
Let us unwind the local data from the above definitions 
looks like for the case of $n=2$, hence for circle 2-bundle / bundle gerbes; 
It is immediate to generalize from this description to higher $n$'s. In the $n=2$ case we have:
\begin{itemize}
\item 2-forms $K_{2;i}$ on $U_i$;
\item 1-forms $A_{1;ij}$ on $U_{ij}$;
\item ${\mathrm{U}(1)}$-valued functions $g_{ijk}$ on $U_{ijk}$.
\end{itemize}
These data are subject to the following constraints:
\begin{itemize}
\item $dA_{1;ij}=K_{2;j}-K_{2;i}$ on $U_{ij}$;
\item $d\log g_{ijk}=A_{1;jk}-A_{1;ik}+A_{1;ij}$ on $U_{ijk}$;
\item $g_{jkl}~g_{ikl}^{-1}~g_{ijl}~g_{ijk}^{-1}=1$ on $U_{ijkl}$. 
\end{itemize}
These data are conveniently depicted on a 3-simplex as follows:
\(
\xymatrix{
&K_{2;i}\ar[rrd]^{A_{1;ij}}\ar[ddd]_{A_{1;ik}}\ar[ldd]_{A_{1;ij}}&&\\
&&&K_{2;j}~~\;.
\ar[lldd]^{A_{1;jk}}\ar[llld]|(.32)\hole|(.35)\hole|(.625)\hole|(.767)\hole^{A_{1;jl}}\\
K_{2;l}&&&\\
&K_{2;k}\ar[lu]^{A_{1;kl}}&&
\ar@{=>}_{g_{ijk}}(32,-20);(28,-21)
\ar@{=>}_{g_{i k l}}(11,-28);(10,-24)
\ar@{=>}_{g_{j k l}}(20,-32);(21,-28)
\ar@{=>}^{g_{ijl}}(20,-7);(20.5,-10)
}
\)
It is evident from the above description that the datum of a trivialization of the ${\mathrm{U}(1)}$-gerbe underlying an $X$-point in $\mathbf{B}^2{\mathrm{U}(1)}_{\mathrm{conn}}$
consists of 
\begin{itemize}
\item 2-forms $\mathcal{H}_{2;i}$ on $U_i$;
\item 1-forms $\mathcal{H}_{1;ij}$ on $U_{ij}$;
\item 1-forms $B_{1;i}$ on $U_i$;
\item ${\mathrm{U}(1)}$-valued functions $\rho_{ij}$ on $U_{ij}$
\end{itemize}
such that
\begin{itemize}
\item $\mathcal{H}_{2;i}=dB_{1;i}+K_{2;i}$ on $U_i$;
\item $\mathcal{H}_{1;ij}=d\log \rho_{ij}+ B_{1;j}-B_{1;i}+A_{1;ij}$ on $U_{ij}$;
\item $\mathcal{H}_{2;i}=dB_{1;i}+K_{2;i}$ on $U_i$;
\item $d\mathcal{H}_{1;ij}=\mathcal{H}_{2;j}-\mathcal{H}_{2;i}$ on $U_{ij}$;
\item $g_{ijk}=\rho_{jk}~\rho_{ik}^{-1}~\rho_{ij}$ on $U_{ijk}$.
\end{itemize}

An immediate consequence of these local equations is that $dK_{2;i}-dK_{2;j}$ vanishes on $U_{ij}$ and so $(dK_{2;i})_{i\in I}$ defines a global closed 3-form on $X$. Moreover, this 3-form is globally exact and so its cohomology class vanishes (this is the triviality on the underlying grebe read in terms of de Rham cohomology). Exhibiting a global primitive for $(dK_{2;i})_{i\in I}$ is an easy exercise in sheaf cohomology. Namely, $\mathcal{H}_{1;jk}-\mathcal{H}_{1;ik}+\mathcal{H}_{1;ij}=0$ on $U_{ink}$ so $(\mathcal{H}_{1;ij})_{i,j\in I}$ is a \v{C}ech 1-cocycle on $X$ with coefficients in the sheaf $\Omega^1$ of smooth 1-forms. Since this sheaf is fine,  $(\mathcal{H}_{1;ij})_{i,j\in I}$ is a 1-coboundary, and so there exist 1-forms $\alpha_i$ on $U_i$ with $\mathcal{H}_{1;ij}=\alpha_j-\alpha_i$. Then $(\mathcal{H}_{2;i}-d\alpha_{1;i})_{i\in I}$ is a global 2-form on $X$ which is a primitive of $(dK_{2;i})_{i\in I}$.

%%%%%%%%%%%%%%%%%%%%%%%%%%%%%%%%%%%%%%%%%%%%%%%%%%%%%%%%%%%%%%%%%%%%%%
\subsection{Multiple D-branes in nonabelian differential cohomology}
\label{Section D-branes}
%%%%%%%%%%%%%%%%%%%%%%%%%%%%%%%%%%%%%%%%%%%%%%%%%%%%%%%%%%%%%%%%%%%%%

%\attn{\color{blue} Suggest move this section to C-field}

In type II string theory in the presence of D-branes, 
the background $B$-field on spacetime $X$
is accompanied by nonabelian gauge fields on the branes satisfying there a compatibility
condition with the restriction of the $B$-field to the branes. This
turns out to be analogous, in a precise fashion, to the situation with the 
$C$-field in 11-dimensional supergravity, and its restiction to 
Ho{\v r}ava-Witten boundaries of spacetime. In both cases the total moduli
stack of field configurations is given by a nonabelian and twisted version of 
\emph{relative} (differential) cohomology. 
For the $C$-field we discuss this in section \ref{CFieldConfigurations}
(and in full detail in \cite{FiSaSc}).
Here we 
%warm up by giving 
give the analogous discussion for the 
$B$-field in terms of the stacky structures that we have already introduced above.
Where for the $B$-field the trivialization on the brane makes a nonabelian
1-form appear, for the $C$-field the trivialization on the brane makes
a nonabelian 2-form appear. 

\medskip

Let $X$ be a 10-dimensional spacetime. By the discussion in 
section \ref{HigherU1Connections}, the $B$-field on $X$ is given by a morphism
of smooth 2-stacks
$\hat B : X \to \mathbf{B}^2 {\mathrm{U}(1)}_{\mathrm{conn}}$. Let then $Q \hookrightarrow X$
be a single $\mathrm{Spin}^c$ $D$-brane in $X$. 
Freed-Witten anomaly cancellation \cite{FW} requires that the restriction of 
$\hat B$ to $Q$ has trivial integral class, hence that there it is,
up to a gauge transformation, 
in the image of the moduli $\Omega^2(-) \to \mathbf{B}^2 {\mathrm{U}(1)}_{\mathrm{conn}}$.
This situation is concisely captured by saying that the field configurations
form a homotopy-commuting diagram
\(
  \raisebox{20pt}{
  \xymatrix{
    Q \ar[r]_>{\ }="s"^{\hspace{-4mm}\mathcal{F}} \ar[d] & \Omega^2(-) \ar[d]
	\\
	X \ar[r]_{\hspace{-7mm}\hat B}^<{\ }="t" & \mathbf{B}^2 {\mathrm{U}(1)}_{\mathrm{conn}}
	\ar@{=>}^{\hat A}_\simeq "s"; "t"
  }
  }
  \label{TrivializationOfBFieldOverDBrane}
\)
of smooth 2-stacks. The bottom morphism is the $B$-field on $X$. Its composite with
the left morphism is its restriction to the brane. The top morphism is a
globally defined 2-form $\mathcal{F}$ on the brane, and the homotopy in the middle is
a gauge transformation from this 2-form, regarded as a connection on a trivial 2-bundle,
to the restriction of the $B$-field. Notice that this means that $\hat A$ is locally,
on a patch $U_i \hookrightarrow Q$, a 1-form with curvature $F_i = d A_i$, such that
equation (\ref{Combined2FormOnBrane}) holds
\(
  \mathcal{F}_i = B_i + F_i
  \,.
\)
This 1-form is the Chan-Paton gauge field on the $D$-brane.
Moreover, the collections of all such triples of field configurations naturally
form the mapping 2-groupoid, denoted
\(
  \mathbf{H}^I(Q \stackrel{i}{\to} X\;,\; \Omega^2(-) \to \mathbf{B}^2 {\mathrm{U}(1)}_{\mathrm{conn}})
  \,,
\)
whose cocycles are homotopy-commuting squares as above, and whose coboundaries
are the corresponding relative gauge transformations. 
We have a variant of this when discussing the boundary $C$-field configurations
in section \ref{CFieldConfigurations}.

\vspace{3mm}
More generally, there may be $N$ coincident $D$-branes with $\mathrm{Spin}^c$-worldvolume $Q$.
In this case (see for instance \cite{Laine} for a clean account) the trivialization
$\hat A$ in the above is to be replaced by a \emph{twisted $\mathrm{U}(N)$-bundle} on $Q$,
whose twist is the restriction of $\hat B$ to $Q$.
In our context, this is formulated as follows.
The short exact sequence of groups (a finite-dimensional counterpart of
sequence \eqref{U U}) 
\(
  {\mathrm{U}(1)} \to \mathrm{U}(N) \to \mathrm{PU}(N)
\)
that exhibits the unitary group as a central extension of the projective unitary 
gives rise to a long sequence of smooth 2-stacks
\(
  \xymatrix{
    \mathbf{B}{\mathrm{U}(1)} 
	  \ar[r] 
	&
	\mathbf{B}\mathrm{U}(N) 
  	  \ar[r]
	& 
	 \mathbf{B}\mathrm{PU}(N)
     \ar[r]^{\mathbf{dd}}
	 &
	 \mathbf{B}^2 {\mathrm{U}(1)}
  }
  \,.
\)
The characteristic map $\mathbf{dd}$ here may be understood as presenting the 
universal class on projective bundles that obstructs their lift to genuine 
unitary bundles. This has an evident differential refinement
\(
  \hat {\mathbf{dd}} 
   : 
   \mathbf{B}\mathrm{PU}_{\mathrm{conn}} 
    \to 
	\mathbf{B}^2 {\mathrm{U}(1)}_{\mathrm{conn}}
	\,.
\)
where now on the left we have the moduli stack of projective unitary bundles 
with projective connections. Underlying such a (nonabelian twisted) connection
is a globally defined abelian curvature 1-form (induced, locally, by the trace operation, 
as in section \ref{DeterminantLineBundles}).
Therefore, we can now consider relative cohomology twisted by $\mathbf{dd}$ on
the brane inclusion $Q \to X$.
Its cocycles are homotopy-commuting diagrams of 2-stacks
\(
  \raisebox{20pt}{
  \xymatrix{
    Q \ar[r]_>{\ }="s"^{\hspace{-5mm}\hat A} \ar[d] & \mathbf{B}\mathrm{PU}(N)_{\mathrm{conn}} 
	 \ar[d]^{\hat {\mathbf{dd}}}
	\\
	X \ar[r]_{\hspace{-7mm}\hat B}^<{\ }="t" & \mathbf{B}^2 {\mathrm{U}(1)}_{\mathrm{conn}}~~\;. 
	\ar@{=>}^{}_\simeq "s"; "t"
  }
  }
  \label{BFieldRelative}
\)
Here the top morphism now characterizes a twisted Chan-Paton gauge field on  
$N$ coincident $D$-branes, whose $\mathbf{dd}$-class trivializes the restriction of the 
spacetime $B$-field to the branes.  
When the lower morphism is presented in terms of bundle gerbes, then the top
morphism is presented by the corresponding \emph{gerbe modules} as in \cite{CBMMS}.
A component-discussion of this relative nonabelian differential cohomology
describing D-brane gauge fields
is for instance in section 4 of \cite{Ruffino}. The $C$-field analog of this
diagram is discussed below as (\ref{CFieldRelativeSimplified})
and (\ref{CFieldRelative}).

\medskip
The groups $\mathrm{PU}(N)$ all embed into the group $\mathrm{PU}(\mathbb{H})$,
of projective unitary operators on any separable infinite-dimensional complex
Hilbert space $\mathbb{H}$, and we have a morphism of classifying stacks 
\(
  \mathbf{dd} : \mathbf{B} \mathrm{PU}(\mathbb{H}) \to \mathbf{B}^2 U(1)
  \label{ddOnPUH}
\)
that classifies the $U(1)$-extension $\mathrm{U}(\mathbb{H})$. 
In this limit of ``arbitrary numbers of D-branes'' something interesting happens:
under geometric realization (\ref{GeometricRealization}) 
the two moduli stacks $\mathbf{B} \mathrm{PU}(\mathbb{H})$ of projective
unitary bundles and $\mathbf{B}^2 U(1)$ of circle 2-bundles both become
the Eilenberg-MacLane space $K(\mathbb{Z},3)$, and the class $\mathbf{dd}$
simply becomes the identity. This is related to Kuiper's theorem,
which asserts that the topological space underlying
$\mathrm{U}(\mathbb{H})$ is a contractible space. 
This says that, while the geometry and differential geometry of 
twisted nonabelian 1-form connections is very different from that of 
abelian 2-form connections, their instanton sectors may be identified.

\medskip

We discuss an analog of all these statements for the supergravity $C$-fields in 
section \ref{CFieldConfigurations}.

%%%%%%%%%%%%%%%%%%%%%%%%%%%%%%%%%%%%%%%%%%%%%%%%%%%%%%%%
\subsection{Higher holonomy}
\label{holonomy}
%%%%%%%%%%%%%%%%%%%%%%%%%%%%%%%%%%%%%%%%%%%%%%%%%%%%%%%%

It is a classical fact that a connection on a circle bundle
induces a notion of \emph{holonomy} along 1-dimensional curves, and that 
this holonomy is the gauge coupling action for a charged particle
in the background of the gauge field presented by the connection.
This fact has a higher analog for the higher circle bundles from
section \ref{HigherU1Connections}. A circle $n$-bundle with 
connection, equivalently: an abelian $n$-form gauge field, 
has holonomy over $n$-dimensional trajectories and this 
holonomy is the gauge coupling action  
of the $(n-1)$-brane charged
under the corresponding form field (the Wess-Zumino-Witten term). 
We recall how this comes about in terms of the Deligne complex and
thus as a map on higher stacks.

\vspace{3mm}
Let $X$ be a smooth manifold. Then the set of connected components of the $n$-groupoid $\mathbf{H}(X,\mathbf{B}^n{\mathrm{U}(1)}_{\mathrm{conn}})$ of ${\mathrm{U}(1)}$-$n$-bundles with connection on $X$ is naturally isomorphic to the $(n+1)$-th ordinary differential cohomology group of $X$:
\(
\pi_0\mathbf{H}(X,\mathbf{B}^n{\mathrm{U}(1)}_{\mathrm{conn}})\cong \hat{H}^{n+1}(X;\mathbb{Z})\;.
\)
Also, the set of  connected components of the $n$-groupoid $\mathbf{H}(X,\mathbf{B}^n{\mathrm{U}(1)})$ of ${\mathrm{U}(1)}$-$n$-bundles on $X$ is naturally isomorphic to the $(n+1)$-th singular cohomology group of $X$:
\(
\pi_0\mathbf{H}(X,\mathbf{B}^n{\mathrm{U}(1)})\cong {H}^{n+1}(X;\mathbb{Z})\;.
\)
and the forgetful map $\mathbf{B}^n{\mathrm{U}(1)}_{\mathrm{conn}}\to \mathbf{B}^n{\mathrm{U}(1)}$,
that ``forgets the connection", induces the natural morphism
\(
\hat{H}^{n+1}(X;\mathbb{Z})\to {H}^{n+1}(X;\mathbb{Z})
\)
from differential cohomology to singular cohomology. 
Moreover, it is well known 
(see, e.g., \cite{HopkinsSinger}) that if $X$ is a smooth oriented manifold, then the above morphisms fit into 
a short exact sequence
\(
0\to \Omega^n(X)/\Omega^n_{\mathrm{cl},0}(X)\to \hat{H}^{n+1}(X;\mathbb{Z})\to {H}^{n+1}(X;\mathbb{Z})\to 0\;,
\)
where $\Omega^n(X)/\Omega^n_{\mathrm{cl},0}(X)$ is the group of differential $n$-forms on $X$ modulo those $n$-forms which are closed and have integral periods. For the reader's convenience, let us briefly recall how the above short exact sequence originates.
Consider the complex of sheaves $\Omega^{1\leq\bullet\leq n}[n]$, i.e.,
\(
\Omega^1\to \Omega^2\to\cdots \to\Omega^n
\)
with $\Omega^n$ in degree zero. Then we have a short exact sequence of complexes of sheaves
\(
0\to \Omega^{1\leq\bullet\leq n}[n]\to {\mathrm{U}(1)}[n]^\infty_D\to {\mathrm{U}(1)}[n]\to 0
\)
inducing a long exact sequence in hypercohomology
\(
\cdots \to H^{-1}(X,{\mathrm{U}(1)}[n])\to H^{0}(X,\Omega^{1\leq\bullet\leq n}[n])\to H^0({\mathrm{U}(1)}[n]^\infty_D)\to H^0(X,{\mathrm{U}(1)}[n])\to 0\;.
\)
Since all the sheaves $\Omega^i$ are acyclic, the usual immediate spectral sequence argument shows that 
\newline
$H^{0}(X,\Omega^{1\leq\bullet\leq n}[n])\cong \Omega^n(X)/d\Omega^{n-1}(X)$ and so the above long exact sequence reads
\(
\cdots \to H^{n}(X;\mathbb{Z})\to \Omega^n(X)/d\Omega^{n-1}(X)\to \hat{H}^{n+1}(X;\mathbb{Z})\to {H}^{n+1}(X;\mathbb{Z})\to 0\;.
\)
From this we get the short exact sequence
\(
\cdots 0\to A\to \hat{H}^{n+1}(X;\mathbb{Z})\to {H}^{n+1}(X;\mathbb{Z})\to 0\;,
\)
where $A$ is the image of $\Omega^n(X)/d\Omega^{n-1}(X)$ inside $ \hat{H}^{n+1}(X;\mathbb{Z})$, and so is naturally isomorphic to the quotient of $\Omega^n(X)/d\Omega^{n-1}(X)$ by the image of $H^{n}(X;\mathbb{Z})$ into $\Omega^n(X)/d\Omega^{n-1}(X)$. Since this image is precisely $\Omega^n_{\mathrm{cl},0}(X)/d\Omega^{n-1}(X)$, we have $A\cong \Omega^n(X)/\Omega^n_{\mathrm{cl},0}(X)$.

\vspace{3mm}
In particular, if $\Sigma$ is an $n$-dimensional smooth oriented manifold, we get a canonical isomorphism
\(
\hat{H}^{n+1}(\Sigma;\mathbb{Z})\xrightarrow{\sim} \Omega^n_{\mathrm{cl},0}(\Sigma)/\Omega^n_{\mathrm{cl},0}(\Sigma)\;,
\)
i.e., each ${\mathrm{U}(1)}$-$n$-bundle with connection on $\Sigma$ is equivalent to a trivial ${\mathrm{U}(1)}$-$n$-bundle with an $n$-connection given by a globally defined $n$-form on $\Sigma$. Moreover, this 
$n$-form is uniquely determined up to an $n$-form with integral periods. By definition of $\Omega^n_{\mathrm{cl},0}(\Sigma)$, the integration over $\Sigma$ induces a well defined group homomorphism
\(
e^{2\pi i \int_\Sigma}\colon \Omega^n_{\mathrm{cl},0}(\Sigma)/\Omega^n_{\mathrm{cl},0}(\Sigma)\to {\mathrm{U}(1)}\;,
\)
and so finally we get the following result.

\begin{observation}

For $\Sigma$ a compact $n$-dimensional smooth manifold, there is a canonical morphism
\(
  \exp(2 \pi i\int_\Sigma(-))
  :
  \mathbf{H}(\Sigma, \mathbf{B}^n {\mathrm{U}(1)}_{\mathrm{conn}})
  \to
  {\mathrm{U}(1)}
  \,.
  \label{Integration}
\)
This is a map that sends $n$-form gauge fields on $\Sigma$ to elements in ${\mathrm{U}(1)}$
and is gauge invariant.

\vspace{3mm}
More generally, let be $X$ a (spacetime) manifold of any dimension or,
in fact, any orbifold or more general smooth stack or higher smooth stack.
Then with $\Sigma$ as before, there is a canonical morphism
\(
\mathrm{hol}_\Sigma\colon \mathbf{H}(\Sigma,X)\times \mathbf{H}(X,\mathbf{B}^n{\mathrm{U}(1)}_{\mathrm{conn}})\to {\mathrm{U}(1)}\;,
\)
where $\mathbf{H}(\Sigma,X)$ denotes the space (or $\infty$-groupoid) of maps from $\Sigma$ to $X$.
This morphis reads in an $n$-form gauge field $\nabla$ on $X$ as well 
as a smooth trajectory $\phi : \Sigma\to X$ and produces the 
\emph{$n$-dimensional holonomy}
\(
  \mathrm{hol}_\Sigma(\phi, \nabla) \in {\mathrm{U}(1)}
\)
of $\nabla$ around $\Sigma$ under the map $\phi$.
Formally, the map $\mathrm{hol}_\Sigma$ is just the composition
\(
 \mathrm{hol}_\Sigma
  :
 \mathbf{H}(\Sigma,X)\times \mathbf{H}(X,\mathbf{B}^n{\mathrm{U}(1)}_{\mathrm{conn}})
   \stackrel{\circ}{\to}
 \mathbf{H}(\Sigma,\mathbf{B}^n{\mathrm{U}(1)}_{\mathrm{conn}})
   \xrightarrow{e^{2\pi i \int_\Sigma(-)}}
  {\mathrm{U}(1)}\;.
\)
\end{observation} 
\noindent An explicit expression for $\mathrm{hol}_\Sigma(\phi,\nabla)$ in terms of local differential forms data can be found in \cite{gomi-terashima1,gomi-terashima2,dupont-ljungmann}.

%\vspace{3mm}
\paragraph{Holonomy of M-branes.}
For illustration in the case at hand,
we will spell this out for the case of M-branes, and rewrite 
the above results in this special, but important situation. 
The M2-brane and the M5-branes correspond to the case
$n=3$ and $n=6$, respectively. 
We have 
\begin{enumerate}
\item Consider the M2-brane with worldvolume a smooth oriented 3-manifold 
$\Sigma_3$. On the 11-dimensional target space $Y^{11}$ 
we have a $C$-field representing a connection on a 
${\mathrm{U}(1)}$-principal 3-bundle. Consider the $\sigma$-model for the M2-brane 
$\phi: \Sigma_3 \to Y^{11}$ with 
space of maps $\mathbf{H}(\Sigma_3, Y^{11})$. 
Then the holonomy of the C-field is given by 
\(
{\rm hol}_{\rm M2}: {\bf H}(\Sigma_3, Y^{11}) \times 
{\bf H}(Y^{11}, \mathbf{B}^3 {\mathrm{U}(1)}_{\rm conn}) \to {\mathrm{U}(1)}\;.
\)
\item For the case of the M5-brane we consider the dual $C_6$ of the 
C-field on the 6-dimensional smooth oriented worldvolume $\Sigma_6$. 
We again have sigma model maps $\phi: \Sigma_6 \to Y^{11}$ 
which form the space $\mathbf{H}(\Sigma_6, Y^{11})$. The holonomy 
of the dual of the C-field is then 
\(
{\rm hol}_{\rm M5}: {\bf H}(\Sigma_6, Y^{11}) \times 
{\bf H}(Y^{11}, \mathbf{B}^6 {\mathrm{U}(1)}_{\rm conn}) \to {\mathrm{U}(1)}\;.
\)
\end{enumerate}
In the special case that the $C$-field happens to be given by a globally defined
3-form $C_3$ on $Y^{11}$, we have the explicit formula
\( 
 \mathrm{hol}_{\mathrm{M2}}(\phi, C_3)
  =
  \exp(2 \pi i \int_{\Sigma_3} \phi^* C_3)
  \,.
\)
This is the familiar higher gauge coupling or \emph{Wess-Zumino-Witten} term 
in the M2-brane action. The above construction generalizes this to 
general $C$-field configurations (and with just slight adaption to branes with boundary).
The description of holonomy for the M5-brane is analogous. 
The partition functions in this setting are described in \cite{Sati10}.

%%%%%%%%%%%%%%%%%%%%%%%%%%%%%%%%%%%%%%%%%%%%%%%%%%%
\subsection{Differential characteristic maps}
\label{Sec CS elements}
%%%%%%%%%%%%%%%%%%%%%%%%%%%%%%%%%%%%%%%%%%%%%%%%%%%

We have seen abelian differential cohomology in sections \ref{HigherU1Connections}
and \ref{holonomy} and are about to consider higher nonabelian cohomology in 
the next section \ref{Section stacks}. The link between the two is given by 
\emph{characteristic classes}, or rather by their refinement to smooth and
differential characteristic maps between smooth moduli stacks.
These differentially refined characteristic maps are the structures from 
which we obtain higher Chern-Simons action functionals in section
\ref{7dTheory}. Here we briefly review some basic ideas. Details
are in \cite{FSS} and in sections 2.3.18 and 3.3.14 of \cite{survey}.

\medskip

Let $G$ be a Lie group and $[c]\in H^{n+1}(BG; \mathbb{Z})$ 
an integral characteristic class on its classifying space. 
If $G$ is compact, then by theorem 3.3.29 in \cite{survey}
there is an isomorphism
\(
  H^{n+1}(BG; \mathbb{Z})
    \cong 
  \pi_0\mathbf{H}(\mathbf{B}G;\mathbf{B}^n {\mathrm{U}(1)})
  \;,
\)
between the integral cohomology in degree $n+1$ of the topological space
$B G$, and the gauge equivalence classes of circle $n$-bundles over the
moduli stack $\mathbf{B}G$. In other words, there is, up to equivalence,
a unique morphism
\(
  \mathbf{c}
   :
   \mathbf{B}G 
     \to 
   \mathbf{B}^n {\mathrm{U}(1)}
\) 
of smooth stacks, such that for $E : X \to \mathbf{B}G$ the map
of stacks classifying a $G$-principal bundle on $X$, the integral class
$c(E) \in H^{n+1}(X, \mathbb{Z})$ is that classifying the circle $n$-bundle
given by the composite $\mathbf{c}(E) 
: X \to \mathbf{B}G \stackrel{\mathbf{c}}{\to} \mathbf{B}^n {\mathrm{U}(1)}$.

\medskip
A simple example of this that we have already seen
was the smooth refinement $\mathbf{c}_1$ of the
first Chern class in (\ref{DeterminantAsSmoothC1}).
For this example we obtained in (\ref{DifferentialC1})
furthermore a \emph{differential refinement}, namely a morphism
$\hat {\mathbf{c}}$ between the corresponding moduli stacks for 
bundles with connection, which refines $\mathbf{c}$ by fitting into 
a homotopy commuting diagram
\(
  \raisebox{20pt}{
  \xymatrix{
    \mathbf{B}G_{\mathrm{conn}}	
    \ar[r]^{\hat{\mathbf{c}}}
	\ar[d]& \mathbf{B}^n {\mathrm{U}(1)}_{\mathrm{conn}}
    \ar[d]
    \ar[r]^<<<<<{\mathrm{curv}}
    & \Omega^{n+1}_{\mathrm{cl}} 
    \\
    \mathbf{B}G\ar[r]^{{\mathbf{c}}}& \mathbf{B}^n {\mathrm{U}(1)}
  }
  }
  \label{DiagramForDifferentialRefinement}
  \,.
\)
This diagram in particular 
says that for $(E,\nabla)$ a $G$-principal bundle $E$
with connection $\nabla$, then the $(n+1)$-form curvature
of the circle $n$-bundle $\hat {\mathbf{c}}(E,\nabla)$
is a de Rham representative of the integral class of the underlying
circle $n$-bundle $\mathbf{c}(E)$. 

\medskip
When restricted to gauge equivalence classes and to cohomology, this is 
a construction that is provided by classical \emph{Chern-Weil theory},
simply by evaluating the curvature form $F_\nabla$ in an 
\emph{invariant polynomial} $\langle -\rangle$ on the Lie algebra of 
$\mathfrak{g}$. We need to refine this construction away from 
gauge equivalence classes to the full higher moduli stacks.
A step in this direction has been provided by Brylinski and McLaughlin in 
\cite{brylinski-mclaughlin}, who showed how to construct for
every single {\v C}ech-Deligne cocycle for a $G$-principal bundle 
representing a map $(E, \nabla) : X \to \mathbf{B}G_{\mathrm{conn}}$, a corresponding 
{\v C}ech-Deligne cocycle for the the circle $n$-bundle with 
connection $\hat {\mathbf{c}}(E) : X \to \mathbf{B}^n {\mathrm{U}(1)}_{\mathrm{conn}}$.
Based on this and on $L_\infty$-algebraic resolutions considered in 
\cite{SSSIII}, we gave in \cite{FSS} a construction of genuine
morphisms $\hat {\mathbf{c}}$ of higher moduli stacks by a procedure that
may be understood as a higher analog of Lie integration of Lie algebra
homomorphism, generalized to Lie $n$-algebras. 
However, the morphisms of higher stacks obtained this way typically go 
out of \emph{higher connected covers} of moduli stacks for Lie groups, 
and extra work is required in pushing them down again. This 
phenomenon is, in low degree, already familiar from classical Lie theory: 
there, morphisms of 
(finite dimensional) Lie algebras correspond to morphisms between the
\emph{connected} and \emph{simply connected} Lie groups integrating these. In order to 
get morphisms between non-simply connected Lie groups from Lie integration,
extra quotienting is required, which may or may not be respected by the morphism.

\paragraph{
%\color{blue}
Example.}
The simplest instance of this phenomenon occurs with $G=O(2)$, the two-dimensional orthogonal group. Namely, the Lie algebra of $O(2)$ is $\mathfrak{so}_2$ and there is a (unique up to a scalar factor) nontrivial real valued Lie algebra 1-cocycle $\mu_1$ on $\mathfrak{so}_2$ (i.e., a nontrivial Lie algebra morphism $\mu_1:\mathfrak{so}_2\to \mathbb{R}$). Yet, $\mu_1$ cannot be integrated to a Lie group 1-cocycle (i.e., to a morphism of Lie groups) $\rho:O(2)\to {\mathrm{U}(1)}$; indeed, the only nontrivial Lie group homomorphism from $O(2)$ to ${\mathrm{U}(1)}$ is $\det:O(2)\to\{\pm1\}\subseteq {\mathrm{U}(1)}$. Here the topological obstruction to the integration of $\mu_1$ is the nonconnectedness of $O(2)$. And indeed, $\mu_1$ can be integrated if we pass to the ``connected cover'' $SO(2)$ of $O(2)$; its integration is nothing but the well known isomorphism $SO(2)\xrightarrow{\sim}{\mathrm{U}(1)}$. In terms of integral characteristic classes, this discussion is summarized by the fact that $H^2(BO(2);\mathbb{Z})=\mathbb{Z}/2\mathbb{Z}$ and so there are, up to equivalence, only two morphisms of stacks $\mathbf{B}O(2)_{\mathrm{conn}}\to \mathbf{B}{\mathrm{U}(1)}_{\mathrm{conn}}$ corresponding to characteristic classes for $O(2)$: a principal $O(2)$ bundle with connection given by local data $(g_{ij},A_i)$ can be mapped either to the trivial ${\mathrm{U}(1)}$-bundle with connection $(1_{ij},0_i)$ or to the ${\mathrm{U}(1)}$-bundle with connection locally given by $(\det(g_{ij}),0_i)$. Notice that in both cases the map at the level of local connection data is the zero morphism $\mathfrak{so}_2\to \mathbb{R}$.

\medskip
The general statement is the natural generalization of the example just presented: a real valued Lie algebra $n$-cocycle on $\mathfrak{g}$ universally integrates to a morphism of
moduli $n$-stacks
\(
\mathbf{B}G\langle n\rangle\to \mathbf{B}^n {\mathrm{U}(1)}\;,
\)
where $G\langle n \rangle$ is an $(n-1)$-connected cover of the Lie group $G$. 
For instance, if $O$ denotes the infinite orthogonal group and $\mathfrak{so}$ its Lie algebra, then there is no nontrivial real valued Lie algebra 1-cocycle on $\mathfrak{so}$; the canonical  Lie algebra 3-cocycle on $\mathfrak{so}$ integrates to the first fractional Pontrjagin class
\(
\tfrac{1}{2}\mathbf{p}_1:\mathbf{B}\mathrm{Spin} \to \mathbf{B}^3{\mathrm{U}(1)}
  \label{bold p1}
  \,,
\)
where $\mathrm{Spin}(n)$ is the 1-connected cover of $\mathrm{O}(n)$.

\medskip
While this machine goes through for all $n$, a crucial subtlety 
but also a 
source for much of the structure that we discuss here, is that further
higher connected covers of Lie groups \emph{do not exist as Lie groups},
in general. Instead, they only exists as \emph{higher Lie groups}.
Notably, the canonical  Lie algebra 7-cocycle on $\mathfrak{so}$  
integrates to the second fractional Pontrjagin class. But this is no longer
defined on the stack $\mathbf{B}\mathrm{Spin}$ itself, but on a higher
nonabelian stack (a 2-stack, in this case), which we denote $\mathbf{B}\mathrm{String}$.
The reader may think of this as a \emph{twisted product} of the nonabelian 1-stack
$\mathbf{B}\mathrm{Spin}$ with the abelian 2-stack 
$\mathbf{B}^2 {\mathrm{U}(1)}$ discussed in section \ref{HigherU1Connections}.
We describe this in more detail in a moment, below in section \ref{Section stacks}.
So in terms of this String 2-stack, the degree-7 cocycle on $\mathfrak{so}$ turns out 
to integrate the second fractional Pontrjagin class refined to a higher stack
morphism of the form
\(
\tfrac{1}{6}\mathbf{p}_2:\mathbf{B}\mathrm{String} \to \mathbf{B}^7{\mathrm{U}(1)}\;.
\)
Moreover, both these Pontrjagin classes are induced by Lie algebra $n$-cocycles, 
and have differential refinements
\(
\tfrac{1}{2}\hat{\mathbf{p}}_1:\mathbf{B}\mathrm{Spin}_{\mathrm{conn}} \to \mathbf{B}^3{\mathrm{U}(1)}_{\mathrm{conn}} 
\label{p1 conn}
\)
and 
\(
\tfrac{1}{6}\hat{\mathbf{p}}_2:\mathbf{B}\mathrm{String}_{\mathrm{conn}}  
\to \mathbf{B}^7{\mathrm{U}(1)}_{\mathrm{conn}}~\;.
\label{String conn}
\)
A proof of these results, together with a construction of $G\langle n\rangle$ as an higher smooth group via (a version of) the Sullivan construction, and a treatment of the integration of Lie algebra $n$-cocycles can be found in \cite{FSS}. The stacks $\mathbf{B}\mathrm{String}$ and $\mathbf{B}\mathrm{String}_{\mathrm{conn}}$ mentioned above will be defined in Section \ref{Section stacks}.

\begin{remark} One may wonder whether the $n$-cocycles $\mathbf{B}G\langle n\rangle\to \mathbf{B}^n {\mathrm{U}(1)}$ and $\mathbf{B}G\langle n\rangle_{\mathrm{conn}}\to \mathbf{B}^n {\mathrm{U}(1)}_{\mathrm{conn}}$ mentioned above descend to  $n$-cocycles $\mathbf{B}G\to \mathbf{B}^n {\mathrm{U}(1)}$ and $\mathbf{B}G_{\mathrm{conn}}\to \mathbf{B}^n {\mathrm{U}(1)}_{\mathrm{conn}}$. As we mentioned above, the obstruction to this descent is of a topological kind: one has to see whether the characteristic class in $H^{n+1}(BG\langle n\rangle;\mathbb{Z})$ defined by the cocycle is the pullback of an integral characteristic class in $H^{n+1}(BG;\mathbb{Z})$ or not.  For instancee,  the first Pontrjagin class is an element in $H^4(BSO;\mathbb{Z})$. It refines to a morphism of stacks $\mathbf{p}_1:\mathbf{B}SO\to \mathbf{B}^3{\mathrm{U}(1)}$, whose infinitesimal version is twice the standard Lie algebra 3-cocycle $\mu_3$ on $\mathfrak{so}_3$. As mentioned 
above, the 3-cocycle $\mu_3$ integrates to $\frac{1}{2}\hat{\mathbf{p}}_1:\mathbf{B}\mathrm{Spin}_{\mathrm{conn}}\to \mathbf{B}^3{\mathrm{U}(1)}_{\mathrm{conn}}$, and so $2\mu_3$ integrates to $\hat{\mathbf{p}}_1:\mathbf{B}\mathrm{Spin}_{\mathrm{conn}}\to \mathbf{B}^3{\mathrm{U}(1)}_{\mathrm{conn}}$. Since the underlying characteristic class $p_1$ actually lives on $BSO$, there are no obstructions to the descent of $\hat{\mathbf{p}}_1$ to $\mathbf{B}\mathrm{SO}_{\mathrm{conn}}$, and so one has a differential refinement of the first Pontrjagin class to a morphism of stacks
\(
\hat{\mathbf{p}}_1:\mathbf{B}\mathrm{SO}_{\mathrm{conn}} \to \mathbf{B}^3{\mathrm{U}(1)}_{\mathrm{conn}}\;. 
\)
For $n=3$ and $G$ a compact simple Lie group, one can find a full detailed treatment of the descent problem for differential cocycles in \cite{gawedzki-waldorf}.
\end{remark}
\par
As anticipated in the previous section, we will only be concerned with top degree 
local differential forms data for $n$-connections 
given by morphisms $X \to \mathbf{B}^n{\mathrm{U}(1)}_{\mathrm{conn}}$. 
So the relevant question for us now is: what is the degree-$n$ differential form $A_{n;i}$ associated with the 
$\mathfrak{g}$-valued local 1-form $\omega_i$ of a $\mathfrak{g}$-connection and with a real valued Lie algebra $n$-cocycle $\mu$ inducing the characteristic map $\hat{\mathbf{c}}$? And the answer is everything but unexpected:
\(
A_{n;i} = \mathrm{CS}_n(\omega_i)\;,
\)
where $\mathrm{CS}_n$ is a \emph{Chern-Simons element} 
for the Lie algebra cocycle $\mu$. 
Again, see \cite{SSSI},\cite{FSS} or \cite{survey} for details
and for the generalization of the notion of Chern-Simons elements
to Lie $n$-algebras.
 
 \medskip
In what follows, we will need 
the full local data for the ${\mathrm{U}(1)}$-principal 3-bundle 
with connection that is associated by $\frac{1}{2}\hat{\mathbf{p}}_1$ 
to a $\mathrm{Spin}$-principal bundle with connection given by 
local data $(\omega_i,g_{i j})$. These are  obtained as follows: 
\begin{itemize}
  \item
    since the group Spin is connected, over each double intersections one can lift the transition function $g_{ij}$ to  a smooth family of based paths in $G$,
    $\hat g_{i j} : U_{ij} \times\Delta^1 \to G$, with $\hat g_{i j}(0) = e$ and $\hat g_{i j}(1) = g_{i j}$;   \item
    since Spin is 1-connected, over each triple intersection one can find a smooth family of based 2-simplices in $G$,
    $\hat g_{i j k} : U_{ijk}\times\Delta^2 \to G$,
    with boundaries labeled by the based paths on double
    overlaps:
    \(
      \xymatrix{
        & g_{i j} \ar[dr]^{g_{i j} \cdot \hat g_{j k}}
        \\
        e 
          \ar@{-}[rr]_{\hat g_{i k}}^{{\ }}="t" 
          \ar@{-}[ur]^{\hat g_{i j}}
          && 
         g_{i k}~~\;.
        \ar@{}^{\hat g_{i j k }} "t"+(3,2); "t"+(3,2)  
      }
    \)
       \item since Spin is 2-connected,
     on each quadruple intersection one can find a smooth family of based 3-simplices in $G$,
     $\hat g_{i j k l} : U_{ijkl} \times \Delta^3 \to G$, cobounding the union of the 2-simplices corresponding to the triple intersections; 
\item for $n=1,2,3$, let $\hat{\omega}_{i_0\dots i_n}$  be the $\Delta^n$-family of $\mathfrak{so}$-valued 1-form on $U_{i_0,\dots i_n}$ obtained via the gauge action of $\hat{g}_{i_0,\dots,i_n}$ on $\omega_{i_0}\vert_{U_{i_0,\dots,i_n}}$, i.e., $\hat{\omega}_{i_0\dots i_n}\in \Omega^1(U_{i_0,\dots i_n}\times\Delta^n;\mathfrak{so})$ is defined as
\(
\hat{g}_{i_0,\dots,i_n}\omega_{i_0}\hat{g}_{i_0,\dots,i_n}^{-1}+ \hat{g}_{i_0,\dots,i_n}d_U\hat{g}_{i_0,\dots,i_n}^{-1}\;,
\)
 where $d_U$ denotes the de Rham differential in the $U_{i_0,\dots,i_n}$-direction on $U_{i_0,\dots,i_n}\times\Delta^n$.     
\end{itemize}

 These pieces of data are sent to the \v{C}ech-Deligne 3-cocycle
\(
   (C_i, B_{i j}, A_{i j k}, g_{i j k l})
   :=
    \left(\mathrm{CS}_3(\omega_i), \int_{\Delta^1} \mathrm{CS}_3(\hat{\omega}_{i j}) , \int_{\Delta^2} \mathrm{CS}_3(\hat{\omega}_{i j k}),
    \int_{\Delta^3} \mu_3(\hat{\omega}_{i j k l}\wedge \hat{\omega}_{i j k l}\wedge \hat{\omega}_{i j k l})\,~~\mathrm{mod}\,\,\mathbb{Z}\right),
	\label{CD3cocycle}
\)
where $\mu_3$ and $\mathrm{cs}_3$ are the standard 3-cocycle and Chern-Simons element for $\mathfrak{so}$. Notice that the integration over the 3-simplex $\Delta^3$ in the above formula only reads the vertical part of the 3-form $\mu_3(\hat{\omega}_{i j k l}\wedge \hat{\omega}_{i j k l}\wedge \hat{\omega}_{i j k l})$ with respect to the projection $U_{ijkl}\times \Delta^k \to U_{ijkl}$, which, by construction, is $\hat{g}_{ijkl}^*\mu_3(\theta\wedge\theta\wedge\theta)$, where $\theta\in \Omega^1(\mathrm{Spin};\mathfrak{so})$ is the Maurer-Cartan form on the group Spin.
%  \end{itemize}

%%%%%%%%%%%%%%%%%%%%%%%
\subsection{The 2-stacks $\mathbf{B}\mathrm{String}$ and $\mathbf{B}\mathrm{String}_{\mathrm{conn}}$ }
\label{Section stacks}
%%%%%%%%%%%%%%%%%%%%%%%

We now discuss the moduli 2-stacks $\mathbf{B}\mathrm{String}$ and 
$\mathbf{B}\mathrm{String}_{\mathrm{conn}}$ of $\mathrm{String}$-principal 2-bundles and of 
$\mathrm{String}$-principal 2-bundles with 2-connection 
(\cite{FSS} and \cite{survey}, section 4.1.3). 
The definitions and discussion here proceeds in direct analogy
with the discussion in \ref{DeterminantLineBundles}, simply by replacing 
the first Chern class with the first fractional Pontrjagin class.

\medskip

Let $\mathbf{B}\mathrm{Spin}$ be the smooth stack of smooth 
$\mathrm{Spin}$-principal bundles. As remarked above 
around \eqref{bold p1}, 
the first fractional Pontrjagin class $\frac{1}{2}p_1$ refines to a morphism of 
smooth 3-stacks 
$
\tfrac{1}{2}\mathbf{p}_1:\mathbf{B}{\mathrm{Spin}}\to \mathbf{B}^3{\mathrm{U}(1)}
$
and so we can define $\mathbf{B}\mathrm{String}$ as the
homotopy fiber of this morphism, by analogy with (\ref{SUNHomotopyPullback}), 
hence as the homotopy pullback
\(
\raisebox{20pt}{
\xymatrix{
  \mathbf{B}\mathrm{String}\ar[r]_>{\ }="s"\ar[d]&{*}\ar[d]
   \\
   \mathbf{B}\mathrm{Spin}
   \ar[r]^{
     \hspace{-3mm}\frac{1}{2}\mathbf{p}_1
	 }
	 &
	 \mathbf{B}^3{\mathrm{U}(1)}~~
   %
   %\ar@{=>}^b_\simeq "s"; "t"
}
}
\label{diag SS}
\)
of smooth higher stacks. 
One way to characterize what this means in 
more concrete terms (another one we get to in a moment) 
is to say that that a $\mathrm{String}$-principal 2-bundle over 
a smooth manifold $X$, hence a stack morphism $X \to \mathbf{B}\mathrm{String}$, is 
\begin{itemize}
\item a {\v C}ech cocycle datum $(g_{ij})$ of a $\mathrm{Spin}$-principal bundle on $X$; 
\item 
  together with a 
  \emph{choice} of a trivialization of the induced \v{C}ech 3-cocycle 
$(\int_{\Delta^3} \hat{g}_{ijkl}^*\mu_3(\theta\wedge\theta\wedge\theta)\,~~\mathrm{mod}\,\,\mathbb{Z})$ given in (\ref{CD3cocycle}). 
\end{itemize}

\vspace{3mm}
Homotopy pullbacks may be more familiar in the context of topological spaces.
There, for $f : X \to Y$ any continuous map of pointed topological spaces, 
its homotopy fiber may be defined, up to weak homotopy equivalence, as the 
ordinary pullback of the \emph{based path space} projection $P Y \to Y$ of $Y$ along $f$.
But there are many other constructions that give the same result up to 
weak homotopy equivalence.

\vspace{3mm}
To relate the homotopy pullbacks of higher stacks with those of 
topological spaces, we invoke the geometric realization map
from (\ref{GeometricRealization}). 
By the discussion there, 
the image
of $\frac{1}{2}\mathbf{p}_1$ under geometric realization is a continuous map
of topological spaces
\(
  \tfrac{1}{2}p_1 : B \mathrm{Spin} \to K(\mathbb{Z}, 4)
  \,,
\)
hence a class in degree-4 integral cohomology $H^4(B \mathrm{Spin}, \Z)$. 
This is the ordinary Ponrtryagin class of which $\frac{1}{2}\mathbf{p}_1$ is the 
smooth refinement.

\vspace{3mm}
While, in general, geometric realization of higher stacks does not preserve
homotopy fibers, one can show that in some special cases, such as the one above, 
it does (theorem 3.3.25 in \cite{survey}). This means that 
\(
  \vert \mathbf{B}\mathrm{String}\vert
  \simeq B \mathrm{String}
\)
is indeed the classifying space of the topological String group as traditionally 
defined, fitting into a homotopy pullback 
\(
  \raisebox{20pt}{
  \xymatrix{
     B \mathrm{String} \ar[r] \ar[d] & {*}\ar[d]
	 \\
	 B \mathrm{Spin}
	 \ar[r]^{\hspace{-2mm}\frac{1}{2}p_1}
	 &
	 K(\mathbb{Z}, 4)
  }
  }
  \label{BStringTopological}
\)
of topological spaces. In this form we can see more directly than otherwise
in which way the String group is indeed related to something \emph{stringy}. 
For if we form topological loop spaces twice on this last homotopy pullback
diagram, we get the homotopy pullbacks
\(
 \raisebox{20pt}{
  \xymatrix{
    {\mathrm{U}(1)} \ar[r]\ar[d] & \Omega \mathrm{String} \ar[rr]\ar[d]
     &&
	{*}
	\ar[d]
	\\
    {*} \ar[r] & \Omega \mathrm{Spin} \ar[rr]^{\Omega^2\frac{1}{2}p_1\phantom{m}}
	&&
    B {\mathrm{U}(1)}~~
}
}
\)
and hence the long fiber sequence
\(
  \xymatrix{
    {\mathrm{U}(1)} \ar[r] &  \Omega \mathrm{String} \ar[r] &  \Omega \mathrm{Spin}
	\ar[r]^{\frac{1}{2}p_1}
	&
	K(\mathbb{Z}, 4)
  }
  \,.
\)
This exhibits an equivalence between the loop space of the topological String group 
and the level-1 Kac-Moody
central extension $\hat {\Omega} \mathrm{Spin}$  of the topological loop group of $\mathrm{Spin}$
\(
  \Omega \mathrm{String}
  \simeq
  \hat \Omega \mathrm{Spin}
  \,.
\) 
It is in this reduced form that 
String bundles first appeared in the string theory literature \cite{Witten86}: as 
central extensions of the structure loop group on the loop space of spacetime
(hence the configuration space of the closed string). But some torsion 
information is lost in this double looping ($\Omega \frac{1}{2}p_1(\Omega X)$
may vanish even if $\frac{1}{2}p_1(X)$ does not), and the precise structure 
needed in the heterotic string are genuine (twisted) String 2-bundles \cite{SSSIII}
as discussed here.

\medskip
While the above simple argument, derivng the topological Kac-Moody loop group
from String,
does not go through quite in this form for the fully-fledged
\emph{smooth} (stacky) version of $\mathrm{String}$, it turns out that
a  refinement of this state of affairs does hold true:
In \cite{BCSS} it is shown that when the Kac-Moody loop group
$\hat \Omega \mathrm{Spin}$ is regarded as a smooth group
(a Fr{\'e}chet group), then the evident smooth map
\(
  \hat \Omega \mathrm{Spin} \to P \mathrm{Spin}
\)
to the smooth group $P \mathrm{Spin}$ of based paths in $\mathrm{Spin}$
is equipped with an action of the path group, such that it becomes what is called
a \emph{crossed module} of smooth groups. This may naturally be thought 
of as defining a smooth 2-stack (def. 1.3.5 in \cite{survey}), written 
$\mathbf{B}(\hat \Omega \mathrm{Spin} \to P \mathrm{Spin})$. By 
theorem 4.1.29 in \cite{survey}, this turns out to be an equivalent
incarnation of the moduli 2-stack of $\mathrm{String}$-2-bundles
\(
  \mathbf{B}\mathrm{String} 
    \simeq
  \mathbf{B}(\hat \Omega \mathrm{Spin} \to P \mathrm{Spin})
  \,.
\)
Having two such seemingly different descriptions for the stack of principal String bundles should not be surprising: $\mathbf{B}\mathrm{String}$ 
is defined as a homotopy fiber, and so it is well defined only up to weak equivalence. 

\medskip
Analogous considerations apply for the infinitesimal 
approximation to the smooth $\mathrm{String}$-2-group: its
Lie 2-algebra $\mathfrak{string}$ (see \cite{SSSI}). 
This Lie 2-algebra is abstractly defined (def. 4.1.15 in \cite{survey}) 
as the $L_\infty$-algebra homotopy fiber of the canonical
degree-3 Lie algebra cocycle $\mu_3$ on $\mathfrak{so}$, which 
we may identify with a map of Lie 3-algebras
\(
  \mu_3
    :
  \mathfrak{so}\to b^2\mathfrak{u}(1)
\)
from $\mathfrak{so}$ to the \emph{line Lie 3-algebra}, the two-fold delooping
of $\mathfrak{u}(1) \simeq \mathbb{R}$. Again, by the very definition
of homotopy pullbacks, this homotopy fiber may be presented by a Lie 2-algebra,
denoted $\mathfrak{so}_{\mu_3}$, defined by the fact that a morphism
of Lie 2-algebroids (see \cite{SSSI})
\(
  (\omega,B) : T X \to b (\mathfrak{so}_{\mu_3})
\)
from the tangent Lie algebroid of a smooth manifold $X$ is
\begin{itemize}
  \item a flat $\mathfrak{so}$-valued 1-form $\omega$;
  \item
    equipped with a choice of 2-form $B$ which trivializes the 
	Chern-Simons 3-form of $\omega$, in that 
	\(
	  d B = \mathrm{tr}(\omega \wedge \omega \wedge \omega)\,.
	  \label{flatgmuformdata}
	\)
\end{itemize}
But, as before for the finite $\mathrm{String}$-2-group, there is,
now by theorem 30 in \cite{BCSS}, a quite different looking
but equivalent incarnation of the Lie 2-algebra $\mathfrak{string}$.
This is given by a pair of ordinary non-abelian Lie algebras, 
the Kac-Moody loop Lie algebra $\hat \Omega \mathfrak{so}$, 
and the based path Lie algebra $P \mathfrak{so}$, 
equipped with the canonical Lie algebra homomorphism
$(\hat{\Omega}\mathfrak{so} \stackrel{h}{\to} P_*\mathfrak{so})$, which defines
what is called a ``strict'' Lie 2-algebra 
(all this is reviewed in the Appendix). This is an equivalent 
presentation (an equivalent higher gauge incarnation) of $\mathfrak{string}$: 
\(
  \mathfrak{so}_{\mu_3}  \; \simeq \; (\hat{\Omega}\mathfrak{so} \stackrel{h}{\to} P_*\mathfrak{so})\;.
  \label{EquivalentIncarnationsOfFrakString}
\)

\medskip
Using this infinitesimal description of the String-2-group, we obtain
now, by \cite{FSS}, 
a differential refinement of $\mathbf{B}\mathrm{String}$, namely the
moduli 2-stack of String-2-bundles equipped with 2-connections.
As remarked in Section \ref{Sec CS elements} (see eq. \eqref{p1 conn}), 
using this, the 
morphism of 3-stacks $\frac{1}{2}\mathbf{p}_1$ can be further refined to a morphism of 
differentially refined 3-stacks
$
\tfrac{1}{2}\hat{\mathbf{p}}_1:\mathbf{B}\mathrm{Spin}_{\mathrm{conn}}\to \mathbf{B}^3{\mathrm{U}(1)}_{\mathrm{conn}}
$.
Therefore, we could define, analogous to (\ref{SUConnHomotopyPullback}), 
the moduli 2-stack of String bundles with connections as the homotopy fiber of $\tfrac{1}{2}\hat{\mathbf{p}}_1$ over the trivial ${\mathrm{U}(1)}$-2-gerbe endowed with the trivial connection.
However, this turns out to be too a strict notion, so
one prefers to adopt a more flexible one, i.e. to define the stack $\mathbf{B}\mathrm{String}_{\mathrm{conn}}$ by analogy with the homotopy fiber of (\ref{CompositeMapc1}),  to  
be the homotopy fiber of $\tfrac{1}{2}\hat{\mathbf{p}}_1:\mathbf{B}\mathrm{Spin}_{\mathrm{conn}}\to \mathbf{B}^3{\mathrm{U}(1)}$ over the the trivial ${\mathrm{U}(1)}$-2-gerbe. In other words, one asks that the underlying Chern-Simons ${\mathrm{U}(1)}$-2-gerbe 
$\frac{1}{2}{\mathbf{p}}_1(P)$ of a $\mathrm{Spin}$-principal bundle $P$
with connection trivializes, but without requiring that the induced connection 
on it be trivial. String  structures of this kind have been considered in \cite{Waldorf}; they are 
particular examples of the more general notion of \emph{twisted String structure} 
considered in \cite{SSSIII}, to which we turn in section
\ref{SmoothStringC2}.

\medskip
In more detail, $\mathbf{B}\mathrm{String}_{\mathrm{conn}}$ is defined by the homotopy pullback
\(
  \raisebox{20pt}{
  \xymatrix{
     \mathbf{B}\mathrm{String}_{\mathrm{conn}}
	   \ar[r]
	   \ar[d]
	   &
	   {*}
	   \ar[d]
	   \\
       \mathbf{B}\mathrm{Spin}_{\mathrm{conn}}
	   \ar[r]^{\frac{1}{2}{\mathbf{p}}_1}
	   &
	   \mathbf{B}^3{\mathrm{U}(1)}~~,
  }
  }
\label{StringConnVar}
\)
which is the higher analog of (\ref{CompositeMapc1}).
Since we have a homotopy pullback
\(
  \raisebox{20pt}{
  \xymatrix{
    \Omega^{1\leq\bullet\leq 3}\ar[r]\ar[d]
	&{*}
	\ar[d]
	\\
    \mathbf{B}^3{\mathrm{U}(1)}_{\mathrm{conn}}
	\ar[r]
	&\mathbf{B}^3{\mathrm{U}(1)}
 }
 }
\)
analogous to (\ref{FiberOfForget}),
the 2-stack $\mathbf{B}\mathrm{String}_{\mathrm{conn}}$ is equivalently 
given by the homotopy pullback
\(
  \raisebox{20pt}{
  \xymatrix{
    \mathbf{B}\mathrm{String}_{\mathrm{conn}}
	\ar[r]
	\ar[d]
	&
	{\Omega^{1\leq\bullet\leq 3}}
	\ar[d]
	\\
    \mathbf{B}\mathrm{Spin}_{\mathrm{conn}}
	\ar[r]^{\frac{1}{2}\hat{\mathbf{p}}_1}
	&
	\mathbf{B}^3{\mathrm{U}(1)}_{\mathrm{conn}}~~,
  }
  }
\)
in analogy with (\ref{BSUconnVariant}). 
As we have seen in Section \ref{Sec CS elements}, a realization of $\frac{1}{2}\hat{\mathbf{p}}_1$ 
maps the local data $(\omega_i,g_{ij})$ of a Spin-bundle with connection on a 
manifold $X$ to the  \v{C}ech-Deligne 3-cocycle
\(
  (C_i, B_{ij}, A_{i j k}, g_{i j k l})
   :=
        \left(\mathrm{CS}_3(\omega_i), \int_{\Delta^1} \mathrm{CS}_3(\hat{\omega}_{i j}) , \int_{\Delta^2} \mathrm{CS}_3(\hat{\omega}_{i j k}),
       \int_{\Delta^3} \mu_3(\hat{\omega}_{i j k l}\wedge \hat{\omega}_{i j k l}\wedge \hat{\omega}_{i j k l})\,~~\mathrm{mod}\,\,\mathbb{Z}\right)~.
\)
So one can derive local expressions for local differential form data of a String connection in perfect analogy with formulae at the end of Section \ref{HigherU1Connections} . Alternatively, one can derive all these local differential forms data and the equations they satisfy by simplicial integration of the string Lie 2-algebra $\mathfrak{so}_{\mu_3}$ (as in section 6.3 of \cite{FSS}, based on the $L_\infty$-algebraic resolutions in \cite{SSSIII}). Whichever of these equivalent presentations one adopts, one finds that on each $U_i$ of the chosen 
open cover $\mathcal{U}$ of $X$ the datum of a 
String 2-connection is the datum of an $\mathfrak{so}$-valued 
1-form $\omega_i$  and of a real valued 2-form $B_i$ on  $U_i$, with 3-form curvature
\(
  \mathcal{H}_i := dB_i +\mathrm{CS}_3(\omega_i)
  \label{3FormCurvatureForsomuConnection}
 \, ,
\)
satisfying a system of compatibility conditions on double and triple
overlaps of the patches in the cover. 
On the other hand, due to the equivalence of Lie 2-algebras
(\ref{EquivalentIncarnationsOfFrakString}), there is an equivalent but 
rather different looking higher gauge in which a String 2-connection is
locally on any $U_i$ given by a pair
$(A_i, \hat{B}_i)$ of nonabelian differential forms, 
with $A_i \in \Omega^1(U_i, P \mathfrak{g})$, 
and $\hat{B}_i \in \Omega^2(U_i, \Omega \mathfrak{g}\oplus \mathbb{R})$. 
Notice that this has correspondingly a pair $(\mathcal{F}_i, \mathcal{H}_i)$ of cuvature forms, with 
\begin{eqnarray}
  \mathcal{F}_i &=& d A_i + \tfrac{1}{2} [A_i \wedge A_i] + h(B_i)
  \;\;\;
  \in \Omega^2(U_i, P_* \mathfrak{so})
  \label{2FormCurvature}
\\
  \mathcal{H}_i &=& d B_i + [A_i \wedge B_i]
  \;\;\;
  \in \Omega^3(U_i, \hat \Omega \mathfrak{so})
  \label{3FormCurvature}
  \,
\end{eqnarray}
(see \cite{SchreiberWaldorfII} for details). Here the bracket in the first line is the Lie bracket on the $P_* \mathfrak{so}$-components
of the 1-form $A$, and in the second line it is the action of the $P_* \mathfrak{so}$-components
of $A$ on the $\hat \Omega \mathfrak{so}$-components of $\hat B$
(see \cite{BCSS} for details of this action). The map $h$ in the first line
sends the $\hat \Omega \mathfrak{so}$-valued 2-form to the underlying $P_* \mathfrak{so}$-valued
2-form obtained by forgetting the central extension and regarding a loop as a special based path.

\medskip
A higher nonabelian curvature structure of this form was also 
proposed in \cite{SSW} for a description of nonabelian 2-forms on 5-branes,
as recalled above around (\ref{SSWProposalForCurvatures}). There, also 
a Chern-Simons term was added, as in (\ref{3FormCurvatureForsomuConnection}). 
We will see further Chern-Simons terms appear
and acts as twists of  the above curvature relations when we pass
to \emph{twisted} String-2-connections  in section \ref{SmoothStringC2} below,
(see equation (\ref{String2aConnection3Form}) there),
though it seems that the details that we derive differ from 
the proposal in \cite{SSW}.
Precisely in the special case that 
the 2-form curvature (\ref{2FormCurvature}) vanishes, 
there is a natural notion of 
non-Abelian Wilson-\emph{surface} observables for 2-connections. 
This and the full local data for such 2-connections is 
derived and spelled out in 
\cite{SchreiberWaldorfII, SchreiberWaldorfIII}.

\medskip

%%%%%%%%%%%%%%%%%%%%%%%%%%%%%%%%%%%%%%%%%%%%%%%%%%%%%%%%%%%%%%%%%%%
\subsection{The 2-stack $\mathbf{B}\mathrm{String}^{2\mathbf{a}}_{\mathrm{conn}}$ of twisted String 2-connections}
\label{SmoothStringC2}
\index{higher $\mathrm{Spin}^c$-structures!$\mathrm{String}^{\mathbf{c}_2}$}
%%%%%%%%%%%%%%%%%%%%%%%%%%%%%%%%%%%%%%%%%%%%%%%%%%%%%%%%%%%%%%%%%%%

We discuss now a notion of \emph{twisted} $\mathrm{String}$-2-connections.
These relate to the String-2-connections from section \ref{Section stacks}
as \emph{twisted vector bundles with connection} (as in twisted K-theory)
relate to ordinary vector bundles. 
\footnote{For more context in string theory see \cite{SSSIII}, for general theory see 
sections 2.3.5, 3.3.7 and 4.4 of \cite{survey}.}
In \cite{FiSaSc} we find twisted $\mathrm{String}$-2-connections in boundary
field configurations of $C$-fields, reviewed below in section \ref{CFieldConfigurations},
and in this form we will identify them as the field configurations 
for 7d Chern-Simons theories in section \ref{CSOnStacksWithCharges}.

\medskip

The construction of the refinement of the 
first fractional Pontrjagin class to a morphism of stacks described in 
section \ref{Section stacks} rests only on the fact 
that $\mathrm{Spin}$ is a compact and simply connected simple Lie group, and so 
the same argument applies to the exceptional
Lie group $E_8$. By classical results \cite{BottSamelson} 
its first non-vanishing homotopy group is
$\pi_3(E_8) \simeq \mathbb{Z}$
and so it follows by the Hurewicz theorem that 
$
  H^4(B E_8, \mathbb{Z}) \simeq \mathbb{Z}
  \,.$ 
  Therefore the generator of this group  is, 
up to sign, a canonical characteristic class, which we write
$
  [a]\in H^4(B E_8, \mathbb{Z}),
$
corresponding to a characteristic map 
\(
a : B E_8 \to K(\mathbb{Z},4)\;.
\)
For any integer $k$, the characteristic class 
  $
    k[a] \in H^4(BE_8,\mathbb{Z})
  $
  has an essentially unique refinement 
  \(
     k\mathbf{a} : \mathbf{B}E_8 \to \mathbf{B}^3 {\mathrm{U}(1)}
  \)
  to a morphism of smooth stacks,   
  a representative of which is provided by the Lie integration of  
  $k \mu_3^{\mathfrak{e}_8}$
  according to \cite{FSS}, where $\mu_3^{\mathfrak{e}_8}$ denotes the canonical Lie algebra 3-cocycle on $\mathfrak{e}_8$.
  Therefore, we can consider 
  the smooth 2-groups $\mathrm{String}^{k\mathbf{a}}$, defined to be the loop space objects 
  of the homotopy pullback in the top left corner of
  \(
    \raisebox{20pt}{
    \xymatrix{
	  \mathbf{B}\mathrm{String}^{k\mathbf{a}}
	  \ar[r]
	  \ar[d]
	  &
	  \mathbf{B} E_8
	  \ar[d]^{k\mathbf{a}}
	  \\
	  \mathbf{B} \mathrm{Spin}
	  \ar[r]^{\hspace{-4mm}\frac{1}{2}\mathbf{p}_1}
	  &
	  \mathbf{B}^3 {\mathrm{U}(1)}~~\;.
	}
	}
  \)
  This is the construction alluded to in the introduction around 
  (\ref{HomotopyPullbackFirstAppearance}).
  By using the (higher) abelian group structure on $ \mathbf{B}^3 {\mathrm{U}(1)}$, the stacks $\mathbf{B}\mathrm{String}^{k\mathbf{a}}$ can be equivalently seen as the homotopy fibers of the difference
  $\tfrac{1}{2}\mathbf{p}_1 - k\mathbf{a}$, via a stacky generalization of the 
  description in \cite{SSSIII},
  \(
    \raisebox{20pt}{
    \xymatrix{
	  \mathbf{B}\mathrm{String}^{k\mathbf{a}}
	  \ar[rr]
	  \ar[d]
	  &&
	  {*}
	  \ar[d]
	  \\
	  \mathbf{B}(
	   \mathrm{Spin}
	  \times
	  E_8
	  )
	  \ar[rr]^{~\frac{1}{2}\mathbf{p}_1 - k\mathbf{a}}
	  &&
	  \mathbf{B}^3 {\mathrm{U}(1)}~~\;.
	}
	}
	\label{Stringka}
  \)
  By the defining nature of homotopy pullback, this means,
  in generalization of the discussion below (\ref{diag SS}),  that 
  a $\mathrm{String}^{k\mathbf{a}}$-principal 2-bundle, classified by 
  a morphism of 2-stacks $X \to \mathbf{B}\mathrm{String}^{k \mathbf{a}}$,
  is equivalently the data of
  \begin{itemize}
    \item 
	   an ordinary $\mathrm{Spin}$-principal bundle $P$ and an 
	   ordinary $E_8$-principal bundle $E$;
	\item
	   equipped with a choice of gauge transformation
	   \(
	     h : \tfrac{1}{2}\mathbf{p}(P) \stackrel{\simeq}{\to} k \mathbf{a}(E)
		 \label{StringkaRefinedCohomologyCondition}
	   \)
	   between their Chern-Simons circle 3-bundles.
  \end{itemize}
  The image of this last condition in integral cohomology is
  \(
    \tfrac{1}{2}p_1 = k a \;\; \in H^4(X, \mathbb{Z})
	\label{StringkaCohomologyCondition}
	\,.
  \)
  For $k = 2$ this is the ``quantization condition'' for supergravity $C$-field
  configurations on a 5-brane boundary. We come back to this in section \ref{CFieldConfigurations}.

\medskip
  The 2-group $\mathrm{String}^{k \mathbf{a}}$ is related
  to the 2-group $\mathrm{String}$ in higher analogy of how 
  the ordinary group $\mathrm{Spin}^c$ is related to $\mathrm{Spin}$.
  This is explained in \cite{SSSIII}.
  As also discussed there, the Freed-Witten anomaly cancellation mechanism
  for type II strings on D-branes
  implies twisted $\mathrm{Spin}^c$-structures on D-branes.
  Here, the (twisted) $\mathrm{String}^{2 \mathbf{a}}$-structures that
  we find on M5-branes can therefore be understood as a direct higher
  generalization of this to higher dimension.
  Notice that for $k = 0$ we recover the untwisted string 2-group 
  from section \ref{Section stacks} together with a factor of $E_8$:
  \(  
    \mathrm{String}^{0 \mathbf{a}} \simeq \mathrm{String} \times E_8
	\,.
  \)
  This is the higher analog of the fact that the ``untwisted version'' of
  $\mathrm{Spin}^c(N)$ is $SO(N) \times {\mathrm{U}(1)}$.
  
  \medskip
  Also notice that, by a classical fact \cite{BottSamelson} which 
  explains much of the role of $E_8$ in 11-dimensional supergravity, 
  $E_8$ is 14-connected, so that for $X$ of dimension 11 
  (and more generally for $X$ of dimension $\leq 15$ ) $E_8$-bundles
  on $X$ have the same classification as circle 3-bundles / 2-gerbes on $X$
  in that there is precisely one equivalence class of them for each element
  of $H^4(X, \mathbb{Z})$. Accordingly, configurations on such $X$ that satisfy
  (\ref{StringkaCohomologyCondition}) for \emph{some} $E_8$ bundle with class
  $a$ are precisely the $\mathrm{Spin}$-structures for which $\frac{1}{2}p_1$
  -- hence $\lambda$ from expression (\ref{Lambda}) -- is further divisible by $k$.
  To amplify this, observe that the identity morphism 
  \(
    \mathbf{DD}_2 : \mathbf{B}^3 {\mathrm{U}(1)} \to \mathbf{B}^3 {\mathrm{U}(1)}
	\label{SmoothDD2}
  \)
  is the canonical smooth refinement of the canonical 4-class of 
  circle 3-bundles / 2-gerbes (the higher \emph{Dixmier-Douady class}), which 
  induces for each $k \in \mathbb{Z}$ the smooth 2-group 
  $\mathrm{String}^{k \mathbf{DD}_2}$. This is such that a lift
  from a $\mathrm{Spin}$-structure to a $\mathrm{String}^{2 \mathbf{DD}_2}$-structure
  exists precisely if $\lambda$ is further divisible by $k$, irrespective of
  the dimension of $X$.

\vspace{3mm}
  As before for $\mathrm{String}$ itself, with the methods of \cite{FSS}
  we obtain a differential refinement to a moduli 2-stack 
  $\mathbf{B}\mathrm{String}^{k \mathbf{a}}_{\mathrm{conn}}$ of 
  $\mathrm{String}^{2\mathbf{a}}$-connections.  
   This has a presentation by differential Lie integration of a
   Lie 2-algebra that extends the direct sum $\mathfrak{so} \oplus \mathfrak{e}_8$
   via its canonical 3-cocycle.
 From this one finds, in generalization of the discussion 
 around (\ref{3FormCurvatureForsomuConnection}), 
 that there is a higher gauge in
which  $\mathrm{String}^{k\mathbf{a}}$-connections are locally given by
 \begin{itemize}
   \item an $\mathfrak{so}$-valued 1-form $\omega_i$;
   \item an $\mathfrak{e}_8$-valued 1-form $A_i$;
   \item a 2-form $B_i$;
 \end{itemize}
 with local curvature 3-form
 the sum of the (opposite of the) de Rham differential of $B$ with the
 difference of the Chern-Simons form of $\omega$
 and $k$ times the Chern-Simons form of $A$, respectively: 
 \(
   \mathcal{H}_i = d B_i + \mathrm{CS}_3(\omega_i) - k\mathrm{CS}_3(A_i)
   \label{String2aConnection3Form}
   \,.
 \)
Note that this implies the equation 
\(
  d\mathcal{H}_i=\langle F_{\omega_i}\wedge F_{\omega_i}\rangle-k\langle F_{A_i}\wedge F_{A_i}\rangle
  \,,
\)
which is the de Rham image of the characteristic relation
(\ref{StringkaRefinedCohomologyCondition}).
It is no coincidence that these are the formulae known from the heterotic 
Green-Schwarz mechanism. See \cite{SSSIII} for more on that.

\vspace{3mm}
Again, beware that these local formulae are a little deceptive, in that
on the one hand there are other higher gauges in which they look rather different,
and also the formulae on single patches $U_i$ do not reflect the complexity of the
data and its conditions on double and triple overlaps.
As before for bare $\mathrm{String}$-2-connections in 
section \ref{Section stacks} we have: due to the higher gauge freedom,
there are other -- very different looking local formulae -- that are however 
higher gauge equivalent via an equivalence explained in the Appendix. In that 
other gauge, the 2-form $B$ above is instead non-abelian and valued in 
a Kac-Moody loop Lie algebra; accordingly, the 3-form curvature is nonabelian
and is given by a twisted version of equation (\ref{3FormCurvature}).

\vspace{3mm}
We discuss the role of these $\mathrm{String}^{k \mathbf{a}}$-2-connections
in 5-brane physics below in section \ref{7dTheory}.

%%%%%%%%%%%%%%%%%%%%%%%%%%%%%%%%%%%%%%%%%%%%
\subsection{The differential second Pontrjagin class}  
\label{SecondPontrjaginOnStacks}
%%%%%%%%%%%%%%%%%%%%%%%%%%%%%%%%%%%%%%%%%%%%%

We discuss now the smooth and differential refinement of the
\emph{second} Pontrjagin class $p_2$ from \cite{SSSI, FSS},
defined on the 2-stack of $\mathrm{String}$-2-connections. Below, in 
section \ref{InfinCS7CS}, this will give the indecomposable part of the 
nonabelian 7-dimensional Chern-Simons theory.

\medskip

In Section \ref{Sec CS elements}
we saw, following \cite{FSS}, 
how the topology of the $\mathrm{Spin}$ group induces a natural morphism 
of higher smooth stacks $\frac{1}{2}\mathbf{p}_1:\mathbf{B}\mathrm{Spin}\to \mathbf{B}^3{\mathrm{U}(1)}$
which differentially refines the first fractional Pontrjagin class. 
Recall that this was constructed in terms of systems of smooth functions
\(
  U_{i j k l} \times \Delta^3 \to \mathrm{Spin}
\)
on patches of space $U$ times a 3-simplex, which serve as a big resolution of
cocycle data $g_{i j} : U_{i j} \to \mathrm{Spin}$ for $\mathrm{Spin}$-principal bundles.
In order to extend this kind of construction directly to one that
supports a construction of $p_2$, we would need to pass all the way up to
7-simplices.
But the nontriviality of the third homotopy group of $\mathrm{Spin}$  
says that there is a topological obstruction to further extending a smooth function
\(
  U_{ijklm}\times \partial\Delta^4 \to \mathrm{Spin}\;,
\) 
%defined by the $\hat{g}_{i_0,\dots,i_3}$ with $i_j\in\{i,j,k,l,m\}$, 
to a map
\(
  U_{ijklm}\times \Delta^4\to \mathrm{Spin}
  \,.
\)
Specifically, we have $\pi_3(\mathrm{Spin})=\mathbb{Z}$, 
and, via the Hurewicz isomorphism, 
\(
  \pi_3(\mathrm{Spin})\cong H^4(B\mathrm{Spin},\mathbb{Z})
  \,.
\)
Therefore, the 
generator of $\pi_3(\mathrm{Spin})$ is canonically 
identified with the class $\frac{1}{2}p_1$, and so the vanishing of the first fractional Pontrjagin class of a Spin bundle precisely means that the above obstruction to further lifting 
of the nonabelin cocycles vanishes. 

\vspace{3mm}
We need to be more precise here, and recall a bit of obstruction theory: 
the vanishing 
of the cohomology class for an  obstruction cocycle does not mean that the cocycle 
is unobstructed, but that in the same cohomology class we can find {\it an} 
unobstructed cocycle. Moreover, the datum of a trivialization of the obstruction 
cocycle precisely tells us how to modify the cocycle in order to get an unobstructed 
cocycle. So the data of a String bundle can be read as the data of a Spin bundle together 
with the ``instructions'' to overcome the first topological obstruction to extend 
the transition functions of the bundle to higher and higher dimensional simplices. 
Once the first obstruction is passed, the construction will go on until the second 
obstruction is met. Since $\pi_i(\mathrm{Spin})$ vanishes for $4\leq i\leq 6$, while 
$\pi_7(\mathrm{Spin})=\mathbb{Z}$, the second topological obstruction for a 
Spin bundle is represented by the generator $Q_2$ of $H^8(B\mathrm{Spin},\mathbb{Z})$, 
which involves the \emph{second} Pontrjagin class $p_2$ as
$Q_2:=\frac{1}{2}(p_2 -\lambda^2)$, which is a power of 2 multiple of the one-loop 
polynomial \cite{KSpin}. Moreover, 
when further refined from $\mathrm{Spin}$ to $\mathrm{String}$,
the generator is simply $\frac{1}{6} p_2$ (see \cite{SSSII}).

\vspace{3mm}
This can be elegantly expressed in terms of classifying spaces: let $B\mathrm{String}$ 
be the \emph{classifying space of String bundles} defined as the homotopy fiber 
of the characteristic map
\(
\tfrac{1}{2}p_1\colon B\mathrm{Spin}\to K(\mathbb{Z},4)\;.
\)
The long exact homotopy sequence and Hurewicz theorem then tell us that $H^i(B\mathrm{String},\mathbb{Z})=0$ for $0\leq i\leq 7$ and $H^8(B\mathrm{String},\mathbb{Z})=\mathbb{Z}$, so that there is a distinguished map (unique up to homotopy)
\(
\tfrac{1}{6}p_2\colon B\mathrm{String} \to K(\mathbb{Z},8)
\)
representing the generator of the eighth singular cohomology group. Hence, for $X$ a topological space, we have a second fractional Pontrjagin class
\(
[X,B\mathrm{String}] \xrightarrow{\frac{1}{6}p_2} [X,K(\mathbb{Z},8)]\cong H^8(X,\mathbb{Z})\;.
\)
Moreover, the obstruction theoretic argument presented above tells us that String bundles are precisely those Spin bundles for which we can suitably define the $\hat{g}_{i_0,\dots,i_n}$ lifts of  the transition functions up to $n=7$, showing that $\frac{1}{6}p_2$ refines to a morphism of stacks
\(
\tfrac{1}{6}\mathbf{p}_2\colon\mathbf{B}\mathrm{String} \to \mathbf{B}^7{\mathrm{U}(1)}\;.
\)
Furthermore, by looking at the construction of the differential refinement $\frac{1}{2}\hat{\mathbf{p}}_1$ presented in Section \ref{Sec CS elements}, 
one immediately sees that what is crucial are the extensions $\hat{g}_{i_0,\dots,i_n}$, since the 
extensions $\hat{\omega}_{i_0,\dots,i_n}$ are defined in terms of those. This means that also 
$\frac{1}{6}\mathbf{p}_2$ has an analogous differential refinement \cite{FSS}
\(
\tfrac{1}{6}\hat{\mathbf{p}}_2\colon\mathbf{B}\mathrm{String}_{\mathrm{conn}} \to \mathbf{B}^7{\mathrm{U}(1)}_{\mathrm{conn}} 
\)
from the stack of principal String bundles with connection to the stack of ${\mathrm{U}(1)}$-7-bundles with connections.

\vspace{3mm}
Explicit models for both $\frac{1}{6}\mathbf{p}_2$ and $\frac{1}{6}\hat{\mathbf{p}}_2$ 
can be obtained via Lie integration and 7-coskeletization from the canonical 
7-cocycle $\mu_7$ and Chern-Simons element $\mathrm{cs}_7$ on $\mathfrak{so}$, 
as in \cite{FSS}. The ``from the top'' Lie integration approach has the 
remarkable advantage of producing canonical stack morphism out of $n$-cocycles and Chern-Simons
elements for a Lie algebra $\mathfrak{g}$, directly at the level of the $(n-1)$-connected cover of 
a Lie group $G$ of the Lie algebra $\mathfrak{g}$. However, here we prefered to present 
$\frac{1}{6}\mathbf{p}_2$ and $\frac{1}{6}\hat{\mathbf{p}}_2$ in terms of a more classical 
``from the bottom'' construction which is probably more familiar to a wider range of readers.
However, in sections \ref{InfinCS7CS} and \ref{7dCSInSugraOnAdS7} the 
Lie integration of $L_\infty$-cocycles will allow us to get explicit local formulae
for the Chern-Simons functionals induced by these differentially refined characteristic
maps on $\mathrm{String}$ 2-connection fields.

%%%%%%%%%%%%%%%%%%%%%%%%%%%%%%%%%%%%%%%%%%%%%%%%%%%%%%%%
\section{The 7-dimensional nonabelian gerbe theory}
\label{7dTheory}
\label{7dChernSimonsFunctionals}
%%%%%%%%%%%%%%%%%%%%%%%%%%%%%%%%%%%%%%%%%%%%%%%%%%%%%%%%

We indicate in this section a precise definition 
and some properties of a certain 
nonabelian 7-dimensional Chern-Simons theory
whose configuration space is the smooth moduli 2-stack
of boundary $C$-field configurations, identified with that
of twisted $\mathrm{String}$-2-connections, described in section
\ref{Section stacks} and section \ref{SmoothStringC2}.
We show that it has the properties that the arguments
in section \ref{Evidences} suggest the Chern-Simons-dual of 
the 5-brane worldvolume theory should have.
We present this in stages:

\begin{itemize}
  \item in section \ref{NaturalCSFunctional} we present the general 
   construction principle of higher Chern-Simons functionals on higher
   gauge fields;
  \item in section \ref{CupProductTheoryOfTwo3DCSTheories} we consider
    theories induced from cup product classes, such as the abelian
	Chern-Simons theory in 7d as well as the nonabelian theory induced from 
	$(\frac{1}{2}p_1)^2$;
  \item in section \ref{CFieldConfigurations} we very briefly review some aspects of the
    details of $C$-field configurations from \cite{FiSaSc};
  \item 
    in section \ref{CSOnStacksWithCharges} we put these ingredients together
	to form an action functional of ``7d abelian CS with $\frac{1}{4}p_1$-background charge'';
    and keeping also the ``$p_1$-background charge'' dynamical this is refined to a 
	nonabelian higher Chern-Simons theory on \emph{twisted String-2-connection fields};
  \item 
   in section \ref{InfinCS7CS} we consider the ``indecomposable'' 7d theory induced
   from $p_2$ on String-2-connection fields;
  \item 
   finally, in section \ref{7dCSInSugraOnAdS7} we put all the pieces together and
   discuss the full nonabelian Chern-Simons term of 7d supergravity on
   $\mathrm{String}$-2-connection fields.   	
\end{itemize}

%%%%%%%%%%%%%%%%%%%%%%%%%%%%%%%%%%%%%%%%%%
\subsection{Higher Chern-Simons functionals and their level quantization}
\label{NaturalCSFunctional}
%%%%%%%%%%%%%%%%%%%%%%%%%%%%%%%%%%%%%%%%%%

The general mechanism behind all these \emph{natural Chern-Simons functionals}
is the following (see in \cite{survey} sections 2.3.21 and 4.6). 
Let $G$ be any higher smooth group (such as for instance 
an ordinary Lie group or the $\mathrm{String}$-2-group) and write 
$\mathbf{B}G_{\mathrm{conn}}$ for the higher moduli stack of $G$-connections.
Then assume any differential characteristic map is given
\(
  \hat{\mathbf{c}}
   :
  \mathbf{B}G_{\mathrm{conn}}\to \mathbf{B}^n {\mathrm{U}(1)}_{\mathrm{conn}}
  \,.
\)
(The examples that we will shortly turn to are those from
section \ref{Section stacks} and section \ref{SecondPontrjaginOnStacks}.) 
Here we may also think of the stack on the left as the 
the higher moduli stack stack of higher nonabelian gauge fields for 
the higher gauge group $G$,
in that for $\Sigma$ any ($n$-dimensional) smooth manifold, a higher form gauge field data
on $\Sigma$ is characterized by a morphism of higher stacks
$\phi : \Sigma \to \mathbf{B}G_{\mathrm{conn}}$. Simply by composing this
representing map with the above map $\hat {\mathbf{c}}$ we send it to 
a map $\hat{\mathbf{c}} (\phi)  : \Sigma \to \mathbf{B}^n {\mathrm{U}(1)}_{\mathrm{conn}}$.
By the discussion in section \ref{HigherU1Connections},
this now represents an $n$-form gauge field on $\Sigma$. 
Since $\Sigma$ itself is $n$-dimensional,
we may identify this with a differential $n$-fom on $\Sigma$. An assignment of a top-degree form
to a field configuration we may think of as a \emph{Lagrangian} $L$
for a field theory
\(
  L_{\mathbf{c}}
  :=
  \mathbf{H}(\Sigma, \hat {\mathbf{c}})
   : 
   \mathrm{Fields}(\Sigma)
   \simeq
   \mathbf{H}(\Sigma, \mathbf{B}G_{\mathrm{conn}})
   \to
   \mathbf{H}(\Sigma, \mathbf{B}^n {\mathrm{U}(1)}_{\mathrm{conn}})
   \,.
   \label{CanonicalLagrangian}
\)
Here $\mathrm{Fields}(\Sigma)$ is the higher groupoid of field configurations:
its objects are nonabelian higher form fields, its morphisms are 
gauge transformations between these, and its 2-morphisms are gauge-of-gauge transformations,
and so on.
This is the higher groupoid that the BRST-complex / Lie $n$-algebroid of the gauge theory
is the infinitesimal approximation to. 

\vspace{3mm}
In order to make this into an action functional, it remains only to integrate the
Lagrangian over $\Sigma$. By the discussion in section \ref{holonomy}
this may be understood as forming the $n$-volume holonomy 
of the $n$-form gauge fields in $\mathbf{B}^n {\mathrm{U}(1)}_{\mathrm{conn}}$.
Thus the exponentiated action functional on the integrated BRST-complex 
(Lie $n$-algebroid) $\mathrm{Fields}(\Sigma)$ 
of field configurations and gauge transformation on $\Sigma$ 
induced from 
$\hat{\mathbf{c}}$ is
\hspace{-.5cm}
\(
\xymatrix{
   \exp(i S_{\mathbf{c}}(-))
   :=
   \exp(2 \pi i \int_\Sigma L_{\mathbf{c}}(-))
 ~:~ 
  \mathrm{Fields}(\Sigma)
   \simeq
   \mathbf{H}(\Sigma, \mathbf{B}G_{\mathrm{conn}})
   \ar[r]^{\hspace{3.2cm}L_{\mathbf{c}}}
   &
   \mathbf{H}(\Sigma, \mathbf{B}^n {\mathrm{U}(1)}_{\mathrm{conn}})
   \ar[rr]^<<<<<<<<<<{\exp(2\pi i\int_\Sigma(-))}
   &&
   {\mathrm{U}(1)}
   }
   \,,
\)
where the second morphism is the higher holonomy morphism from (\ref{Integration}).

\medskip
The action functionals obtained this way are guaranteed to 
satisfy the requirements on a Chern-Simons functionals associated with 
a class $c$:
\begin{enumerate}
 \item The Lagrangian $L_{\mathbf{c}}$ is locally given by a higher Chern-Simons form 
   for the de Rham image of the integral class $c$;
 \item The action functional $\exp(i S_{\mathbf{c}}(-))$ is (higher) gauge invariant.
\end{enumerate}
The first property follows from the nature of differential characteristic maps 
(\ref{DiagramForDifferentialRefinement}). It is the statement of 
traditional \emph{Chern-Weil theory} refined to higher gauge fields. 
The
higher Chern-Simons forms for higher gauge fields that appear here have been introduced 
in \cite{SSSI}. Their appearance in the higher Lagrangians as above has been
established in \cite{FSS}. For more on the general context of higher
Chern-Weil theory see in \cite{survey} sections 2.3.18 and 3.3.14.
However, just as ordinary degree-3 Chern-Simons forms, also these 
higher Chern-Simons forms by themselves are not gauge invariant.

\vspace{3mm}
The gauge invariance of the integrated and exponentiated Lagrangian,
hence of the action fucntional, follows in the above general construction
by the very nature of what it means 
to give a morphism form a higher groupoid/stack $\mathbf{H}(\Sigma, \mathbf{B}G_{\mathrm{conn}})$
to a set such as ${\mathrm{U}(1)}$. This is what the intrinsic formulation in terms of
higher stacks here accomplishes for us. It is in fact impossible to make a non-gauge-invariant 
construction in higher stack theory as long as one sticks to universal constructions
as in the above (as opposed to direct component presentations by local differential form data). 

\vspace{3mm}
Notice here that the subtle quantization condition on the \emph{level}
that is the familiar condition on the ordinary 3-dimensional Chern-Simons action
to be gauge invariant (see \cite{FreedCS} for a review) is all encoded in 
the initial choice of the characteristic map $c$, which is a \emph{discrete} choice.
In generalization of this, 
the gauge invariance and the \emph{level quantization} 
of the above $\exp(i S_{\mathbf{c}}(-))$ is all in the fact that $\mathbf{c}$
is indeed a morphism of higher stacks to $\mathbf{B}^n {\mathrm{U}(1)}$, and as such 
indeed a smooth refinement of an integral cohomology class.

\paragraph{Example.}
As the archetypical example for this phenomenon, consider the the case $G := \mathrm{Spin}$ and 
$\hat {\mathbf{c}} := \frac{1}{2} \hat {\mathbf{p}}_1$, from (\ref{bold p1}).
Feeding this into the above machine spits out the action
functional of ordinary 3d $\mathrm{Spin}$-Chern-Simons theory 
(this has essentially been observed in \cite{CJMSW}) at 
\emph{level 1} (or at level -1, depending on an inessential convention):
\(
  exp(i S_{\frac{1}{2}p_1}(-)) : \omega \mapsto 
  \exp(2 \pi i \int_\Sigma \mathrm{CS}_3(\omega))
  :=
  \exp(2 \pi i \int_\Sigma \langle\omega \wedge d \omega\rangle + \frac{2}{3}\langle\omega \wedge \omega \wedge \omega\rangle)
  \,.
  \label{3dCSAction}
\)
This is the direct reflection of the fact that $\frac{1}{2}\hat {\mathbf{p}}_1$
is the differential refinement of the \emph{integral} class 
$\frac{1}{2}p_1 \in H^4(B \mathrm{Spin}, \mathbb{Z})$, and that this, in turn, is the
\emph{generator} of $H^4(B \mathrm{Spin}, \mathbb{Z}) \simeq \mathbb{Z}$.
This is the reason for caring about the \emph{fractional} Pontrjagin
class here: precisely for every integer $k \in \mathbb{Z}$ do we get
another class $\frac{k}{2}p_1 \in H^4(B \mathrm{Spin}, \mathbb{Z})$
and its differential refinement $\frac{k}{2}\hat{\mathbf{p}}_1$.
Feeding that into the above machine produces the 
ordinary action functional of 3d $\mathrm{Spin}$-Chern-Simons theory
at level $k$.
\(
  exp(i S_{\frac{k}{2}p_1}(-)) : \omega \mapsto 
  \exp(k\, 2  \pi i  \int_\Sigma \langle\omega \wedge d \omega\rangle + \frac{2}{3}\langle\omega \wedge \omega \wedge \omega\rangle)
  \,.
  \label{3dCSActionkfold}
\)

While this issue of gauge invariance and quantized levels is classical and well understood by 
explicit computation for ordinary Chern-Simons theory, the analogous problem 
in constructing higher dimensional Chern-Simons theories on higher gauge fields
quickly becomes intractable in terms of local differential form data and
its higher gauge transformations. But the above general construction serves 
as an algorithm that tells us how to obtain gauge invariant higher Chern-Simons
functionals and how exactly their levels have to be quantized.

\vspace{3mm}
We go through some examples in the following sections. For instance, for the
indecomposable 7d theory induced by $p_2$ 
on String 2-connection fields (see section \ref{InfinCS7CS}) the
quantization condition on the level is controled by the fact that
$H^8(B \mathrm{String}, \mathbb{Z}) \simeq \mathbb{Z}$ and that 
the generator of this integral group is \cite{SSSII} 
the second fractional Pontrjagin
class $\frac{1}{6}p_2$. Therefore the canonically induced 7d action
$\exp(i S_{\frac{1}{6}p_2}(-))$ is precisely the \emph{level-1} theory
in 7d on nonabelian 2-connection fields, and we get the theory at another 
level precisely for any choice $k \in \mathbb{Z}$ by forming
$\exp(i S_{\frac{k}{6}p_2}(-))$.

\vspace{3mm}
In the following we discuss some 7-dimensional action
functionals of the above form. 
%A longer list of such action 
%functionals with other examples relevant
%to string theory, such as those of \cite{FRSI}, can be 
%found in section 4.6 of \cite{survey}.

%%%%%%%%%%%%%%%%%%%%%%%%%%%%%%%%%%%%%%%%%%%%%%%%%%%%%%%%%%%%
\subsection{The cup product of a 3d CS theory with itself}
  \label{CupProductTheoryOfTwo3DCSTheories}
  \index{$\infty$-Chern-Simons functionals!7d CS theory!cup product of two 3d CS theories}
%%%%%%%%%%%%%%%%%%%%%%%%%%%%%%%%%%%%%%%%%%%%%%%%%%%$$$$$$$$

The action functional of the 7d theory involves decomposable as well as nondecomposable 
terms. In this section we consider the former, which is 
essentially the product of two copies of 3d Chern-Simons theory. 
The latter case is discussed in section \ref{InfinCS7CS}
below.

\medskip
Let $G$ be a compact and simply connected simple Lie group and let $c$ the characteristic class given by the canonical generator of $H^4(G;\mathbb{Z})$. Then we have the cup product of $c$ with itself defining a degree 8 integral cohomology class $c\cup c$. In terms of characteristic maps, this corresponds to the composition
 \(
    c \cup c
	:
	\xymatrix{
	  B G 
	  \ar[r]^<<<<{(c,c)}
	  &
	  K(\mathbb{Z},4)
      \times
	  K(\mathbb{Z},4)
	  \ar[r]^<<<<\cup
	  &
	  K(\mathbb{Z}, 8)~~\;.
	  }
  \)
  Such a structure is utilized in \cite{tw} to define particular such twists to Fivebrane structures
 in relation to the M5-brane.  
Since the characteristic map $c$ is induced by the canonical Lie algebra 3-cocycle on $G$, 
it has, by \cite{FSS}, a differential refinement to a morphism of stacks
\(
  \hat {\mathbf{c}}
   :
  \mathbf{B}G_{\mathrm{conn}}
  \to 
  \mathbf{B}^3 {\mathrm{U}(1)}_{\mathrm{conn}}
  \,.
\)
By itself, this induces ordinary 3d Chern-Simons theory, as discussed around
(\ref{3dCSAction}). But using a differential refinement of the 
cup product, its cup square induces a 7d theory.
Indeed, under the Dold-Kan correspondence (see section \ref{HigherSmoothStacks}), the 
\emph{Beilinson-Deligne cup product},
which refines the cup product on differential cohomology classes to Deligne complexes,
naturally gives a corresponding morphism of moduli stacks
\(
\hat{\cup}: \mathbf{B}^k {\mathrm{U}(1)}_{\mathrm{conn}}\times  \mathbf{B}^l {\mathrm{U}(1)}_{\mathrm{conn}}\to  \mathbf{B}^{k+l+1} {\mathrm{U}(1)}_{\mathrm{conn}}
  \,.
\)
This cup product is such that for $\hat C : \Sigma \to \mathbf{B}^k {\mathrm{U}(1)}_{\mathrm{conn}}$
a $k$-connection with local connection $k$-forms $\{C_i\}$ and globally
defined curvature $(k+1)$-forms $G$, and for 
$\hat B : \Sigma \to \mathbf{B}^l {\mathrm{U}(1)}_{\mathrm{conn}}$
an $l$-connection with with local connection $l$-forms $\{B_i\}$ and globally
defined curvature $(l+1)$-form $H$, the cup product
\(
  \hat C ~\hat \cup~ \hat B
  :
  \xymatrix{
    \Sigma 
	 \ar[rr]^{\hspace{-2cm}(\hat C, \hat B)}
	 &&
	 \mathbf{B}^k {\mathrm{U}(1)}_{\mathrm{conn}}\times  \mathbf{B}^l {\mathrm{U}(1)}_{\mathrm{conn}}
	 \ar[r]^{\hspace{7mm}\hat \cup}
	 &
	 \mathbf{B}^{k+l+1} {\mathrm{U}(1)}_{\mathrm{conn}}
  }
\)
is a $k+l+1$-connection whose local connection $(k+l+1)$-forms can be taken to be
$\{C_i \wedge H\}$ or $\{G \wedge B_i\}$, and hence whose curvature $(k+l+2)$-form is
$G \wedge H$. Moreover, the underlying integral $(k+l+2)$-class of 
$\hat G ~\hat \cup ~\hat H$ is the ordinary cup product of the integral classes 
underlying $\hat G$ and $\hat H$.
Therefore, this induces a differential refinement of the 
ordinary integral cup square $c\cup c$ to a morphism of stacks
  \(
    \hat {\mathbf{c}}~ \hat {\mathbf{\cup}}~ \hat {\mathbf{c}}
	: 
	\xymatrix{
	  \mathbf{B}G_{\mathrm{conn}}
	  \ar[r]^<<<<{(\hat {\mathbf{c}},\hat {\mathbf{c}})}
	  &
	  \mathbf{B}^3 {\mathrm{U}(1)}_{\mathrm{conn}}
	  \times
	  \mathbf{B}^3 {\mathrm{U}(1)}_{\mathrm{conn}}
	  \ar[r]^<<<<{\hat {\mathbf{\cup}}}
	  &
	  \mathbf{B}^7 {\mathrm{U}(1)}_{\mathrm{conn}}
	}
	\,.
  \)
So if $\Sigma$ is a compact oriented smooth manifold of dimension 7, we have a \emph{cup product Chern-Simons theory} induced by
  $\mathbf{c}$: its Chern-Simons functional is
  \( 
    \exp(i S_{c\cup c})
	:
    \xymatrix{	
      \mathbf{H}(\Sigma,\mathbf{B}G_{\mathrm{conn}})
	  \ar[r]^<<<<{\hat {\mathbf{c}} ~\hat {\mathbf{\cup}}~ \hat {\mathbf{c}}}
	  &
	  \mathbf{H}(\Sigma,\mathbf{B}^7 {\mathrm{U}(1)}_{\mathrm{conn}})
	  \ar[r]^<<<<{\int_\Sigma}
	  &
	  {\mathrm{U}(1)}
	}.
  \)
For ordinary Chern-Simons theory,
  the assumption that $G$ is simply connected implies that 
  $B G$ is 3-connected, hence that every $G$-principal bundle
  on a 3-dimensional $\Sigma$ is trivializable, so that 
  $G$-principal connections on $\Sigma$ can be identified with
  $\mathfrak{g}$-valued differential forms on $\Sigma$. This
  is no longer in general the case over a 7-dimensional
  $\Sigma$. Therefore, no simple explicit expression of  the action 
  $\exp(i S_{c\cup c}(\nabla))$ can be given in general 
  (one can always describe in in terms of nonabelian Cech cocycles). 
  However, if the underlying $G$-bundle of a field configuration 
  happens to be trivial, then we do have such a simple expression. Namely, once a trivialization is chosen, the $G$-connection $\nabla$ is given by a globally defined $\mathfrak{g}$-valued 1-form $A$ on $\Sigma$ and the explicit expression of the Beilinson-Deligne cup product
  mentioned above implies  that the cup product Chern-Simons action is
\(
  \exp(i S_{\mathrm{c\cup c}}(\nabla))
   = 
    \exp\left(2\pi i  \int_\Sigma
    \mathrm{CS}_3(A) \wedge d\mathrm{CS}_3(A) \right)
	=
  \exp\left(2\pi i  \int_\Sigma
    \mathrm{CS}_3(A) \wedge \langle F_A \wedge F_A\rangle\right)
	\,,
\)
  where $\mathrm{CS}(A)$ is the usual Chern-Simons 3-form, and $\langle F_A \wedge F_A\rangle$ is the canonical de Rham representative for the cohomology class $c$. 

\vspace{3mm}
Let, for instance, $c = \frac{1}{2}p_1$ be the first fractional Pontrjagin class on 
$\mathrm{Spin}$-connections. Its smooth differential refinement
\(
  \frac{1}{2}\hat{\mathbf{p}}_1 : \mathbf{B} \mathrm{Spin}_{\mathrm{conn}}
  \to 
  \mathbf{B}^3 {\mathrm{U}(1)}_{\mathrm{conn}}
\)
was constructed in \cite{FSS}. The corresponding cup product action
$\exp(i S_{(\frac{1}{2}p_1^2)}(-))$ is the refinement to moduli stacks of the 
Chern-Simons term induced by the $\lambda^2$-summand (\ref{I8})
in the quantum corrected supergravity action.

\vspace{3mm}
In direct analogy we obtain standard abelian 7d-Chern-Simons theory
refined to its moduli 3-stack of field configuratiojns $\mathbf{B}^3 {\mathrm{U}(1)}_{\mathrm{conn}}$.
The \emph{identity morphism}
\(
  \hat {\mathbf{DD}}_2 : \mathbf{B}^3 {\mathrm{U}(1)}_{\mathrm{conn}} \to \mathbf{B}^3 {\mathrm{U}(1)}_{\mathrm{conn}}
  \label{DifferentialDD2}
\)
we may think of as the differential refinement of the 
smooth refinement (\ref{SmoothDD2}) of the higher 
Dixmier-Douady class on circle 3-bundles / bundle 2-gerbes. The corresponding cup product
7d Chern-Simons action functional is the composite
\(
    \exp(i S_{\mathrm{DD}_2\cup \mathrm{DD}_2}(-))
	:
    \xymatrix{	
      \mathbf{H}(\Sigma,\mathbf{B}^3{\mathrm{U}(1)}_{\mathrm{conn}})
	  \ar[rrr]^<<<<<<<<<<<<{\hat {\mathbf{DD}}_2 ~\hat {\mathbf{\cup}}~ \hat {\mathbf{DD}}_2 }
	  &&&
	  \mathbf{H}(\Sigma,\mathbf{B}^7 {\mathrm{U}(1)}_{\mathrm{conn}})
	  \ar[r]^<<<<{\int_\Sigma}
	  &
	  {\mathrm{U}(1)}
	}.
\)
Again, this has in general a complicated expression in terms of local data. 
But when restricted to fields $C_3$ 
in the inclusion $\Omega^3(-) \to \mathbf{B}^3{\mathrm{U}(1)}_{\mathrm{conn}}$, the
action has the simple expression
\(
  \exp(i S_{\mathrm{DD}_2\cup \mathrm{DD}_2}) (C)
  =
  \exp(2 \pi i \int_\Sigma C_3 \wedge d C_3)
  \,.
\)

%%%%%%%%%%%%%%%%%%%%%%%%%%%%%%%%%%%%%%%%
\subsection{The moduli stack of supergravity $C$-field configurations}
\label{CFieldConfigurations}
%%%%%%%%%%%%%%%%%%%%%%%%%%%%%%%%%%%%%%%%%

The 7d Chern-Simons functionals that we consider will be defined
on the fields of 11-dimensional supergravity, their reduction to 
7-dimensions and their restriction to 5-brane boundaries. 
The collection of these fields consists locally of a 3-form
(the $C$-field), an $\mathfrak{so}$-valued 1-form (the field of gravity),
and on the boundary also of an $\mathfrak{e}_8$-valued 1-form, the gauge field,
and a $B$-field that witnesses the gauge identifications on the boundary.
Globally, however, these fields are interrelated and 
arrange to certain nonabelian twisted differential cocycles. 

\vspace{3mm}
In \cite{FiSaSc} we give a detailed discussion of a smooth moduli 
3-stack $\mathbf{CField}$ of $C$-field configurations, as well
as of a morphism $\mathbf{CField}^{\mathrm{bdr}} \to \mathbf{CField}$
that exhibits the relative cohomology that classifies boundary $C$-field
configurations. For convenient reference in the following discussion,
we briefly list some basic statements from \cite{FiSaSc} here.

\medskip

There is a close analogy with the discussion of $B$-field
configurations on D-branes in section \ref{Section D-branes}.
Recall that there the moduli of abelian bulk and nonabelian
boundary field configurations were those of differential cohomology
\emph{relative} to the brane inclusion $Q \hookrightarrow X$ and
relative to the characteristic map 
$\mathbf{dd} : \mathbf{B}\mathrm{PU}(\mathcal{H}) \to \mathbf{B}^2 U(1)$.
The analog of this characteristic map for the $C$-field is the 
canonical (second Chern) map
\(
  \mathbf{a} : 
  \mathbf{B} E_8
    \to 
 \mathbf{B}^3 U(1)
\)
from the moduli stack of $E_8$-bundles to that of circle 3-bundles / bundle 2-gerbes,
which is constructed as a morphism of smooth 3-stacks as in \cite{FSS}.
Under geometric realization (\ref{GeometricRealization}) this becomes 
morphism $a : B E_8 \to K(\mathbb{Z},4)$ 
of topological spaces representing a generating degree-4 integral 
cohomology class in $H^4(B E_8) \simeq \mathbb{Z}$. 
By the higher connectedness of $E_8$ \cite{BottSamelson}
this $a$ is an equivalence on 15-coskeleta, which means that, while
nonabelian $E_8$-gauge fields have a very different differential geoemtry than
abelian 3-form connections, the instanton sectors on both sides may be identified. 
All this is analogous to $\mathbf{dd}$ and the situation for the $B$-field on 
D-branes around (\ref{ddOnPUH}).

So let $Q \to X$ be a 5-brane worldvolume embedded into 11-dimensional spacetime $X$. 
A corresponding cocycle in $\mathbf{a}$-twisted relative differential cohomology
is a homotopy commuting diagram of higher stacks
\(
  \raisebox{20pt}{
  \xymatrix{
     Q \ar[r]^<<<<<{\hat B} \ar[d]^>{\ }="t" 
	  & 
	  \mathbf{B} (E_8)_{\mathrm{conn}} 
	  \ar[d]^{\hat {\mathbf{a}}}_<{\ }="s"
	 \\
	 X \ar[r]_<<<<<<{\hat C} & \mathbf{B}^3 U(1)_{\mathrm{conn}}
	 \ar@{=>}^\simeq "s"; "t"
  }
  }
  \label{CFieldRelativeSimplified}
\)
analogous to (\ref{BFieldRelative}).
For fixed bulk field $\hat C$, this is equivalently an element in the homotopy pullback
\(
  \raisebox{20pt}{
  \xymatrix{
    \mathbf{a}\mathrm{Struc}_{\hat C|_{Q}}(Q)
	\ar[d]
	\ar[r]
	&
	{*}
	\ar[d]^{\hat C|_Q}
    \\
    \mathbf{H}(Q, \mathbf{B}(E_8)_{\mathrm{conn}})
	\ar[r]^{\mathbf{H}(Q,\hat {\mathbf{a}})}
	&
	\mathbf{H}(Q, \mathbf{B}^3 U(1)_{\mathrm{conn}})
  }
  }
\)
of $\hat C|_Q$-twisted differential $\mathrm{String}(E_8)$-structures \cite{SSSIII},
which are twisted $\mathrm{String}(E_8)$-2-connections on $Q$. Therefore, 
where the restriction of the abelian $B$-field on a D-brane gives rise to 
a nonabelian 1-form gauge field, the restriction of the $C$-field relative 
the $\mathbf{a}$-class gives rise to a nonabelian 2-form gauge field.
This sets the main mechanism for the supergravity $C$-field boundary moduli. But 
there the situation is a little bit more complex, due to the additional presence 
of the dynamical $\mathrm{Spin}$-connection.

\medskip

We consider throughout the case of $\mathrm{Spin}$-structures
for which the class $\frac{1}{2}p_1$ is
further divisible by 2. As discussed in \cite{SSSIII}, this 
may naturally be understood as given by 
\emph{$\mathrm{String}^{2\mathbf{DD}_2}$-structures}, where, for 
$\mathbf{c}$ some universal 4-class, $\mathrm{String}^{\mathbf{c}}$
is a higher analog of $\mathrm{Spin}^c$ \cite{Sati10Twist}, and where $\mathbf{DD}_2$
is the universal Dixmier-Douady class for 2-gerbes / circle 3-bundles.
Over the spacetimes of dimension $\leq 14$ that we care about here, this are equivalently 
$\mathrm{String}^{2\mathbf{a}}$-structures, where $\mathbf{a}$ is the universal
4-class of $E_8$-principal bundles.

\vspace{3mm}
The moduli 3-stack of $C$-field configurations for $\frac{1}{2}p_1$
divisible by 2 is then the homotopy pullback
\(
  \label{CFieldByPullback}
  \raisebox{20pt}{
  \xymatrix{
    \mathbf{CField} \ar[rr] 
	\ar[d]
	   && 
	\mathbf{B}^3 {\mathrm{U}(1)}_{\mathrm{conn}}
	\ar[d]^{\cdot 2}
	\\
	\mathbf{B}\mathrm{Spin}_{\mathrm{conn}}
	\times \mathbf{B}E_8
	\ar[rr]^{\tfrac{1}{2}\mathbf{p}_1 + 2 \mathbf{a}}
	&&
	\mathbf{B}^3 {\mathrm{U}(1)}~~\;,
  }
  }
\)
where $\frac{1}{2} \mathbf{p}_1$ is the smooth refinement of the 
first fractional Pontrjagin class from \cite{FSS}. 
This is a further refinement of the situation discussed in the introduction
around (\ref{HomotopyPullbackFirstAppearance}).
Along the lines of the discussion there, one finds that a field configuration
$\phi : \Sigma \to \mathbf{CField}$ has an underlying circle 3-connection 
$\hat C$, an underlying $\mathrm{Spin}$-connection $\hat F_\omega$, and
an underlying $E_8$-principal bundle with class $a$, as well as a 
choice of gauge transformation
\(
  H : G
    \stackrel{\simeq}{\longrightarrow} 
  a - \tfrac{1}{4} {p}_1
\)
between the underlying circle 3-bundle of $\hat G$ and the difference between the
Chern-Simons circle 3-bundles of the $\mathrm{Spin}$- and the $E_8$-bundle.

\vspace{3mm}
There are two stages of boundary conditions for this data we consider,
exhibited by a sequence of maps of moduli stacks
\(
  \mathbf{CField}^{\mathrm{bdr}_0}
  \to
  \mathbf{CField}^{\mathrm{bdr}}
  \to
  \mathbf{CField}
  \,.
\)
For boundary field configurations 
$\phi : \Sigma \to \mathbf{CField}^{\mathrm{bdr}}$ 
the integral cohomology class of $\hat G_4$ is required to vanish
and a differential 3-form part may remain (this is the condition for 
restriction to an M5-brane), while for $\mathbf{CField}^{\mathrm{bdr}_0}$ the full differential cohomology class 
of $\hat G_4$ is required 
to vanish (this is the condition for restriction to the orbifold fixed point
of a Ho{\v r}ava-Witten bounday). In both cases the $E_8$-bundle picks up
a connection over the boundary. 

\medskip
In more detail, the boundary moduli $\mathbf{CField}^{\mathrm{bdr}}$
are given by the homotopy pullback of smooth 2-stacks
\(
  \raisebox{20pt}{
  \xymatrix{
    \mathbf{CField}^{\mathrm{bdr}}
	\ar[rr]
	\ar[d]^>{\ }="t"
	&&
	\Omega^{1 \leq \bullet \leq 3}(-)
	\ar[d]^{\mathcal{H}}_<{\ }="s"
	\\
	\mathbf{B}(\mathrm{Spin} \times E_8)_{\mathrm{conn}}
	\ar[rr]_{\tfrac{1}{2}\hat{\mathbf{p}}_1 + 2 \hat{\mathbf{a}}}
	&&
	\mathbf{B}^3 U(1)_{\mathrm{conn}}~~\;,
	\ar@{=>}_{\hat B}^\simeq "s"; "t"
  }
  }
\)
where the right morphism includes the moduli 3-stack for globally defined 3-forms and their
gauge transformation canonically into the moduli 3-stack for 3-form field configurations.
This is the analog for the $C$-field of the situation in (\ref{TrivializationOfBFieldOverDBrane})
for the $B$-field. Over a patch $U_i$ of a brane $Q\hookrightarrow X$ it encodes a 2-form $B_i$
with differential $H_i = d B_i$ such that
\(
  \mathcal{H}_i = H_i + (\mathrm{CS}_3(\omega) - 2 \mathrm{CS}_3(A))
  \,.
\)
This corresponds ot equation (\ref{Combined3FormOnBrane}) in the introduction.

\medskip
The 2-stack $\mathbf{CField}^{\mathrm{bdr}_0}$ is defined analogously, but
with the right morphism being the inclusion ${*} \to \mathbf{B}^3 U(1)_{\mathrm{conn}}$ 
of the entirely trivial 3-form connection. 
Comparison with (\ref{Stringka})
shows that therefore these $C$-field boundary moduli
are equivalent to those of $2\mathbf{a}$-twisted
String 2-connections discuss in section \ref{SmoothStringC2}
\(
  \mathbf{CField}^{\mathrm{bdr}}
  \simeq
  \mathbf{B}\mathrm{String}^{2\mathbf{a}}_{\mathrm{conn}}
  \,.
  \label{CFieldBoundaryIsString2a}
\)
Then for $Q \to X$ a brane in an 11-dimensional spacetime $X$, the 3-stack
of bulk and boundary field configurations is that of homotopy commuting 
squares of 3-stacks
\(
  \raisebox{20pt}{
  \xymatrix{
    Q \ar[r]^<<<<{\hat B} \ar[d] & \mathbf{CField}^{\mathrm{bdr}} \ar[d]
	 \ar[d]
	\\
	X
	\ar[r]^<<<<<<{\hat C} 
	&
	\mathbf{CField}~~\;.
  }
  }
  \label{CFieldRelative}
\)
This is the $C$-field analog of what for the $B$-field is (\ref{BFieldRelative}).
More details are discussed in \cite{FiSaSc}.
 
%%%%%%%%%%%%%%%%%%%%%%%%%%%%%%%%%%%%%%%%%%%%%%%%%%%
\subsection{7d CS theory with charges on the supergravity $C$-field}
 \label{CSOnStacksWithCharges}
%%%%%%%%%%%%%%%%%%%%%%%%%%%%%%%%%%%%%%%%%%%%%%%%%%%

We can directly combine the above two kinds of Chern-Simons theories to their
\emph{direct product theory}, given by the action functional

$$
\hspace{-11cm} \exp(i S_{\frac{1}{2}p_1 \cup \frac{1}{2}p_1})
  \exp(i S_{\mathrm{DD}_2 \cup \mathrm{DD}_2})
  :
  $$
  %\nonumber \\
  \(
  \xymatrix{
    \mathbf{H}(
	  \Sigma,
      \mathbf{B}\mathrm{Spin}_{\mathrm{conn}}
    \times 
    \mathbf{B}^3 {\mathrm{U}(1)}_{\mathrm{conn}}
	)
	\ar[rrr]^<<<<<<<<<<<<<<{(
	  \frac{1}{2}\hat{\mathbf{p}}_1 \hat \cup \frac{1}{2}\hat{\mathbf{p}}_1
	  + 
	  \hat {\mathbf{DD}} \hat \cup \hat {\mathbf{DD}}
	)}
	&&&
	\mathbf{H}(\Sigma, \mathbf{B}^7 {\mathrm{U}(1)}_{\mathrm{conn}})
	\ar[rr]^{\hspace{9mm}\exp(2 \pi i\int_\Sigma(-))}
	&&
	{\mathrm{U}(1)}
  }\;,
\)
defined on pairs consisting of a $\mathrm{Spin}$-connection and 
a 3-form field. 
To incorporate the interrelation between these fields 
in supergravity, we precompose this with the canonical projection map
from the moduli of supergravity $C$-field configurations, as discussed in
section \ref{CFieldConfigurations}.
Precomposing with the paired projections out of the
defining homotopy pullback (\ref{CFieldByPullback}) 
gives the action functional
$$
  \xymatrix{
    \mathbf{H}(\Sigma, \mathbf{CField})
	\ar[r]
	&
    \mathbf{H}(
	  \Sigma,
      \mathbf{B}\mathrm{Spin}_{\mathrm{conn}}
    \times
    \mathbf{B}^3 {\mathrm{U}(1)}_{\mathrm{conn}}
	)
	\ar[rrr]^<<<<<<<<<<<<<<{(
	  \frac{1}{2}\hat{\mathbf{p}}_1 \hat \cup \frac{1}{2}\hat{\mathbf{p}}_1
	  + 
	  \hat {\mathbf{DD}} \hat \cup \hat {\mathbf{DD}}
	)}
	&&&
	\mathbf{H}(\Sigma, \mathbf{B}^7 {\mathrm{U}(1)}_{\mathrm{conn}})
	\ar[r]^{\hspace{1cm}\int_\Sigma}
	&
	{\mathrm{U}(1)}
  }
  \,.
$$
This is now a functional defined on 11-dimensional supergravity fields 
whose local form data is $\{C_i\}$ (the $C$-field) and $\{\omega_i\}$
(the spin connection),
and whose integral classes satisfy the quantization condition 
\(
  G_4 = a - \tfrac{1}{4}p_1
  \;\;
  \in
  \;\;
  H^4(\Sigma, \mathbb{Z})
  \,.
\)
The action functional is locally given by
\(
  (\omega_i, C_i) \mapsto \int (C_i \wedge d C_i - 
    \mathrm{CS}(\omega_i) \wedge \mathrm{tr}(F_{\omega} \wedge F_\omega))
	\,.
	\label{LocalActionForChargeShiftedCS}
\)
This may be further restricted to the $C$-field \emph{boundary} configuration
along the morphism of moduli stacks $\mathbf{CField}^{\mathrm{bdr}} \to \mathbf{CField}$
to an action functional
$$
  \xymatrix{
    \mathbf{H}(\Sigma, \mathbf{CField}^{\mathrm{bdr}})
	\ar[r]
	&
    \mathbf{H}(
	  \Sigma,
      \mathbf{B}\mathrm{Spin}_{\mathrm{conn}}
    \times
    \mathbf{B}^3 {\mathrm{U}(1)}_{\mathrm{conn}}
	)
	\ar[rrr]^<<<<<<<<<<<<<<{(
	  \frac{1}{2}\hat{\mathbf{p}}_1 \hat \cup \frac{1}{2}\hat{\mathbf{p}}_1
	  + 
	  \hat {\mathbf{DD}} \hat \cup \hat {\mathbf{DD}}
	)}
	&&&
	\mathbf{H}(\Sigma, \mathbf{B}^7 {\mathrm{U}(1)}_{\mathrm{conn}})
	\ar[r]^{\hspace{1cm}\int_\Sigma}
	&
	{\mathrm{U}(1)}
  }
  \,.
$$

By  the equivalence (\ref{CFieldBoundaryIsString2a}) the fields in 
$\mathbf{H}(\Sigma,\mathbf{CField}^{\mathrm{bdr}})$
are \emph{twisted $\mathrm{String}$-2-connections}.  As discussed in 
section \ref{SmoothStringC2}, there 
is a higher gauge in which these are locally given by form data
$(\omega_i, A_i, B_i, \mathcal{H}_i)$ subject to some constraints. In terms of this 
local data the local value of the action functional still reads as in the functional
(\ref{LocalActionForChargeShiftedCS}). 
However, the local data is insufficient to accurately judge the nature of the field
content. For one, the relations (\ref{String2aConnection3Form})
%and (\ref{dOfString2aConnection3Form}) 
satisfied by the local form
data means that equivalently the local action is given by expression such as
\(
  (\omega_i, A_i, B_i, \mathcal{H}_i)
  \mapsto
	\int (C_i \wedge d C_i - 
	\mathcal{H}_i \wedge d \mathcal{H}_i
	-
	2 \mathcal{H}_i \wedge \langle F_{A} \wedge F_A\rangle
	-
    \mathrm{CS}(A_i) \wedge \mathrm{tr}(F_{A} \wedge F_A))
	\,.
\)
Moreover, as discussed in 
section \ref{Section stacks}, there are other gauges in which 
equivalently a nonabelian 2-form $B_i$ appears. Finally, in either case
the global value of the action functional involves in general contributions from
gauge transformations between local patches, and gauge-of-gauge
transformations between these, which are not immediately evident from the 
above local formulae. A general account of the complete formulae 
in terms of nonabelian {\v C}ech cocycles with coefficients in
$L_\infty$-algebra valued forms \cite{SSSI} is in \cite{FSS}. 

\vspace{3mm}
Whichever way one uses to derive (or guess, should that indeed be possible) 
these correct explicit formulae for the higher nonabelian gauge field actions,
the above simple constructions in terms of canonical Chern-Simons action functionals
on mapping stacks guarantees that these formulae exist, are indeed gauge invariant,
and are controled by the defining characteristic classes in the way they should.
It is then a straightforward matter of applying the general machinery to obtain
any level of explicit detail as desired.

%%%%%%%%%%%%%%%%%%%%%%%%%%%%%%%%%%%%%%%%%%%%%%%%%%%
\subsection{7d indecomposable CS theory on String 2-connection fields}
 \label{InfinCS7CS}
  \index{$\infty$-Chern-Simons functionals!7d CS theory!on string 2-connection fields}
%%%%%%%%%%%%%%%%%%%%%%%%%%%%%%%%%%%%%%%%%%%%%%%%%%%

We now turn to the 7-dimensional Chern-Simons theory that is induced by the 
second Pontrjagin class $p_2$ on $B \mathrm{Spin}$ and its fractional
refinement $\frac{1}{6}p_2$ on $B \mathrm{String}$.
As discussed in Section \ref{Sec CS elements}
(eq. \eqref{String conn})
we have a canonical differential characteristic map
\(
  \tfrac{1}{6}
  \hat{\mathbf{p}}_2
  : 
  \mathbf{B}\mathrm{String}_{\mathrm{conn}}
  \to 
  \mathbf{B}^7 {\mathrm{U}(1)}_{\mathrm{conn}}
\)
from the moduli 2-stack of $\mathrm{String}$-2-connections  to 
the moduli 7-stack of  ${\mathrm{U}(1)}$-6-gerbes
with connection. 
By the general mechanism 
described at the beginning of section \ref{7dTheory}, this induces a 7-dimensional Chern-Simons theory: for $\Sigma$ a compact 7-dimensional oriented smooth manifold,
  define $\exp(i S_{\frac{1}{6}p_2}(-))$ to be the 
  Chern-Simons action functional 
  \(
    \exp(i S_{\frac{1}{6}p_2}(-))
	:
	\xymatrix{
  	  \mathbf{H}(\Sigma, \mathbf{B} \mathrm{String}_{\mathrm{conn}})
	  \ar[r]^{\frac{1}{6}\hat{\mathbf{p}_2}}
	  &
	   \mathbf{H}(\Sigma, \mathbf{B}^7 {\mathrm{U}(1)}_{\mathrm{conn}})
	  \ar[r]^<<<<{\int_\Sigma}
	  &
	  {\mathrm{U}(1)}
	}
	\,.
  \)
This can be explicitly described as follows. To begin with notice that since the classifying space 
$B{\rm String}$ of principal String bundles is 8-connected, the underlying String bundle to an object in 
$\mathbf{H}(\Sigma, \mathbf{B} \mathrm{String}_{\mathrm{conn}})$ is trivial, for any 7-dimensional $\Sigma$. Therefore the local differential forms data defining a String connection can actually be chosen to be globally defined.
(But this is true only as long as we ignore here for the moment 
the twist that arises when passing to $\mathrm{String}^{2\mathbf{a}}$-connections, 
as in the previous section.)
 
 \medskip
Then, recall from section \ref{Section stacks}
the different incarnations of the local differential form data
for string 2-connections. With this in mind we have:
\begin{proposition}
%  \begin{itemize}
  %\item 
 {\bf (i)} in terms of the strict $\mathfrak{string}$ Lie 2-algebra
  $(\hat{\Omega}\mathfrak{so}\to P_*\mathfrak{so})$, an object in $\mathbf{H}(\Sigma, \mathbf{B} \mathrm{String}_{\mathrm{conn}})$  is the datum of
  a pair of nonabelian differential forms 
  $A \in \Omega^1(\Sigma, P_* \mathfrak{so})$,
  $B \in \Omega^2(\Sigma, \hat \Omega_* \mathfrak{so})$and  $\exp(i S_{\frac{1}{6}p_2}(-))$ takes this to
  \(
    \begin{aligned}
    \exp(i S_{\frac{1}{6}p_2}(A,B))
	& =
	\exp\left(2\pi i\int_{\Sigma} \mathrm{CS}_7(A(1))\right)
  \end{aligned}
	\,,
  \)
  where $A(1) \in \Omega^1(\Sigma, \mathfrak{so})$ is the 1-form obtained by evaluating on the endpoint 1 the path Lie algebra-valued 1-form $A$, and $\mathrm{CS}_7$ is the standard 
  degree-7 Chern-Simons element on $\mathfrak{so}$ from (\ref{Degree7ChernSimonsForm}).
  %\item

\vspace{1mm}
\noindent {\bf  (ii)} in terms of the skeletal $\mathfrak{string}$ Lie 2-algebra $\mathfrak{so}_{\mu_3}$,
  an object in $\mathbf{H}(\Sigma, \mathbf{B} \mathrm{String}_{\mathrm{conn}})$ is the datum of  a pair of differential forms 
  $A \in \Omega^1(\Sigma, \mathfrak{so})$,
  $B \in \Omega^2(\Sigma, \mathbb{R})$, and 
  $\exp(i S_{\frac{1}{6}p_2}(-))$ takes this to
  \(
    \exp(i S_{\frac{1}{6}p_2}(A,B))
	=
	\exp\left(2\pi i\int_{\Sigma} \mathrm{CS}_7(A)\right)
	\,.
  \)
%  \end{itemize}
\end{proposition}
Notice that, while the 2-form $B$ does not appear explicitly in the 
integrands on the right, it does nevertheless affect the kinematics of the theory.
Its presence forces the connection $A$ to be such that the first Pontrjagin term
$\langle F_A \wedge F_A\rangle$ is exact (see \cite{SSSI} and \cite{FSS} for details).

\vspace{3mm}
Note also that the universal differential  map $\frac{1}{6}\hat {\mathbf{p}}_2$
plays a role already on the 10-dimensional boundary of spacetime, 
as the differential twist that induces the Green-Schwarz mechanism
in magnetic heterotic String theory \cite{SSSII,SSSIII}. In
dimension 10 a String 2-connection
field configuration $\phi : X \to \mathbf{B}\mathrm{String}_{\mathrm{conn}}$
is in general far from being given by globally defined differential form data,
and the 2-form $B$ appears more prominently in the relevant formulae, 
see \cite{SSSIII,FSS}.

%%%%%%%%%%%%%%%%%%%%%%%%%%%%%%%%%%%%%%%%%%%%%%%%%%%
\subsection{7d CS theory in 11d supergravity on $\mathrm{String}$-2-connection fields}
 \label{7dCSInSugraOnAdS7}
 \index{$\infty$-Chern-Simons functionals!7d CS theory!in 11d supergravity}
%%%%%%%%%%%%%%%%%%%%%%%%%%%%%%%%%%%%%%%%%%%%%%%%%%%

In this section we can put the ingredients together and construct a 
7-dimensional Chern-Simons theory induced by the quantum corrected Chern-Simons 
term in 11-dimensional supergravity (\ref{The7dLagrangianInMotivation}).
This is a twisted combination of the two 7-dimensional
Chern-Simons action functionals from \ref{CupProductTheoryOfTwo3DCSTheories} and 
\ref{InfinCS7CS} which naturally lives on (a higher conected cover of) the moduli 2-stack
$C \mathrm{Field}(-)^{\mathrm{bdr}}$ of boundary $C$-field
configurations from section \ref{CFieldConfigurations}. 

\medskip

Recall from the introduction that 
our task is to find a 7d Chern-Simons functional that is induced from
a characteristic class of the form (\ref{I8}) 
$I_8 = \frac{1}{48}(p_2 + (\tfrac{1}{2}p_1)^2)$ on fields that satisfy
a quantization condition \eqref{cond} given by $\tfrac{1}{2}p_1 = 2 a$.
For the \emph{universal} characteristic class $p_2 + (\tfrac{1}{2}p_1)^2$
on the universal $\mathrm{Spin}$-bundle over $B \mathrm{Spin}$,
the divisibility is (see \cite{KSpin})
\(
  24 I_8
  =
  \tfrac{1}{2}(p_2 - (\tfrac{1}{2}p_1)^2) \in 
  H^8(B \mathrm{Spin}, \mathbb{Z})
  \label{24I6}
  \,.
\)
If, however, we fix an 8-dimensional oriented manifold $T X : X \to B \mathrm{SO}(8)$
which admits a $\mathrm{Spin}$-structure
\(
  \raisebox{20pt}{
  \xymatrix{
    & B \mathrm{Spin}(8)
	 \ar[d]
    \\
    X \ar[r]^{T X} \ar@{-->}[ur]^{} & B \mathrm{SO}(8)
  }
  }
  \,,
\)
then  index theory shows (equation (3.2) in \cite{WittenFluxQuantization})
that the pullback of the universal class to $X$ has now divisibility by 6
\(
  8 I_8(X)
  =
  \frac{1}{6}(p_2(X) - (\tfrac{1}{2}p_1(X))^2)
  \in 
  H^8(X, \mathbb{Z})
  \,.
\)
In order to eventually formulate this is the form of canonical Chern-Simpns theories
defined on moduli stacks as in section \ref{NaturalCSFunctional},
observe that this means that we have the following diagram of differential cohomology
sets/groups
\(
  \raisebox{40pt}{
  \xymatrix{
    &&&& H(\Sigma, \mathbf{B}^7 U(1)_{\mathrm{conn}})
	\ar[d]^{\cdot 3}
    \\
    \mathrm{Fields}(\Sigma) 
     \ar[r]
	 \ar[dd]
	 \ar@/^1pc/[urrrr]^{H(\Sigma, \tfrac{1}{6}(\hat p_2 - (\tfrac{1}{2}\hat p_1)^2))}
     &
    H(\Sigma, \mathbf{B} \mathrm{Spin}_{\mathrm{conn}})
    \ar[d]
    \ar[rrr]^{H(\Sigma, \tfrac{1}{2}\hat p_2 - (\tfrac{1}{2}\hat p_1)^2)}
    &&&
    H(\Sigma, \mathbf{B}^7 U(1)_{\mathrm{conn}})
    \\
	& H(\Sigma, B \mathrm{Spin}(8))
	\ar[d]
	\\
    {*} \ar[r]^{T \Sigma } & H(\Sigma, \mathbf{B} SO(8))
   }
   }
   \,.
\)
Here $\mathrm{Field}(\Sigma)$ 
denotes the connected comonents of the homotopy pullback of
$\mathbf{H}(\Sigma, \mathbf{B}\mathrm{Spin}_{\mathrm{conn}}) \to 
\mathbf{H}(\Sigma, \mathbf{B} \mathrm{SO})$ along $T \Sigma$,
which  is encodes the gauge equivalence classes of $\mathrm{Spin}$-connections
on all possible $\mathrm{Spin}$-structures on the tangent bundle $T X$. 
The existence of the top
arrow expresses the fact that restricted to such fields, the universal differential
class $\tfrac{1}{2}(\hat p_2 - (\tfrac{1}{2}p_1)^2)$ is further divisible by 3,
where up to here the differential refinements (denoted by the hats)
can still be given by classical Chern-Weil theory.

\medskip
However, this is still not the refined Chern-Simons functional (so far on gauge equivalence classes)
that we need. While this cannot be further divided, also $\mathrm{Fields}(\Sigma)$
here is not the correct field configuration space yet. The correct configuration space
for the boundary values of $C$-field configurations has to satisfy the 
quantization condition (\ref{QuantizationCondition})
saying that $\tfrac{1}{2}p_1 $ is further divisible by 2. 
By (\ref{BStringTopological}) the classifying space for such configurations
is $B \mathrm{String}^{2\mathrm{DD}}$, whose smooth and differential refinement is the 
moduli 2-stack $\mathbf{B} \mathrm{String}^{2\mathbf{DD}}_{\mathrm{conn}}$,
which is such that a map $X \to \mathbf{B} \mathrm{String}^{2\mathbf{DD}}_{\mathrm{conn}}$
is a $\mathrm{Spin}$-connection $\omega$ and a choice of divisibility morphism
of the corresponding Pontrjagin class.
So the quantization condition may be implemented in 
a gauge equivariant way by replacing in the above the moduli stack 
$\mathbf{B}\mathrm{Spin}_{\mathrm{conn}}$ with the moduli 2-stack 
$\mathbf{B}\mathrm{String}^{2\mathbf{DD}}_{\mathrm{conn}}$. 
Then by the discussion on p. 9 of 
\cite{WittenFluxQuantization} we then have
\(
  \raisebox{40pt}{
  \xymatrix{
    &&&&
    H(\Sigma, \mathbf{B}^7 U(1)_{\mathrm{conn}})
	\ar[d]^{\cdot 24}
    \\
    \mathrm{Fields}_{\tfrac{1}{2}\lambda}(\Sigma) 
     \ar[r]
	 \ar[d]
	 \ar@/^1pc/[rrrru]^{H(\Sigma, \tfrac{1}{48}(\hat p_2 - (\tfrac{1}{2}\hat p_1)^2))}
     &
    H(\Sigma, \mathbf{B} \mathrm{String}^{2\mathbf{DD}}_{\mathrm{conn}})
    \ar[d]
    \ar[rrr]^{H(\Sigma, \tfrac{1}{2}\hat p_2 - (\tfrac{1}{2}\hat p_1)^2)}
    &&&
    H(\Sigma, \mathbf{B}^7 U(1)_{\mathrm{conn}})~~\;.
    \\
    {*} \ar[r]^{T \Sigma } & H(\Sigma, \mathbf{B} SO(8))
   }
   }
\)
Here $\mathrm{Fields}_{\tfrac{1}{2}\lambda}$ is the connected components of the
homotopy pullback of 
$\mathbf{H}(\Sigma, \mathbf{B}\mathrm{String}^{2\mathbf{DD}}) \to 
\mathbf{H}(\Sigma, \mathbf{B}\mathrm{SO}(8))$ along $T \Sigma$, which encodes
the gauge equivalence classes of $\mathrm{String}^{2\mathbf{DD}}$-2-connections
on the tangent bundle of $X$. The curved morphism indicates that on this
configuration space now the 7d Chern-Simons action has the full divisibility 
with prefactor $\tfrac{1}{48}$.
In order to add dynamical $E_8$-gauge fields on the boundary 
we invoke the canonical morphism 
$\mathbf{B}\mathrm{String}^{2\mathbf{a}} \to \mathbf{B}\mathrm{String}^{2\mathbf{DD}}$
which implements the condition that $\tfrac{1}{2}\hat p_1$ is not just divisible further
by 2, but that half of it is the class of a given $E_8$-bundle with given $E_8$-connection.

\medskip
In conclusion, the canonical boundary 7d Chern-Simons functional 
induced by the quantum correction term (\ref{The7dLagrangianInMotivation})
and consistent at any given level, hence consistent for any number of 5-branes $N \in \mathbb{N}$
is the composite functional
$$
  \hspace{-1cm}
  \raisebox{20pt}{
  \xymatrix{
    &&& 
	\mathbf{H}(\Sigma, \mathbf{B}^7 {\mathrm{U}(1)}_{\mathrm{conn}})
	\ar[d]^{\cdot 24}
    \\
    \mathbf{H}(\Sigma, \mathbf{CField}^{\mathrm{bdr}})
	\ar[r]^\simeq
	\ar@/^1pc/[rrru]^{L_{I_8}}
	&
	\mathbf{H}(\Sigma, \mathbf{B}\mathrm{String}^{2\mathbf{a}}_{\mathrm{conn}})
	\ar[r]
	&
	\mathbf{H}(\Sigma, \mathbf{B}\mathrm{String}^{2\mathbf{DD}_2}_{\mathrm{conn}})
	\ar[r]^{L_{24 I_8}}
	&
	\mathbf{H}(\Sigma, \mathbf{B}^7 {\mathrm{U}(1)}_{\mathrm{conn}})
	\ar[rr]^<<<<<<<<<<{\exp(2 \pi i\int_\Sigma(-))}
	&&
	{\mathrm{U}(1)}\;.
  }
  }
$$
Notice here that, by the high connectedness of $E_8$, we have over a 
7-dimensional $\Sigma$ that
\(
  \mathbf{H}(\Sigma, \mathbf{B}E_8)
    \stackrel{}{\to}
  \mathbf{H}(\Sigma, \mathbf{B}^3 {\mathrm{U}(1)})
\)
is an isomorphism on gauge equivalence classes, identifying $E_8$-instantons
sectors with higher magnetic charge sectors of 3-bundles / 2-gerbes.
But since as \emph{smooth stacks} $\mathbf{B}E_8$ is different from
$\mathbf{B}^3 {\mathrm{U}(1)}$ (the gauge transformations are very different!)
this is not an equivalence of the 2-groupoids of gauge transformations
and gauge-of-gauge transformations. Moreover, the differential refinement
\(
  \mathbf{H}(\Sigma, (\mathbf{B}E_8)_{\mathrm{conn}})
    \stackrel{}{\to}
  \mathbf{H}(\Sigma, \mathbf{B}^3 {\mathrm{U}(1)}_{\mathrm{conn}})
\)
is not even an isomorphism on gauge equivalence classes: 
$E_8$-gauge fields are much more refined than the 3-form 
Chern-Simons gauge fields induced from them. For these reasons,
also the morphism
\(
  \mathbf{H}(\Sigma, \mathbf{B}\mathrm{String}^{2\mathbf{a}}_{\mathrm{conn}})
  \to
  \mathbf{H}(\Sigma, \mathbf{B}\mathrm{String}^{2\mathbf{DD}}_{\mathrm{conn}})
\)
appearing in the above composition is far from being an equivalence of of gauge
field configuration data, even though the gauge equivalence classes of the
underlying instanton/charge sectors are canonically identified.

\medskip
By the discussion in section \ref{Sec CS elements}
we have the Lagrangian $L_{I_8} := \mathbf{H}(\Sigma, \hat {\mathbf{I}}_8)$
universally defined, by higher Lie integration, on the 
universal 7-connected cover 
moduli 2-stack $\mathbf{B} \widehat{\mathrm{String}^{2\mathbf{a}}}_{\mathrm{conn}}$
of the moduli 2-stack $\mathbf{B}\mathrm{String}^{2\mathbf{a}}_{\mathrm{conn}}$,
and the Lagrangian down on the latter is obtained by consistent quotienting
from that. 
By the discussion in section \ref{NaturalCSFunctional}
the action $\exp(i S_{\mathbf{I}_8})$ 
is guarateed to locally be of the required form (\ref{main action}), while
at the same time having the correct global properties and be gauge invariant
under higher gauge transformations of nonabelian $\mathrm{String}^{2\mathbf{a}}$
2-form connections fields, by level quantization.

\medskip
An explicit presentation of the full Lagrangian on  
$\mathbf{B} \widehat{\mathrm{String}^{2\mathbf{a}}}_{\mathrm{conn}}$ can be
constructed directly with the tools in \cite{FSS}. Instead of 
going here through the full details, which are spelled out there in general, 
we close by pointing out how the full functional $\exp(i S_{\mathbf{I}_8})$
restricts to the special cases which we discussed before.
Namely, by the universal property of the homotopy pullback, and by the 
``pasting law'' for homotopy pullbacks we have a canonical morphism of 
moduli 2-stacks
\(
  \mathbf{B} \mathrm{String} \to \mathbf{B}\mathrm{String}^{2\mathbf{a}}
\)
given as the universal dashed arrow in  the diagram
\(
  \raisebox{30pt}{
  \xymatrix{
    \mathbf{B}\mathrm{String}
	\ar[r]
	\ar@{-->}[d]
	&
	{*}
	\ar[d]
    \\
    \mathbf{B}\mathrm{String}^{2 \mathbf{a}}
	\ar[r]
	\ar[d]
	&
	\mathbf{B} E_8
	\ar[d]^{2 \mathbf{a}}
	\\
	\mathbf{B}\mathrm{Spin} 
	  \ar[r]^{\tfrac{1}{2}\mathbf{p}_1}
	&
	\mathbf{B}^3 {\mathrm{U}(1)}~~\;.
  }
  }
\)
The action functional $\exp(i S_{\mathbf{I}_8})$ may be restricted
along this morphism to the configuration 2-stack of untwisted
$\mathrm{String}$-2-connections. This restriction then coincides with the
``indecomposable'' 7-dimensional 
Chern-Simons functional $\exp(i S_{\tfrac{1}{6}\mathbf{p}_2})$
discussed in section \ref{InfinCS7CS}.

\medskip

\vspace{0.5cm}
\noindent {\large \bf Acknowledgements}

\vspace{2mm}
 The research of H.S. is supported by NSF Grant PHY-1102218.
 U.S. gratefully acknowledges an invitation 
 to the Department of Mathematics of Pittsburgh University
 in September 2011, which led to the work presented here.

\vspace{1cm}
%%%%%%%%%%%%%%%%%%%%%%%%%%%%%%%%%%%%%%%%%%%%%%%%%%%%%%%%%%%%%%%%%
\section*{Appendix: Two models for $\mathfrak{string}$, seven models for $\mathrm{String}$}
\label{TheFields}
\addcontentsline{toc}{section}{Appendix: Two models for $\mathfrak{string}$, six models for $\mathrm{String}$}
%%%%%%%%%%%%%%%%%%%%%%%%%%%%%%%%%%%%%%%%%%%%%%%%%%%%%%%%%%%%%%%%
 
\setcounter{equation}{0} 
\renewcommand{\theequation}{A.\arabic{equation}}

In ordinary gauge theory, any two different but equivalent incarnations of,
say, a Lie algebra are related by an \emph{isomorphism}. In practice this
typically appears simply as a linear transformation between two choices of 
basis of the underlying vector space, and  is typically easily recognized
and known as such.
In contrast, there is a somewhat subtle new phenomenon that appears when 
passing beyond ordinary nonabelian
gauge theory to higher nonablian gauge theory. Due to the higher gauge freedom,
we have in general a plethora of possibly very different looking but 
nevertheless equivalent incarnations of a given
higher gauge group, its Lie $n$-algebra and hence of the local differential
form data of higher gauge fields. 
This accounts for a good bit of the subtlety of higher nonabelian gauge theory,
the effects of which have not always been dealt with appropriately in existing
proposals, in particular concerning the characterization and identification
of what it means to have a ``nonabelian 2-form theory''.

\medskip
Discussing this requires a slightly higher level of mathematical detail
than we wanted to use in the main text here, since these can be found 
discussed extensively in our previous publications. But because 
awareness of this phenomenon helps to put the higher nonabelian gauge theories
discussed here in the proper perspective, we here briefly review some relevant facts
for the case of the $\mathrm{String}$-2-group and its Lie 2-algebra 
$\mathfrak{string}$ (see here section \ref{Section stacks} for more discussion).
Analogous comments would apply to their twisted versions $\mathrm{String}^{2\mathbf{a}}$
and $\mathfrak{string}^{\mathfrak{e}_8}$ from section \ref{SmoothStringC2}.

\medskip

First we consider two different incarnations of the Lie 2-algebra $\mathfrak{string}$,
from one of main theorems in \cite{BCSS}.
Let $\mathfrak{g}$ be a semisimple Lie algebra.
Write $\langle -,-\rangle : \mathfrak{g}^{\otimes 2} \to \mathbb{R}$
for its Killing form and 
\(
  \mu = \langle -, [-,-]\rangle : \mathfrak{g}^{\otimes 3} \to \mathbb{R}
  \label{Canonical3Cocycle}
\)
for the canonical 3-cocycle Lie algebra cocycle.

\begin{definitionapp}[skeletal version of $\mathfrak{string}$]
  Write $\mathfrak{g}_\mu$ for the Lie 2-algebra
  whose underlying graded vector space is
  $$
    \mathfrak{g}_\mu = \mathfrak{g} \oplus \mathbb{R}[-1]
	\,,
  $$
  and whose nonvanishing brackets are defined as follows.
  \begin{itemize}
    \item 
	  The binary bracket is that of $\mathfrak{g}$ when both 
	  arguments are from $\mathfrak{g}$ and 0 otherwise.
	\item
	  The trinary bracket is the 3-cocycle
	  $$
	    [-,-,-]_{\mathfrak{g}_\mu}
		:=
		\langle -, [-,-]\rangle
		: 
		\mathfrak{g}^{\otimes 3} \to \mathbb{R}
		\,.
	  $$
  \end{itemize}
\end{definitionapp}
\begin{definitionapp}[strict version of $\mathfrak{string}$]
  Write $(\hat \Omega \mathfrak{g} \stackrel{h}{\to} P_* \mathfrak{g})$
  for the Fr{\'e}chet Lie 2-algebra whose underlying vector space is
  $$
    (\hat \Omega \mathfrak{g} \to P_* \mathfrak{g})
	= 
	P_* \mathfrak{g} \oplus (\Omega \mathfrak{g} \oplus \mathbb{R})[-1]
	\,,
  $$
  where $P_* \mathfrak{g}$ is the vector space of smooth 
  maps $\gamma : [0,1] \to \mathfrak{g}$ such that $\gamma(0) = 0$,
  and where $\Omega \mathfrak{g}$ is the subspace for which also
  $\gamma(1) = 0$, and whose non-vanishing brackets are defined as follows
  \begin{itemize}
    \item 
	  $[-]_1 = \partial := \Omega \mathfrak{g} \oplus \mathbb{R} \to 
	 \Omega \mathfrak{g} \hookrightarrow P_* \mathfrak{g}$;
	\item
	  $[-,-] : P_* \mathfrak{g} \otimes P_* \mathfrak{g} \to 
	   P_* \mathfrak{g}$
	   is given by the pointwise Lie bracket on $\mathfrak{g}$ as
	   $$
	     [\gamma_1, \gamma_2] = (\sigma \mapsto [\gamma_1(\sigma), \gamma_2(\sigma)])
		 \,;
	   $$
	\item 
	  $[-,-] : P_* \mathfrak{g} \otimes (\Omega \mathfrak{g} \oplus \mathbb{R}) \to 
	   \Omega \mathfrak{g} \oplus \mathbb{R}$
	   is given by pairs
	   \(
	     \label{CocycleInStrictStringLie2Algebra}
	     [\gamma, (\ell, c)]
		 :=
		 \left(
		   [\gamma,\ell],
		   \;
		   2 \int_0^1 \langle \gamma(\sigma), \frac{d \ell}{d \sigma}(\sigma)\rangle
		   d \sigma
		 \right)
		 \,,
	   \)
		 where the first term is again pointwise the Lie bracket in 
		 $\mathfrak{g}$.
  \end{itemize}
\end{definitionapp}
\begin{propositionapp}
   \label{EquivalenceOfTheTwoStringLie2Algebras}
  The linear map
  $$
    P_* \mathfrak{g} \oplus (\Omega \mathfrak{g} \oplus \mathbb{R})[-1]
	\to 
	\mathfrak{g} \oplus \mathbb{R}[-1]
	\,,
  $$
  which in degree 0 is evaluation at the endpoint
  $$
    \gamma \mapsto \gamma(1)
  $$
  and which in degree 1 is projection onto the $\mathbb{R}$-summand,
  induces an equivalence of Lie 2-algebras
  $$
    (\hat \Omega \mathfrak{g} \to P_* \mathfrak{g})
	\simeq
	\mathfrak{g}_\mu\;.
  $$
\end{propositionapp}
This is theorem 30 in \cite{BCSS}.

\medskip
By section 3.4.1 in \cite{survey}, Lie $n$-algebras $\mathfrak{g}$ for all $n$ can 
naturally be understood as \emph{infinitesimal} smooth $n$-stacks 
$b \mathfrak{g} \hookrightarrow \mathbf{B}G$, forming the infinitesimal neighbourhood
of the canonical base point in the moduli $n$-stacks $\mathbf{B}G$ integrating them.
In this context, any Lie algebra $n$-cocycle $\mu$ is a morphism of infinitesimal
$n$-stacks $\mu : b \mathfrak{g} \to b^n \mathbb{R}$, explicitly models for which 
are discussed in 
\cite{SSSI}\cite{FSS}. Therefore we may ask for the \emph{homotopy fiber}
of a Lie algebra cocylce and make the following definition.
\begin{definitionapp}
  For $\mathfrak{g}$ a semisimple Lie algebra as in (\ref{Canonical3Cocycle}),
  the Lie 2-algebra $\mathfrak{string}(\mathfrak{g})$ is, up to weak equivalence, the 
  loop space object of the homotopy fiber $b \mathfrak{string}(\mathfrak{g})$ in 
  $$
    \raisebox{20pt}{
    \xymatrix{
	  b\mathfrak{string}(\mathfrak{g}) \ar[r] \ar[d] & {*} \ar[d]
	  \\
	  b\mathfrak{g}
	  \ar[r]^{\mu}
	  &
	  b^3 \mathbb{R}
	}
	}
	\,.
  $$
For the case $\mathfrak{g} = \mathfrak{so}_N$ we we write for short
  $$
    \mathfrak{string} := \mathfrak{string}(\mathfrak{so})
	\,.
  $$
\end{definitionapp}
Notice the analogy to (\ref{diag SS}). With this we have
\begin{propositionapp}
  With $\mathfrak{g}$ as above,  
  both $\mathfrak{g}_\mu$ as well as $(\hat \Omega \mathfrak{g} \to P_* \mathfrak{g})$
  are equivalent incarnations of $\mathfrak{string}(\mathfrak{g})$.
\end{propositionapp}
\begin{remarkapp}
The two models $\mathfrak{g}_\mu$ and $(\hat \Omega \mathfrak{g} \to P_* \mathfrak{g})$ 
are at two opposite extremes of all possible models: while $\mathfrak{g}_\mu$
is singled out by having trivial unary bracket, $(\hat \Omega \mathfrak{g} \to P_* \mathfrak{g})$
is singled out by having trivial trinary bracket.
Analogous statements apply to the models of the String 2-group, to which we now
turn.
\end{remarkapp}
In direct analogy to how Lie algebras integrated to Lie groups
in classical Lie theory, so higher Lie algebras integrate to 
higher smooth groups in higher Lie theory. 

\begin{propositionapp}
  The degreewise ordinary Lie integration of the differential crossed module
  $(\hat \Omega \mathfrak{so} \to P_* \mathfrak{so})$
  yields the Fr{\'e}chet Lie crossed module
  $(\hat \Omega \mathrm{Spin} \to P_* \mathrm{Spin})$,
  where $\hat \Omega \mathrm{Spin}$ is the level-1 Kac-Moody
  central extension of the smooth loop group of $\mathrm{Spin}$.
  This is naturally a strict Fr{\'e}chet Lie 2-group.
\end{propositionapp}
 The nontrivial part to check is that the action of 
  $P_* \mathfrak{so}$ on $\hat \Omega \mathfrak{so}$
  lifts to a compatible action of $P_* \mathrm{Spin}$
  on $\hat \Omega \mathrm{Spin}$ which lifts the infinitesimal action
  and such as to satisfy the axioms
  of a crossed moduel. This is prop. 24 in \cite{BCSS}.
In the above, the group operation in $P_* G$ and in $\Omega G$ is the 
  \emph{pointwise} multiplication of parameterized paths in $G$,
  which in $\hat \Omega G$ is twisted by the action of a 2-cocycle
  on loops. There are two evident variants of this.
    \begin{itemize}
    \item 
  One may consider forming \emph{thin homotopy} equivalence classes of
  paths in $G$, which form a group not under pointwise multiplication,
  but under composition. The corresponding strict 2-group 
  \(
    (\hat \Omega_{\mathrm{th}}G  \to P_{\mathrm{th}}G)
  \)
  is constructed in def. 4.1.20 of \cite{survey}.
  \item
    One may consider a different cocycle on the loop group, 
	known as \emph{Mickelsson's cocycle}. This yields a strict 2-group
	$\mathrm{String}_{\mathrm{Mick}}$ given in prop. 4.1.26 of \cite{survey}.
  \end{itemize}
  One may also form a universal higher Lie integration of 
  $\mathfrak{g}_\mu$ as in \cite{Henriques}, which in \cite{FSS} was
  put in the context of higher smooth stacks as used here.
  This yields a weak smooth 2-group
  $\Omega \tau_2 \exp(\mathfrak{g}_\mu)$. 
  \begin{theoremapp}
    All these 2-groups equivalent models for the smooth 2-group 
	$\mathrm{String}$ as defined in (\ref{diag SS}), as are their smooth moduli 2-stacks.
  \end{theoremapp}
  This is theorem 4.1.29 in \cite{survey}.

\medskip
  There are further, very different looking models. In \cite{NSW}
  a strict model is given whose degree-0 group is the
  actual topological string 2-group, but equipped with a smooth structure,
  and whose degree-1 group is is contractible group.
  All these models so far are degreewise presented by infinite-dimensional smooth spaces.
  In \cite{Schommer-Pries} an algorithm is given for constructing models
  of $\mathrm{String}$
  by degreewise finite dimensional manifolds. In 
  \cite{WaldorfString} this is explicitly related to the construction of 
  \emph{multiplicative bundle gerbes} on the group.
  This construction has at times
  been motivated as a plausible prerequisite for a tractable discussion of
  the differential geometry of $\mathrm{String}$-geometry. 
  But notice two things
  \begin{enumerate}
    \item Constructions as in \cite{FSS} show that, 
  contrary to that expectation, it is the model $\Omega \tau_2 \exp(\mathfrak{g}_\mu)$
  which is most directly accessible by ordinary differential geometric
  methods and differential form computations as in \cite{SSSI}. 
  Most of the local formulae for String 2-connections and their
  twisted version that we used here are constructed and controled by this model.
  One may understand this from the fact that, while when thought of as 
  a graded (simplicial) space this model is degreewise infinite-dimensional, 
  when thought of as a presentation of a higher stack -- which is what we are
  actually interested in -- then its \emph{probes} by maps 
  $X \to \Omega \tau_2 \exp(\mathfrak{g}_\mu)$ from a smooth manifold $X$
  are characterized by classical and tractable differential form data
  on $X$, which has immediate and useful interpretation in physics. 
   \item
     If one simply drops the requirement that a model
	 has Kan fillers (which is not a necessary requirement) 
	 and allows spaces with arbitrary
	 many connected components, then \emph{every} higher smooth stack 
	 has a model that is degreewise a finite-dimensional smooth manifold. 
	 This is prop. 2.1.49 in \cite{survey}. The construction given there
	 also shows that, while they are guaranteed to exist, these degreewise
	 finite-dimensional models are typically not useful for practical
	 computations with local differential form data. Instead, as shown there, their existence 
	 is most useful for abstract considerations in the homotopy theory
	 of higher stacks, because they serve as \emph{cofibrant models}.
  \end{enumerate}
  On the other hand, of course every concrete model for a higher group has its advantages
  and disadvantages. The more models one has, the more
  aspects one sees of the abstractly defined higher group for which all these models
  are, after all, models. From the point of 
  view of the corresponding higher gauge theory, the equivalences between the
  different models play the role of higher gauge transformations between 
  higher gauges. Already from ordinary gauge theory it is a familiar fact that
  different gauges have their different uses, and the more of them are 
  under control, the better for understanding the intrinsic,
  gauge invariant, nature of the theory.

%%%%%%%%%%%%%%%%%%%%%%%%%%%%%%%%%%%%%%%%%%%%%%%%%%%


\begin{thebibliography}{10}
%%%%%%%%%%%%%%%%%%%%%%%%%%%%%%%%%%%%%%%%%%%%%%%%%%%



 \bibitem[APPS]{APPS}
M. Aganagic, J. Park,
C. Popescu, and J. H. Schwarz, 
{\it World-volume action of the M-theory five-brane},
Nucl. Phys. {\bf B 496} (1997), 191,
 [{\tt arXiv:hep-th/9701166}]. 

\bibitem[ABJM]{ABJM}
O. Aharony, O. Bergman, D. L. Jafferis and J. Maldacena, 
{\it $N=6$ superconformal
Chern-Simons-matter theories, M2-branes and their gravity duals}, 
J. High Energy Phys. {\bf 0810} (2008)
091, [{\tt arXiv:0806.1218}] [{\tt hep-th}].


\bibitem[AGMOO]{AGMOO}
O.~Aharony, S.~Gubser, J.~Maldacena, H.~Ooguri, and Y.~Oz,
\newblock {\it Large $N$ field theories, string theory and gravity},
\newblock Phys. Rept. {\bf 323} (2000), 183--386,
\newblock [{\tt hep-th/9905111}].

%\bibitem[AG]{AG85}
%L. Alvarez-Gaum\'e and P. H. Ginsparg,
%{\it The structure of gauge and gravitational anomalies},
%Annals Phys. {\bf 161} (1985) 423, Erratum-ibid. {\bf 171} (1986) 233.

%\bibitem[AGW]{AW84}
%L. Alvare-Gaum\'e and E. Witten,
%{\it Gravitational anomalies},
%Nucl. Phys. {\bf B234} (1984), 269--330.


\bibitem[Ar]{Ar}
A. Armoni,
{\it Comments on perturbative dynamics of non-commutative Yang-Mills theory},
Nucl. Phys. {\bf B593} (2001), 229--242,
[{\tt arXiv:hep-th/0005208}].
	

 \bibitem[AJ]{AJ}
P. Aschieri and B. Jurco,
{\it Gerbes, M5-brane anomalies and $E_8$ gauge theory},
J. High Energy Phys. {\bf 0410} (2004) 068,
[{\tt arXiv:hep-th/0409200}].


%\bibitem{AM}
%A. Ashtekar and A. Magnon,  
%{\it Asymptotically anti-de Sitter space-times},
% Class. Quant. Grav. {\bf 1} (1984), L39--L44.
% 
 \bibitem[AK]{AK}
 S. D. Avramis and A. Kehagias,
 {\it Gauged $D=7$ supergravity on the $S^1/\Z_2$ orbifold},
 	Phys.Rev. {\bf D71} (2005) 066005,
	[{\tt arXiv:hep-th/0407221}].	

\bibitem[BCSS]{BCSS}
J.~Baez, A.~Crans, U.~Schreiber, and D.~Stevenson,
 {\it From loop groups to 2-groups},
  {Homology, Homotopy Appl.} {\bf 9(2)} (2007), 101--135.


\bibitem[BL]{BL}
J.~Bagger, N.~Lambert, 
\newblock {\it Modeling Multiple M2's}, 
\newblock Phys. Rev. {\bf D75} (2007), 045020, \newline
\newblock [{\tt arXiv:hep-th/0611108}]
  
\bibitem[BM]{BM}
D. Belov and G. W. Moore,
{\it Holographic action for the self-dual field},
[{\tt arXiv:hep-th/0605038}].


\bibitem[BBSS]{BBSS}
E. Bergshoeff, D. S. Berman, J. P. van der Schaar, and P. Sundell,
{\it A noncommutative M-Theory five-brane},
Nucl. Phys. {\bf B590} (2000), 173--197,
[{\tt arXiv:hep-th/0005026}].	
	
\bibitem[BKS]{BKS}
E. Bergshoeff, I. G. Koh and E. Sezgin, 
{\it Yang-Mills/Einstein supergravity in seven-dimensions},
Phys. Rev. {\bf D 32} (1985), 1353--1357.


\bibitem[Be]{B08}
D. Berman, 
{\it M-theory branes and their interactions},
Phys. Rept. {\bf 456} (2008), 89--126,
[{\tt arXiv:0710.170}] [{\tt hep-th}].

\bibitem[BSST]{BSST}
L. Bonora, M. Schnabl, M.M. Sheikh-Jabbari, and A. Tomasiello,
{\it Noncommutative SO($n$) and Sp($n$) gauge theories},
	Nucl. Phys. {\bf B589} (2000), 461--474,
[{\tt arXiv:hep-th/0006091}].	

\bibitem[BoSa]{BottSamelson}
R. Bott and H. Sameslon, 
\newblock {\it Application of the theory of Morse to symmetric spaces}, 
\newblock 
Amer. J. Math. {\bf 80} (1958), 964--1029.
 

\bibitem[BrMc]{brylinski-mclaughlin}
J.-L. Brylinski and D.~A. McLaughlin,
\newblock{\it  {\v C}ech cocycles for characteristic classes},
\newblock {Commun. Math. Phys.} {\bf 178} (1996), 225--236.

\bibitem[CBMMS]{CBMMS}
A.~Carey, P.~Bouwknegt, V.~Mathai, M.~Murray, D. Stevenson, 
\newblock {\it Twisted K-theory and K-theory of bundle gerbes},  
\newblock { Commun. Math. Phys.} {\bf 228} (2002), 17--49,
\newblock [{\tt arXiv:hep-th/0106194}].


\bibitem[CJMSW]{CJMSW}
A.~Carey, S.~Johnson, M.~Murray, D.~Stevenson, B.-L. Wang, 
\newblock {\it Bundle gerbes for Chern-Simons and Wess-Zumino-Witten theories}, 
\newblock Commun. Math. Phys. {\bf 259} (2005) 577-613, 
\newblock [{\tt arXiv:math/0410013}].


\bibitem[CaDAFr]{CDF}
L.~Castellani, R.~D'Auria, and P.~Fr{\'e},
\newblock {\it Supergravity and Superstrings -- A geometric perspective},
\newblock World Scientific , Singapore, 1991.


\bibitem[CS]{CS}
S. S. Chern and J. Simons, {\it Characteristic forms and 
geometric invariants}, Ann. Math. {\bf (2) 99} (1974), 48--69.


\bibitem[DFM]{DFM}
E.~Diaconescu, D.~Freed, G.~Moore,
\newblock {\it The M-theory 3-form and E8 gauge theory}, \newline
\newblock [{\tt arXiv:hep-th/0312069}]

\bibitem[DMW]{DMW}
E. Diaconescu. G.  Moore, and E. Witten,
{\it $E_8$ gauge theory, and a derivation of K-theory from M-theory},
 Adv. Theor. Math. Phys. {\bf 6} (2003), 1031--1134, 
[{\tt  arXiv:hep-th/0005090}].



\bibitem[DLM]{DLM}
M. J. Duff, J. T. Liu, and R. Minasian,
{\it Eleven-dimensional origin of string-string duality: A one loop test},
Nucl. Phys. {\bf B452} (1995) 261-282, 
[{\tt arXiv:hep-th/9506126}].

\bibitem[DL]{dupont-ljungmann}
J. L. Dupont and R. Ljungmann,
{\it  Integration of simplicial forms and Deligne
cohomology}, Math. Scand. {\bf 97(1)} (2005), 11--39.


\bibitem[Ev]{Ev}
J. Evslin, 
{\it  From $E_8$ to $F$ via $T$},
	J. High Energy Phys.  {\bf 0408} (2004) 021,
[{\tt arXiv:hep-th/0311235}].	

\bibitem[Fal]{Falkowski}  
A.~Falkowski, 
\newblock {\it Five dimensional locally supersymmetric theories with branes}, 
\newblock Master Thesis, Warsaw   
\newline
\newblock {\verb"http://www.fuw.edu.pl/~afalkows/Work/Files/msct.ps.gz"}


\bibitem[FKPZ]{FKPZ}
S. Ferrara, A. Kehagias, H. Partouche, and A. Zaffaroni,
{\it Membranes and fivebranes with lower supersymmetry and their AdS supergravity duals},
	Phys. Lett. {\bf B431} (1998), 42--48,
	[{\tt arXiv:hep-th/9803109}].


\bibitem[Fi]{Fi}
V. Filippov,
{\it $n$-Lie algebras}, 
Siberian Math. J. {\bf 26} (1985), 879--891.

\bibitem[FRS11a]{FRSI}
D.~Fiorenza, C.~Rogers, U.~Schreiber, 
\newblock {\it A higher Chern-Weil derivation of AKSZ $\sigma$-models}, \newline
\newblock [{\tt arXiv:1108.4378}]


\bibitem[FiSaSc11]{FiSaSc}
D.~Fiorenza, H. Sati, and U.~Schreiber, 
\newblock {\it The higher moduli stacks of the $C$-field and its dual}, 
%in nonabelian differential cohomology}
 to appear.


\bibitem[FSS10]{FSS}
D.~Fiorenza, U.~Schreiber, and J.~Stasheff,
\newblock {\it {\v C}ech-cocycles for differential characteristic classes},
\newblock preprint [{\tt arXiv:1011.4735}].

\bibitem[Fr]{Freed}
D.~Freed, 
\newblock{\it Dirac charge quantization and generalized differential cohomology},
[{\tt arXiv:hep-th/0011220}].

\bibitem[Fre]{FreedCS}
D.~Freed,
\newblock {\it Classical Chern-Simons theory Part I}, 
\newblock Adv. Math. {\bf 113} (1995), 237--303,
%\newline
\newblock and
\newline
\newblock {\it Classical Chern-Simons theory, part II}, 
\newblock Houston J. Math. {\bf 28} (2002), 293--310, 
\newblock and
\newblock {\it Remarks on Chern-Simons theory}, 
\newblock Bull. AMS (NS) {\bf 46} (2009),  221--254.


\bibitem[FW]{FW}
D. Freed and E. Witten, 
\newblock {\it Anomalies in string theory with D-branes},
\newblock Asian J. Math {\bf 3}
\newblock (1999), 819--851, 
\newblock [{\tt arXiv:hep-th/9907189}].

\bibitem[GW]{gawedzki-waldorf}
K. Gawedzki and K. Waldorf,
\newblock {\it Polyakov-Wiegmann formula and multiplicative gerbes},
\newblock J. High Energy Phys.  {\bf 0909} (2009), 073,
\newblock [{\tt arXiv:0908.1130}].

\bibitem[GK]{GK}
T. Gherghetta and  A. Kehagias,
{\it Anomaly cancellation in seven-dimensional supergravity with a boundary},
	Phys.Rev. {\bf D68} (2003), 065019,
		[{\tt arXiv:hep-th/0212060}].

\bibitem[GPvN]{GPv}
F. Giani, M. Pernici and P. van Nieuwenhuizen, 
{\it Gauged $N=4$ $d = 6$ supergravity},
Phys. Rev. {\bf D 30} (1984), 1680--1687.



\bibitem[GT1]{gomi-terashima1} 
 K. Gomi and Y. Terashima, 
 {\it A fiber integration formula for the smooth Deligne cohomology}, 
 Internat. Math. Res. Notices {\bf 13} (2000), 699--708.

\bibitem[GT2]{gomi-terashima2} 
K. Gomi and Y. Terashima,
{\it Higher-dimensional parallel transports},
Math. Res. Lett. {\bf 8} (2001), 25--33.




%\bibitem[G1]{G1}
%A. Gustavsson,
%{\it The non-Abelian tensor multiplet in loop space},
%J. High Energy Phys. {\bf 0601} (2006) 165,
%[{\tt arXiv:hep-th/0512341}].
%
%
%\bibitem[G2]{G2}
%A. Gustavsson,
%{\it Loop space, $(2,0)$ theory, and solitonic strings},
%J. High Energy Phys. {\bf 0612} (2006) 066,
%[{\tt arXiv:hep-th/0608141}].
%

%\bibitem[G3]{G3}
%A. Gustavsson,
%{\it Selfdual strings and loop space Nahm equations},
%J. High Energy Phys. {\bf 0804} (2008) 083,
%	[{\tt arXiv:0802.3456}] [{\tt hep-th}].

\bibitem[Ha]{Ha}
J. A. Harvey, {\it TASI lectures on anomalies},
[{\tt arXiv:hep-th/0609097}].

\bibitem[HMM]{HMM}
J. A. Harvey, R. Minasian, and G. Moore,
{\it Non-abelian tensor-multiplet anomalies},
J. High Energy Phys.  {\bf 9809} (1998) 004,
[{\tt arXiv:hep-th/9808060}].

%\bibitem{HE}
%S. Hawking and G. Ellis, The large scale structure of spacetime, 
%Cambridge University Press, Cambridge, 1973.

\bibitem[HNS]{HNS}
M. Henningson, B. E. W. Nilsson, and P. Salomonson, 
{\it Holomorphic factorization of correlation functions in $(4k+2)$-dimensional 
$(2k)$-form gauge theory},
J. High Energy Phys.  {\bf 9909} (1999) 008,
[{\tt arXiv:hep-th/9908107}].

\bibitem[He]{He}
 M. Henningson,
 {\it The quantum Hilbert space of a chiral two-form in $d=5 + 1$ dimensions},
  J. High Energy Phys.  {\bf 0203} (2002) 021, [{\tt arXiv:hep-th/0111150}].

\bibitem[Hen]{Henriques}
A.~Henriques,
\newblock {\it Integrating {$L_\infty$}-algebras},
\newblock Compos. Math. {\bf 144} (2008), 1017--1045, \newline
\newblock [{\tt arXiv:math/0603563}] [{\tt math.AT}].

  
\bibitem[HS]{HopkinsSinger}
M. J. Hopkins and I. M. Singer, 
{\it Quadratic functions in geometry, topology, and M-theory},
J. Differential Geom. {\bf 70} (2005), no. 3, 329--452,
[{\tt arXiv:math/0211216}] [{\tt math.AT}].


\bibitem[HoWi]{HoravaWitten}
P.~Horava, E.~Witten, 
\newblock {\it Heterotic and Type I string dynamics from eleven dimensions}, 
\newblock Nucl. Phys. B460 (1996) 506
\newblock [{\tt arXiv:hep-th/9510209}]
\newblock and
\newblock {\it Eleven dimensional supergravity on a manifold with boundary}, 
\newblock Nucl. Phys. B475 (1996) 94,
\newblock [{\tt arXiv:hep-th/9603142}]


\bibitem[HSe]{HSe}
P. S. Howe and E. Sezgin, {\it $D = 11$, $p = 5$}, 
Phys. Lett. {\bf B 394} (1997), 62,
[{\tt arXiv:hep-th/9611008}]. 



\bibitem[Int]{In}
K. Intriligator,
{\it Anomaly matching and a Hopf-Wess-Zumino term in 
$6d$, $N=(2,0)$ field theories},
Nucl. Phys. {\bf B581} (2000), 257--273,
[{\tt arXiv:hep-th/0001205}].

\bibitem[La]{Laine}
K.~Laine,
\newblock {\it Geometric and topological aspects of Type IIB D-branes },
\newblock Master thesis, (2009)


\bibitem[LP]{LP}
N. Lambert and C. Papageorgakis,
{\it Nonabelian (2,0) tensor multiplets and 3-algebras}, \newline
[{\tt arXiv:1007.2982}] [{\tt hep-th}].


\bibitem[LR]{LR}
N. Lambert and P. Richmond,
{\it (2,0) supersymmetry and the light-cone description of M5-branes},
[{\tt arXiv:1109.6454}] [{\tt hep-th}].

\bibitem[Ma]{Ma}
J. M. Maldacena,
{\it The large $N$ limit of superconformal field theories and supergravity},
	Adv. Theor. Math. Phys. {\bf 2} (1998), 231--252,
		[{\tt arXiv:hep-th/9711200}].


\bibitem[MS]{MaSa}
V. Mathai and H. Sati,
{\it Some relations between twisted K-theory and $E_8$ gauge theory},
J. High Energy Phys.  {\bf 0403} (2004) 016, 
[{\tt arXiv:hep-th/0312033}].

\bibitem[Mo]{Moore} G. Moore,
\newblock {\it On the role of six-dimensional (2,0)-theories in recent developments in Physical Mathematics}
\newblock talk at \emph{Strings2011},
\newline
\newblock {\verb"http://www-conference.slu.se/strings2011/presentations/3 Wednesday/930_Moore.pdf"}

\bibitem[Mu]{Mu}
M. K. Murray, {\it Bundle gerbes},
 J. Lond. Math. Soc. {\bf 54} (1996), 403--416, 
[{\tt arXiv:dg-ga/9407015}].

\bibitem[NSW]{NSW}
T.~Nikolaus, C.~Sachse, C.~Wockel,
\newblock {\it A smooth model for the String group},
[{\tt arXiv:1104.4288}].


\bibitem[PaS]{PaS}
C. Papageorgakis and C. Saemann,
{\it The 3-Lie algebra (2,0) tensor multiplet and equations of motion on loop space},
J. High Energy Phys.  {\bf 1105} (2011), 099,
	[{\tt 	arXiv:1103.6192}][{\tt hep-th}].

\bibitem[PST]{PST}
P. Pasti, D. P. Sorokin, and M. Tonin,
{\it Covariant
action for a $D = 11$ five-brane with the chiral field},
 Phys. Lett. {\bf B 398} (1997), 41,
[{\tt arXiv:hep-th/9701037}]. 

\bibitem[PeS]{PeS}
M. Perry and J. H. Schwarz, {\it Interacting chiral gauge fields in six
dimensions and Born-Infeld theory},
 Nucl. Phys. {\bf B 489} (1997), 47,
  [{\tt arXiv:hep-th/9611065}].

\bibitem[PvNT]{PvNT}
K. Pilch, P. van Nieuwenhuizen, and 
P. K. Townsend,
{\it Compactification of $d=11$ supergravity on $S^4$ (or $11= 7 + 4$, too)},
Nucl. Phys. {\bf B 242} (1984), 377--392.


\bibitem[PS]{PresslySegal}
A. Pressly and G. Segal, 
Loop Groups, 
Oxford University Press, Oxford, UK, 1988.

\bibitem[Ru]{Ruffino}
F.~Ruffino, 
\newblock {\it Classifying $A$-field and $B$-field configurations in the presence of D-branes - Part II: Stacks of D-branes},
\newblock to appear in Nucl. Phys. {\bf B} (2012), 
\newblock [{\tt arXiv:1104.2798}] [{\tt hep-th}].

\bibitem[SaSz]{SS} 
C. Saemann and R. J. Szabo,
{\it Quantization of 2-plectic manifolds},
[{\tt arXiv:1106.1890}][{\tt hep-th}]. 

\bibitem[SSW]{SSW}
H. Samtleben, E. Sezgin, and R. Wimmer,
{\it (1,0) superconformal models in six dimensions}, \newline
[{\tt arXiv:1108.4060}] [{\tt hep-th}].





\bibitem[Sa06]{KSpin}
H. Sati, 
{\it An approach to anomalies in M-theory via KSpin},
 J. Geom. Phys. {\bf 58} (2008), 387--401,
 [{\tt 	arXiv:0705.3484}] [{\tt hep-th}].

\bibitem[Sa07]{S-loop}
H. Sati,
{\it The loop group of $E_8$ and targets for spacetime},
	Mod. Phys. Lett. {\bf A24} (2009), 25--40,
	[{\tt 	arXiv:hep-th/0701231}].


\bibitem[Sa10a]{S-gerbe}
H. Sati, 
{\it E8 gauge theory and gerbes in string theory},
Adv. Theor. Math. Phys. {\bf 14} (2010), 1--39,
[{\tt arXiv:hep-th/0608190}].

\bibitem[Sa10b]{Sati10}
H.~Sati,
\newblock {\it Geometric and topological structures related to M-branes},
\newblock Proc. Symp. Pure Math. {\bf 81} (2010), 181-236,
\newblock [{\tt arXiv:1001.5020}] [{\tt math.DG}].

\bibitem[Sa10c]{Sati10Twist}
H.~Sati,
\newblock {\it Geometric and topological structures related to M-branes II:
 Twisted $\mathrm{String}$- and $\mathrm{String}^c$-structures},
  J. Australian Math. Soc. {\bf 90} (2011), 93--108,
\newblock [{\tt arXiv:1007.5419}] [{\tt hep-th}].

\bibitem[Sa11a]{tw}
H. Sati, 
{\it Twisted topological structures related to M-branes},
 Int. J. Geom. Meth. Mod. Phys. {\bf 8} (2011), 1097--1116,
 	[{\tt arXiv:1008.1755}] [{\tt hep-th}].

\bibitem[Sa11b]{S-String}
H. Sati, 
{\it 
Anomalies of $E_8$ gauge theory on String manifolds},
 Int. J. Mod. Phys. {\bf A26} (2011), 2177--2197,
 	[{\tt arXiv:0807.4940}] [{\tt hep-th}].



\bibitem[Sa11c]{II}
H. Sati, 
{\it Twisted topological structures related to M-branes II: Twisted Wu and Wu${}^c$ 
structures },
	[{\tt arXiv:1109.4461}] [{\tt hep-th}].

\bibitem[Sa11d]{NS5}
H. Sati, 
{\it Topological aspects of the partition function of the NS5-brane},
	[{\tt arXiv:1109.4834}] [{\tt hep-th}]. 


\bibitem[Sa12]{Hodge}
H. Sati, 
Duality and cohomology in M-theory with boundary,
to appear in J. Geom. Phys., 
[{\tt arXiv:1012.4495}] [{\tt hep-th}].


\bibitem[SSS09a]{SSSI}
H.~Sati, U.~Schreiber, and J.~Stasheff,
\newblock {\it {L}$_\infty$-algebra connections and applications to String- and Chern-Simons $n$-transport},
\newblock In {\em Recent developments in {Quantum Field Theory}}, Birkh\"auser, 2009,
	[{\tt arXiv:0801.3480}] [{\tt math.DG}].

\bibitem[SSS09b]{SSSII}
H.~Sati, U.~Schreiber, and J.~Stasheff,
\newblock {\it Fivebrane structures},
Rev. Math. Phys. {\bf 21} (2009), 1197--1240,
	[{\tt arXiv:0805.0564}] [{\tt math.AT}].

\bibitem[SSS09c]{SSSIII}
H.~Sati, U.~Schreiber, and J.~Stasheff,
\newblock {\it Twisted differential String- and Fivebrane structures},
\newblock Communications in Mathematical Physics (2012)
\newblock [{\tt{arXiv:0910.4001}}].


\bibitem[Sc-P]{Schommer-Pries}
C.~Schommer-Pries,
\newblock Central extensions of smooth 2-groups and a finite-dimensional string 2-group,
\newblock [{\tt arXiv:0911.2483}].


\bibitem[Sch05]{Urs-thesis}
U. Schreiber, 
{\it From loop space mechanics to nonabelian strings},
PhD thesis,  Universit\"at Duisburg-Essen, 2005.

\bibitem[Sch11]{survey}
U.~Schreiber,
\newblock {\it Differential cohomology in a cohesive $\infty$-topos},
\newblock  Habilitation, Hamburg University (2011)
\newblock \newline {\verb"http://ncatlab.org/schreiber/show/differential+cohomology+in+a+cohesive+topos"}


 \bibitem[SW1]{SchreiberWaldorfII}
U.  Schreiber and K. Waldorf, 
{\it  Smooth functors vs. differential forms},
 Homology, Homotopy Appl., {\bf 13(1)} (2011), 143--203,
[{\tt arXiv:0802.0663}] [{\tt math.DG}].
 
\bibitem[SW2]{SchreiberWaldorfIII} 
 U. Schreiber and K. Waldorf,
{\it Connections on non-abelian gerbes and their holonomy},
[{\tt arXiv:0808.1923}] [{\tt math.DG}]. 


\bibitem[Sch]{Sch}
J. H. Schwarz, {\it Coupling a self-dual tensor to gravity in six dimensions},
 Phys. Lett. {\bf B 395} (1997), 191,
 [{\tt arXiv:hep-th/9701008}]. 


\bibitem[SW]{SW}
N. Seiberg and E. Witten, 
{\it String theory and noncommutative geometry}, 
J. High Energy Phys.  {\bf 09} (1999) 032, 
[{\tt arXiv:hep-th/9908142}].

\bibitem[St]{St}
A. Strominger, 
{\it Open $p$-branes},
Phys. Lett. {\bf B383} (1996), 44--47,
[{\tt arXiv:hep-th/9512059}].


\bibitem[To]{To}
P. K. Townsend, 
{\it D-branes from M-branes}, Phys. Lett. {\bf B373} (1996), 68--75,
[{\tt arXiv:hep-th/9512062}].

\bibitem[VW]{VW}
C. Vafa and E. Witten,
{\it A one-loop test of string duality},
Nucl. Phys. {\bf B447} (1995) 261--270,
[{\tt arXiv:hep-th/9505053}].


\bibitem[Wa09]{Waldorf}
K.~Waldorf,
\newblock {\it String connections and {C}hern-{S}imons theory}, (2009)
\newblock [{\tt{arXiv:0906.0117}}].

\bibitem[Wa12]{WaldorfString}
K.~Waldorf, 
\newblock {\it A Construction of String 2-Group Models using a Transgression-Regression Technique},
\newblock [{\tt arXiv:1201.5052}]

\bibitem[Wi86]{Witten86}
E. Witten, 
\newblock {\it The Index of the Dirac Operator in Loop Space},
\newblock in   Elliptic Curves and Modular Forms in Algebraic Topology, 
\newblock   Lecture  Notes in Mathematics Vol. 1326,  Springer, Berlin,  (1986).

\bibitem[Wi89]{WittenCS}
E.~Witten,
\newblock {\it Quantum field theory and the Jones polynomial},
\newblock Commun. Math. Phys. {\bf 121 (3)} (1989) 351--399.

\bibitem[Wi96]{Witten96}
E.~Witten, 
\newblock {\it Five-brane effective action in M-theory}, 
\newblock J. Geom. Phys. {\bf 22} (1997),  103--133,
\newblock [{\tt arXiv:hep-th/9610234}].

\bibitem[Wi97]{WittenFluxQuantization}
E.~Witten,
\newblock {\it On flux quantization in M-theory and the effective action},
\newblock J. Geom. Phys. {\bf 22} (1997), 1--13, 
\newblock [{\tt arXiv:hep-th/9609122}].

\bibitem[Wi98a]{Witten98a}
E.~Witten, 
\newblock {\it Anti-de Sitter space and holography}, 
\newblock {Adv. Theor. Math. Phys.},
\newblock {\bf 2} (1998),  253--291, 
\newblock [{\tt arXiv:hep-th/9802150}].

\bibitem[Wi98b]{Witten98} 
E.~Witten,
\newblock {\it AdS/CFT correspondence and topological field theory},
\newblock J. High Energy Phys.  {\bf 9812} (1998), 012,
\newblock [{\tt arXiv:hep-th/9812012}].

\bibitem[Wi04]{Witten04}
E.~Witten, 
\newblock {\it Conformal field theory in four and six dimensions},
\newblock U.~Tillmann (ed.),
\newblock {Topology, geometry and quantum field theory}
\newblock LMS Lecture Note Series (2004),
\newblock [{\tt arXiv:0712.0157}].



\bibitem[Wi09]{Witten09}
E.~Witten, 
\newblock {\it Geometric Langlands from six dimensions},
\newblock [{\tt arXiv:0905.2720}] [{\tt hep-th}].

\bibitem[Wi11]{Witten11}
E.~Witten,
\newblock {\it Fivebranes and knots},
\newblock [{\tt arXiv:1101.3216}] [{\tt hep-th}].

%\bibitem[ZWZ]{ZWZ}
%B. Zumino, Y.-S. Wu, and A. Zee, 
%{\it Chiral anomalies, higher dimensions, and differential geometry},
%Nucl. Phys.{\bf B239} (1984) 477-507.



\end{thebibliography}
\end{document}